\documentclass[10pt, final, twocolumn, twoside]{IEEEtran}
\usepackage{cite}
\usepackage{amssymb}
\usepackage{psfrag,graphicx,subfigure,textcomp}
\usepackage[dvips]{color}
\usepackage{epsf,epsfig,pstricks,pst-plot}
\usepackage{stfloats}
\usepackage[cmex10]{amsmath}
\newtheorem{defi}{Definition}
\newtheorem{thm}{Theorem}
\newtheorem{cor}{Corollary}

\newtheorem{pro}{Proposition}
\newtheorem{remrk}{Remark}

\newcommand\Matrix[2]
{   \left[ \begin{array}{#1}
        #2
    \end{array} \right]
}

\newcommand\Diag[1] {   \text{diag} \left\{ #1 \right\}}

\newcommand\A[2]  { \alpha_{#1}^{(#2)} }
\newcommand\D[2]  { \tau_{#1}^{(#2)}  }
\newcommand\BB[2]  { b_{#1}^{(#2)}  }

\newcommand\E     {  \mathbb{E} }
\newcommand\PEB   {  \mathcal{P} }
\newcommand\V[1]  {  \mathbf{#1} }
\newcommand\B[1]  {  \boldsymbol{#1} }

\newcommand\NB  { { \mathcal{N}_\text{b} } }
\newcommand\Nb  { { N_\text{b} } }
\newcommand\NA  { { \mathcal{N}_\text{a} } }
\newcommand\Na  { { N_\text{a} } }

\newcommand\SNR[2]  {   \mathsf{SNR}_{#1}^{(#2)}   }

\newcommand\JTH {  \V{J}_{\B\theta} }
\newcommand\FTR {  f(\V{r}|{\B\theta}) }

\newcommand\gp   {f}

\newcommand\gk   {f}

\newcommand\FIM[4] {\V{F}_{#1}({#2}\,;{#3},{#4})}

\newcommand\JE[1] {  {\V{J}_\text{e}(#1)} }
\newcommand\JD   {  {\V{J}_\text{w}} }
\newcommand\JP   {  {\V{J}_\text{p}} }

\newcommand\I   {  \V{I} }

\newcommand\q   {  \V{q} }
\newcommand\R   {  \V{J}_\text{r} }

\newcommand\NL  { { \mathcal{N}_\text{L} } }
\newcommand\NNL { { \mathcal{N}_\text{NL} }  }
\newcommand\HL  {  \V{T}_{\text{L}} }
\newcommand\HNL {  \V{T}_{\text{NL}} }

\newcommand\LL    {  \B{\Lambda}_{\text{L}} }
\newcommand\BLL   {  \bar{\B{\Lambda}}_{\text{L}} }
\newcommand\LNL   {  \B{\Lambda}_{\text{NL}} }
\newcommand\BLNL  {  \bar{\B{\Lambda}}_{\text{NL}} }

\newcommand\GP  { \B\Xi_{\V{p}} }
\newcommand\Gph { \Xi_\varphi}

\newcommand\Gkk[1] {\B\Xi_{\B\kappa,\B\kappa}^{#1}}
\newcommand\Gpp[1] {\B\Xi_{\V{p,p}}^{#1}}
\newcommand\Gpk[1] {\B\Xi_{\V{p},\B\kappa}^{#1}}
\newcommand\Gvk[1] {\B\Xi_{\varphi,\B\kappa}^{#1}}

\newcommand\GpkT[1]{\left[{\B\Xi_{\V{p},\B\kappa}^{#1}}\right]^{\text{T}}}

\newcommand\tGpk[1] {{\B\Xi}_{d,\B\kappa}^{#1}}
\newcommand\tGpp[1] {{\Xi}_{d,d}^{#1}}

\begin{document}


\title{Fundamental Limits of Wideband Localization---\\Part I: A General Framework}

\author{Yuan~Shen,~\IEEEmembership{Student~Member,~IEEE,}
and Moe~Z.~Win,~\IEEEmembership{Fellow,~IEEE}
\thanks{Manuscript received April 17, 2008; revised October 10, 2008. Current version published Month Day 2010. This research was supported, in part, by the National Science Foundation under Grant ECCS-0901034, the Office of Naval Research Presidential Early Career Award for Scientists and Engineers (PECASE) N00014-09-1-0435, and MIT Institute for Soldier Nanotechnologies. The  paper was presented in part at the IEEE Wireless Communications and Networking Conference, Hong Kong, March, 2007.}
\thanks{The authors are with the Laboratory for Information and Decision Systems (LIDS), Massachusetts Institute of Technology, 77 Massachusetts Avenue, Cambridge, MA 02139 USA (e-mail: \{shenyuan, moewin\}@mit.edu).}
\thanks{Communicated by Holger Boche, Associated Editor for Communications.}
\thanks{Color versions of the figures in this paper are available online at http://ieeexplore.ieee.org.}
\thanks{Digital Object Identifier XXX.XXX}
}

\markboth{IEEE Transactions on Information Theory, Vol.~X, No.~Y,
Month~2010}{Shen and Win: Fundamental Limits of Wideband
Localization---Part I: A General Framework}

\maketitle


\begin{abstract}\label{Sec:Abstract}
The availability of positional information is of great importance in
many commercial, public safety, and military applications. The
coming years will see the emergence of location-aware networks with
sub-meter accuracy, relying on accurate range measurements provided
by wide bandwidth transmissions. In this two-part paper, we determine the fundamental limits of localization accuracy of wideband wireless networks in harsh multipath environments. We first develop a general framework to characterize the localization accuracy of a given node here and then extend our analysis to cooperative location-aware networks in Part II.

In this paper, we characterize localization accuracy in terms of a performance measure called the squared position error bound (SPEB), and introduce the notion of equivalent Fisher information to derive the SPEB in a succinct expression. This methodology provides insights into the essence of the localization problem by unifying localization
information from individual anchors and information from a priori
knowledge of the agent's position in a canonical form. Our analysis
begins with the received waveforms themselves rather than utilizing
only the signal metrics extracted from these waveforms, such as
time-of-arrival and received signal strength. Hence, our framework
exploits all the information inherent in the received waveforms, and
the resulting SPEB serves as a fundamental limit of localization
accuracy.
\end{abstract}


%
%

\section{Introduction}\label{Sec:Intr}

Location-awareness plays a crucial role in many wireless network
applications, such as localization services in next generation
cellular networks \cite{SayTarKha:05}, search-and-rescue operations
\cite{PahLiMak:02, CafStu:98}, logistics
\cite{PatAshKypHerMosCor:05}, and blue force tracking in
battlefields \cite{ChoKum:03}. The Global Positioning System (GPS)
is the most important technology to provide location-awareness
around the globe through a constellation of at least 24 satellites
\cite{Spi:78,Kap:06}. However, the effectiveness of GPS is limited
in harsh environments, such as in buildings, in urban canyons, under
tree canopies, and in caves
\cite{JouDarWin:J08,GezTiaGiaKobMolPooSah:05}, due to the inability
of GPS signals to penetrate most obstacles.  Hence, new localization
techniques are required to meet the increasing need for accurate
localization in such harsh environments \cite{JouDarWin:J08,
GezTiaGiaKobMolPooSah:05}. 

Wideband wireless networks are capable of providing accurate
localization in GPS-denied environments
\cite{GezTiaGiaKobMolPooSah:05, JouDarWin:J08, FalDarMucWin:J06,
SheWin:J10, Mol:09}. Wide bandwidth or ultra-wide bandwidth (UWB) signals are particularly well-suited for localization, since they can
provide accurate and reliable range (distance) measurements due to their fine delay resolution and robustness in harsh environments
\cite{WinSch:L98, WinSch:J02, Win:J02, LowCheLawNgLee:05,  ZhaLawGua:05, DarChoWin:J08, DarConFerGioWin:J09, LeeSch:02}. For more information about UWB, we refer the reader to \cite{Mol:05, WinChrSol:J00, YanGia:04a, QueWin:J05, SuwWinShe:J05, MolCasChoEmaForKanKarKunSchSiwWin:J06}.

\begin{figure}[t]
    \begin{center}
	    \psfrag{1}[c][][1.3]{1}
	\psfrag{2}[c][][1.3]{\hspace{-1mm}2}
	    \psfrag{C}[c][][1.3]{\hspace{-1mm}C}
	    \psfrag{D}[c][][1.3]{ D}
    \psfrag{A}[c][][1.3]{A}
    \psfrag{B}[c][][1.3]{B}
    {\includegraphics[angle=0,width=0.70\linewidth,draft=false]{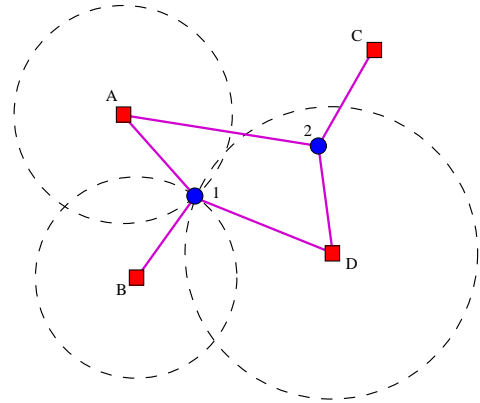}}

    \caption{\label{fig:Sen_Net} Location-aware networks:
    the anchors (A, B, C, and D) communicate with the agents (1 and 2), and each edge denotes a connection link between anchor and agent.}
    \end{center}
\end{figure}

Location-aware networks generally consist of two kinds of nodes: anchors and agents. Anchors have known positions (for example, through GPS or
system design), while agents have unknown positions and attempt to
determine their positions (see Fig.~\ref{fig:Sen_Net}). Each node is
equipped with a wideband transceiver, and localization is
accomplished through the use of radio communications between agents and their neighboring anchors.
Localizing an agent requires a number of signals transmitted from
the anchors, and the relative position of the agent can be inferred
from these received waveforms using a variety of signal metrics.
Commonly used signal metrics include time-of-arrival (TOA)
\cite{DarChoWin:J08, DarConFerGioWin:J09, JouDarWin:J08, LeeSch:02,
Mai:08, GezTiaGiaKobMolPooSah:05, ZhaLawGua:05,QiKob:02a, QiSudKob:04,QiKobSud:06}, time-difference-of-arrival (TDOA) \cite{Caf:00,
RapReeWoe:96}, angle-of-arrival (AOA)
\cite{NicNat:03,GezTiaGiaKobMolPooSah:05}, and received signal
strength (RSS) \cite{PatHer:03,GezTiaGiaKobMolPooSah:05,
PavCosMazConDar:06}.

Time-based metrics, TOA and TDOA, are obtained by measuring the
signal propagation time between nodes. In ideal scenarios, the
estimated distance equals the product of the known propagation speed
and the measured signal propagation time. The TOA metric gives
possible positions of an agent on a circle with the anchor at the
center, and it can be obtained by either the one-way time-of-flight
of a signal in a synchronized network \cite{DarChoWin:J08,
DarConFerGioWin:J09, QiKob:02a, QiSudKob:04}, or the round-trip
time-of-flight in a non-synchronized network
\cite{MolCasChoEmaForKanKarKunSchSiwWin:J06, Mol:B05}.
Alternatively, the TDOA metric provides possible positions of an
agent on the hyperbola determined by the difference in the TOAs from
two anchors located at the foci. Note that TDOA techniques require
synchronization among anchors but not necessarily with the agent.

Another signal metric is AOA, the angle at which a signal arrives at
the agent. The AOA metric can be obtained using an array of
antennas, based on the signals' TOA at each antenna.\footnote{The
AOA metric can be obtained in two ways, directly through measurement
by a directional antenna, or indirectly through TOA measurements
using an antenna array \cite{VanBuc:88, KriVib:96, Eng:83,
ChoTanLauMcLBeaNix:03}. Wideband directional antennas that satisfy
size and cost requirements are difficult to implement, since they
are required to perform across a large bandwidth \cite{Mol:B05}. As
such, antenna arrays are more commonly used when angle measurement
for wide bandwidth signals is necessary.} The use of AOA for
localization has been investigated, and many hybrid systems have
been proposed, including hybrid TOA/AOA systems \cite{DenFan:00,
QiKobSud:06}, and hybrid TDOA/AOA systems \cite{ConZhu:02}. However,
some of these studies employ narrowband signal models, which are not
applicable for wideband antenna arrays. Others are restricted to
far-field scenarios or use far-field assumptions.

RSS is also a useful metric that uses the strength of the received
signal to estimate the propagation distance between nodes
\cite{GezTiaGiaKobMolPooSah:05, PatHer:03, Mol:B05}. This metric can
be measured during the data communications using low-complexity
circuits. Although widely implemented, RSS has limited accuracy due
to the difficulty in precisely modeling the relationship between the
RSS and the propagation distance \cite{GezTiaGiaKobMolPooSah:05,
PatAshKypHerMosCor:05}.

Note that the signal metrics extracted from the received waveforms
may discard relevant information for localization. Moreover, models
for the signal metrics depend heavily on the specific measurement
processes.\footnote{For instance, the error of the TOA metric is
commonly modeled as an additive Gaussian random variable
\cite{ChaSah:04, QiKobSud:06, JouDarWin:J08}. This model contradicts
the studies in \cite{DarChoWin:J08, DarConFerGioWin:J09, LeeSch:02,
HamSch:92, LeeYoo:05}, and the experimental results in
\cite{JouDarWin:J08, LowCheLawNgLee:05}.} Therefore, in deriving the
fundamental limits of localization accuracy, it is necessary to
utilize the received waveforms rather than the signal metrics
extracted from the waveforms \cite{QiKob:02a, QiSudKob:04,
SheWin:C07, SheWin:C08}.

Since the received waveforms are affected by random phenomena such
as noise, fading, shadowing, multipath, and non-line-of-sight (NLOS)
propagations \cite{SilRan:96, CafStu:95}, the agents' position
estimates are subject to uncertainty. The Cram\'{e}r-Rao bound (CRB)
sets a lower bound on the variance of estimates for deterministic
parameters \cite{Tre:68, Poo:B94}, and it has been used as a
performance measure for localization accuracy \cite{Spi:01}.
However, relatively few studies have investigated the effect of
multipath and NLOS propagations on localization accuracy. Multipath
refers to a propagation phenomenon in which a transmitted signal
reaches the receive antenna via multiple paths. The superposition of
these arriving paths results in fading and interference. NLOS
propagations, created by physical obstructions in the direct path,
produce a positive bias in the propagation time and decrease the
strength of the received signal, which can severely degrade the
localization accuracy. Several types of methods have been proposed
to deal with NLOS propagations: 1) treat NLOS biases as additive
noise injected in the true propagation distances
\cite{JouDarWin:J08, GavFog:90, KooGruCed:98};\footnote{In practice,
however, a NLOS induced range bias can be as much as a few
kilometers depending on the propagation environment \cite{SilRan:96,
WylHol:96}, and small perturbation may not compensate for NLOS
induced error.} 2) identify and weigh the importance of NLOS
signals for localization \cite{WylHol:96, WooYouKoh:00, AlJCaf:02,
Che:99, Xio:98, Wan:04}; or 3) consider NLOS biases as parameters to
be estimated \cite{SheWin:C07, SheWin:C08, QiKob:02a, QiSudKob:04, WaxLes:97, QiKobSud:06, BotHosFat:04, Mai:08}. The authors in \cite{GezTiaGiaKobMolPooSah:05, QiKob:02a, QiSudKob:04,
JouDarWin:J08} showed that NLOS signals do not improve localization
accuracy unless a priori knowledge of the NLOS biases is available,
but their results were restricted to specific models or
approximations. Moreover, detailed effects of multipath propagations
on localization accuracy remains under-explored.

In this paper, we develop a general framework to determine the localization accuracy of wireless networks.\footnote{In Part II \cite{SheWymWin:J10}, we extend our analysis to cooperative location-aware networks.} Our analysis begins with the received waveforms themselves rather than utilizing only signal metrics extracted from the waveforms, such as TOA, TDOA, AOA, and RSS. The main contributions of this paper are as follows:

\begin{itemize}
\item
We derive the fundamental limits of localization accuracy for wideband wireless networks, in terms of a performance measure
called the \emph{squared position error bound} (SPEB), in the
presence of multipath and NLOS propagation.
\item
We propose the notion of \emph{equivalent Fisher information} (EFI) to derive the agent's localization information. This approach unifies such information from different anchors in a canonical form as a weighed sum of the direction matrix associated with individual anchors with the weights characterizing the information intensity.
\item
We quantify the contribution of the \emph{a priori knowledge} of the channel parameters and agent's position to the agent's localization information, and show that NLOS components can be beneficial when a priori channel knowledge is available.
\item
We derive the performance limits for localization systems employing \emph{wideband antenna arrays}. The AOA metric obtained from antenna arrays are shown not to further improve the localization accuracy beyond that provided by TOA metric alone.
\item
We quantify the effect of \emph{clock asynchronism} between anchors and agents on localization accuracy for networks where nodes employ a single antenna or an array of antennas.
\end{itemize}

The rest of the paper is organized as follows. Section \ref{Sec:Model} presents the system model, the notion of the SPEB, and the Fisher information matrix (FIM) for the SPEB. In Section \ref{Sec:Evalu}, we introduce the notion of EFI and show how it can help the derivation of the SPEB. In Section \ref{Sec:Array}, we investigate the performance of localization systems employing wideband antenna arrays. Section \ref{Sec:ClockBias} investigates the effect of clock asynchronism between anchors and agents. Discussions are provided in Section \ref{Sec:Discus}. Finally, numerical illustrations are given in Section \ref{Sec:Simu}, and conclusions are drawn in the last section.

\subsubsection*{Notations} The notation $\E_{\V{z}}\{\cdot\}$ is the expectation operator with respect to the random vector $\V{z}$; $\V{A} \succeq \V{B}$ denotes that the matrix $ \V{A} - \V{B} $ is positive semi-definite; $\text{tr}\{\cdot\}$ is the trace of a square matrix; $\left[\, \cdot \, \right]_{n \times n}$ denotes the upper left $n \times n$ submatrix of its argument; $[\, \cdot \,]_{n,m}$ is the element at the $n$th row and $m$th column of its argument; $\|\cdot\|$ is the Euclidean norm of its argument; and the superscripts $[\,\cdot\,]^\text{T}$ represents the transpose of its argument. We denote by $f(\V{x})$ the probability density function (PDF) $f_\V{X}(\V{x})$ of the random vector $\V{X}$ unless specified otherwise, and we also use in the paper the following function for the FIM:
    \begin{align*}
        \FIM{\V{z}}{\V{w}}{\V{x}}{\V{y}} \triangleq
            \E_{\V{z}} \left\{
            \left[\frac{\partial}{\partial \V{x}} \ln f(\V{w}) \right]
            \left[ \frac{\partial}{\partial \V{y}} \ln f(\V{w})
            \right]^\text{T} \right\},
    \end{align*}
where $\V{w}$ can be either a vector or a symbol.\footnote{For example,  $\V{w}$ is replaced by symbol $\V{r}|\B\theta$ in the case that $f(\cdot)$ is a conditional PDF of $\V{r}$ given $\B\theta$.}

%
%

\section{System Model}\label{Sec:Model}

In this section, we describe the wideband channel model \cite{WinSch:J02, QueWin:J05, MolCasChoEmaForKanKarKunSchSiwWin:J06, Mol:05, SalVal:87}, formulate the problem, and briefly review the information inequality and Fisher information. We also introduce the squared position error bound, which is a fundamental limit of localization accuracy.

\subsection{Signal Model}\label{Sec:Model_System}

Consider a wireless network consisting of $\Nb$ anchors and multiple
agents. Anchors have perfect knowledge of their positions, and each agent attempts to estimate its position based on the received waveforms
from neighboring anchors (see
Fig.~\ref{fig:Sen_Net}).\footnote{Agents estimate their positions
independently, and hence without loss of generality, our analysis
focuses on one agent.} Wideband signals traveling from anchors to
agents are subject to multipath propagation.

Let $\V{p} \in \mathbb{R}^2$ denote the position of the
agent,\footnote{We first focus on two dimensional cases and then
extend the results to three-dimensional cases where $\V{p}\in
\mathbb{R}^3$.} which is to be estimated. The set of anchors is
denoted by $\NB = \{1,2,\dotsc,\Nb \} \triangleq \NL \cup \NNL$,
where $\NL$ denotes the set of anchors that provide LOS signals to
the agent and $\NNL$ denotes the set of remaining anchors that
provide NLOS signals to the agent. The position of anchor $k$ is
known and denoted by $\V{p}_k \in \mathbb{R}^2$ ($k \in \NB$). Let
$\phi_{k}$ denote the angle from anchor $k$ to the agent, i.e.,
    \begin{align*}
        \phi_{k} = \tan^{-1}\frac{y - y_k}{x - x_k}  \,,
    \end{align*}
where $\V{p} \triangleq [\,x \;\; y\,]^\text{T}$ and $\V{p}_k
\triangleq [\, x_k \;\; y_k \,]^\text{T}$.

The received waveform at the agent from anchor $k$ can be written as
    \begin{align}\label{eq:Model_Multipath}
        r_k(t)=\sum_{l=1}^{L_k} \A{k}{l} \, s\left(t - \D{k}{l} \right) + z_k(t)\, ,
        \quad  t \in [ \, 0, T_{\text{ob}})\,,
    \end{align}
where $s(t)$ is a known wideband waveform whose Fourier transform is
denoted by $S(f)$, $\A{k}{l}$ and $\D{k}{l}$ are the amplitude and
delay, respectively, of the $l$th path, $L_k$ is the number of
multipath components (MPCs), 
$z_k(t)$ represents the observation noise modeled as additive white
Gaussian processes with two-side power spectral density $N_0/2$, and
$[\, 0,T_\text{ob})$ is the observation interval. The relationship
between the agent's position and the delays of the propagation paths
is
    \begin{align}\label{eq:Model_Tau_L}
        \D{k}{l} = \frac{1}{c} {\Big[} \, \| \V{p} - \V{p}_k \| + \BB{k}{l} \, {\Big ]}
        \,,
    \end{align}
where $c$ is the propagation speed of the signal, and $\BB{k}{l} \geq 0$ is a
range bias. The range bias $\BB{k}{1} = 0$ for LOS propagation, whereas
$\BB{k}{l} > 0$ for NLOS propagation.\footnote{LOS propagation does
not introduce a range bias because there is an unblocked direct
path. NLOS propagation introduces a positive range bias because such
signals either reflect off objects or penetrate through obstacles.
In this paper, received signals whose first path undergoes LOS
propagation are referred to as LOS signals, otherwise these signals
are referred to as NLOS signals.}

\subsection{Error Bounds on Position Estimation}\label{Sec:Model_SPEB}

Our analysis is based on the received waveforms given by
\eqref{eq:Model_Multipath}, and hence the parameter vector
$\B\theta$ includes the agent's position and the nuisance multipath
parameters \cite{GezTiaGiaKobMolPooSah:05, BotHosFat:04}, i.e.,
    \begin{align*}
        \B\theta = \Matrix{ccccc}{\V{p}^\text{T} &
        \B{\kappa}_1^\text{T} & \B{\kappa}_2^\text{T}
        & \cdots & \B{\kappa}_{\Nb}^\text{T} }^\text{T},
    \end{align*}
where $\B{\kappa}_k$ is the vector of the multipath parameters
associated with $r_k(t)$, given by
    \begin{align*}
        \renewcommand{\arraystretch}{1.3}
        \B{\kappa}_k  =
        \begin{cases}
            \qquad \;\; \Matrix{cccccc}{\A{k}{1} & \BB{k}{2} & \A{k}{2}
            & \cdots & \BB{k}{L_k} & \A{k}{L_k}}^\text{T}, \\
            \hfill  k \in \NL \,, \\
            \Matrix{ccccccc}{ \BB{k}{1} & \A{k}{1} & \BB{k}{2} & \A{k}{2}
            & \cdots & \BB{k}{L_k} & \A{k}{L_k}}^\text{T}, \\
            \hfill k \in \NNL \,.
        \end{cases}
    \end{align*}
Note that $\BB{k}{1} = 0$ for $k\in\NL$ and is excluded from
$\B{\kappa}_k$.

We introduce $\V{r}$ as the vector representation of all the
received waveforms $r_k(t)$, given by
    \begin{align*}
        \V{r} = \Matrix{cccc} { \V{r}_1^\text{T}   & \V{r}_2^\text{T} &
        \cdots  & \V{r}_{\Nb}^\text{T}}^\text{T}
        \, ,
    \end{align*}
where $\V{r}_k$ is obtained from the Karhunen-Loeve expansion of
$r_k(t)$ \cite{Tre:68, Poo:B94}. Let $\B{\hat\theta}$ denote an
estimate of the parameter vector $\B{\theta}$ based on observation
$\V{r}$. The mean squared error (MSE) matrix of $\B{\hat\theta}$
satisfies the information inequality \cite{Tre:68, ReuMes:97,
Poo:B94}
    \begin{align}\label{eq:Model_CRLB_FIM}
        \E_{\V{r},\B\theta} \left\{ \B{(\hat\theta-\theta) (\hat\theta-\theta)}^\text{T} \right\}
        \succeq \JTH^{-1} .
    \end{align}
where $\JTH$ is the Fisher information matrix (FIM) for the
parameter vector $\B\theta$.\footnote{When a subset of parameters is
random, $\JTH$ is called the Bayesian information matrix. Inequality
\eqref{eq:Model_CRLB_FIM} also holds under some regularity
conditions and provides lower bound on the MSE matrix of any
unbiased estimates of the deterministic parameters and any estimates
of the random parameters \cite{Tre:68, ReuMes:97}. With a slight
abuse of notation, $\E_{\V{r},\B\theta}\{\cdot\}$ will
be used for deterministic, hybrid, Bayesian cases with the
understanding that the expectation operation is not performed over
the deterministic components of $\B\theta$.} Let $\hat{\V{p}}$ be an
estimate of the agent's position, and it follows from
\eqref{eq:Model_CRLB_FIM} that\footnote{Note that for
three-dimensional localization, we need to consider a $3 \times 3$
matrix ${\big[} \JTH^{-1} {\big]}_{3 \times 3}$.}
    \begin{align*}
        \E_{\V{r}, \B\theta} \left\{ ( \hat{\V{p}} - \V{p} )
        ( \hat{\V{p}} - \V{p} )^\text{T} \right\}
        \succeq  \left[ \JTH^{-1} \right]_{2 \times 2},
    \end{align*}
and hence
    \begin{align}\label{eq:Sing_PEB_def}
        \E_{\V{r},\B\theta} \left\{ \| \hat{\V{p}}-\V{p}\|^2
        \right\} \geq \text{tr} \left\{ \left[ \JTH^{-1} \right]_{2 \times 2} \right\}.
    \end{align}
Therefore, we define the right hand side of \eqref{eq:Sing_PEB_def}
as a measure to characterize the limits of position accuracy as
follows.

\begin{defi}[Squared Position Error Bound]
The squared position error bound (SPEB) is defined to be
    \begin{align*}
        \PEB(\V{p}) \triangleq \text{tr} \left\{ \left[ \JTH^{-1} \right]_{2 \times 2} \right\}.
    \end{align*}
\end{defi}

\subsection{Fisher Information Matrix}\label{Sec:Model_FIM}

In this section, we derive the FIM for both deterministic and random
parameter estimation to evaluate the SPEB.

\subsubsection{FIM without {A Priori} Knowledge} \label{Sec:FIM_NoPrior}

The FIM for the deterministic parameter vector $\B\theta$ is given
by \cite{Tre:68}
    \begin{align}\label{eq:FIM_FIMdef}
        \JTH = \FIM{\V{r}}{\V{r}|\B\theta}{\B\theta}{\B\theta}
        \, ,
    \end{align}
where $\FTR$ is the likelihood ratio of the random vector $\V{r}$ conditioned on ${\B\theta}$.  Since the received waveforms from
different anchors are independent, the likelihood ratio can be
written as \cite{Poo:B94}
    \begin{align}\label{eq:FIM_FTR}
        \FTR =  \prod_{k \in \NB} f(\V{r}_k|{\B\theta})\,,
    \end{align}
where
    \begin{align*}
        f(\V{r}_k|{\B\theta}) & \varpropto \exp \left\{
                \frac{2}{N_0} \int_0^{T_\text{ob}}
                r_k(t)\, \sum_{l=1}^{L_k} \A{k}{l} s\left(t-\D{k}{l}\right) dt
                \right. \nonumber \\
             & \hspace{9mm}  \left.
               - \frac{1}{N_0} \int_0^{T_\text{ob}} \left[\sum_{l=1}^{L_k}
                \A{k}{l} s\left(t-\D{k}{l}\right)\right]^2 dt
            \right\}.
    \end{align*}
Substituting \eqref{eq:FIM_FTR} in \eqref{eq:FIM_FIMdef}, we have
the FIM $\JTH$ as
    \begin{align}\label{eq:FIM_JTH_NPrior}
        \renewcommand{\arraystretch}{1.3}
        \JTH = \frac{1}{c^2} \Matrix{cc}
            {   \HL \LL \HL^\text{T} + \HNL \LNL \HNL^\text{T}      &   \HNL \LNL \\
                \LNL \HNL^\text{T}                                  &  \LNL},
    \end{align}
where $\LL$, $\HL$, $\LNL$, and $\HNL$ are given by
\eqref{eq:Apd_HL_HNL} and \eqref{eq:Apd_Lamda}. In the above
matrices, $\LL$ and $\HL$ are related to the LOS signals, and $\LNL$
and $\HNL$ are related to the NLOS signals.

\subsubsection{FIM with {A Priori} Knowledge}
\label{Sec:FIM_WPrior}

We now incorporate the a priori knowledge of the agent's position
and channel parameters for localization. Since the multipath
parameters $\B\kappa_k$ are independent a priori, the PDF of
$\B\theta$ can be expressed as\footnote{When a subset of parameters
are deterministic, they are eliminated from $f(\B\theta)$.}
    \begin{align}\label{eq:FIM_Prior_Dis}
        f(\B\theta) = \gp(\V{p}) \, \prod_{k \in \NB} \gk(\B{\kappa}_k |
        \V{p})\, ,
    \end{align}
where $\gp(\V{p})$ is the PDF of the agent's position, and
$\gk(\B{\kappa}_k | \V{p})$ is the joint PDF of the multipath
parameter vector $\B\kappa_k$ conditioned on the agent's position.
Based on the models of wideband channels \cite{Mol:B05, SalVal:87,
ChoTanLauMcLBeaNix:03} and UWB channels \cite{WinSch:J02,
QueWin:J05, MolCasChoEmaForKanKarKunSchSiwWin:J06, Mol:05,
Mol:B05}, we derive $\gk(\B{\kappa}_k | \V{p})$ in
\eqref{eq:FIM_Prior_PDF} in Appendix \ref{Apd:FIM_Channel} and show
that
    \begin{align}\label{eq:FIM_prior_pdf}
        \gk(\B\kappa_k | \V{p}) = \gk(\B\kappa_k | d_k) \, ,
    \end{align}
where $d_k = \| \V{p} - \V{p}_k\|$.

The joint PDF of observation and parameters can be written as
    \begin{align*}
        f(\V{r},\B\theta) =  \FTR \, f(\B\theta)
        \, ,
    \end{align*}
where $\FTR$ is given by \eqref{eq:FIM_FTR}, and hence the FIM
becomes
    \begin{align}\label{eq:FIM_JD_JP}
        \JTH = \JD + \JP \, ,
    \end{align}
where $\JD \triangleq \FIM{\V{r},\B\theta}{\V{r}|
\B\theta}{\B\theta}{\B\theta}$ and $\JP \triangleq
\FIM{\B\theta}{\B\theta}{\B\theta}{\B\theta}$ are the FIMs from the
observations and the a priori knowledge, respectively.\footnote{Note
that $\JTH$ in \eqref{eq:FIM_JD_JP} requires averaging over the
random parameters, and hence does not depend on any particular value
of $\B\theta$. In contrast, $\JTH$ in \eqref{eq:FIM_FIMdef} is a
function of a particular value of the deterministic parameter vector
$\B\theta$.} The FIM $\JD$ can be obtained by taking the expectation
of $\JTH$ in \eqref{eq:FIM_JTH_NPrior} over the random parameter
vector $\B\theta$, and $\JP$ can be obtained by substituting
\eqref{eq:FIM_Prior_Dis} in \eqref{eq:FIM_JD_JP} as
    \begin{align}\label{eq:FIM_Prior_Matrix}
    \renewcommand{\arraystretch}{1.5}
        \JP =
        \Matrix{ccccc}{ \GP + \sum_{k \in \NB} \Gpp{k} & \Gpk{1} & \cdots & \Gpk{\Nb} \\
                        \GpkT{1}      &  \Gkk{1}  &   & \V{0}\\
                        \vdots &      &  \ddots   &   &  \\
                        \GpkT{\Nb}  &  \V{0}    &   & \Gkk{\Nb} &  },
    \end{align}
where $\GP$ describes the FIM from the a priori knowledge of
$\V{p}$, given by
    \begin{align*}
        \GP = \FIM{\B\theta}{\V{p}}{\V{p}}{\V{p}} \, ,
    \end{align*}
and $\Gkk{k} =
\FIM{\B\theta}{\B\kappa_k|\V{p}}{\B\kappa_k}{\B\kappa_k}$, $\Gpp{k}
= \FIM{\B\theta}{\B\kappa_k|\V{p}}{\V{p}}{\V{p}}$, and $\Gpk{k} =
\FIM{\B\theta}{\B\kappa_k|\V{p}}{\V{p}}{\B\kappa_k}$ characterize
the joint a priori knowledge of $\V{p}$ and $\B\kappa_k$.

\subsection{Equivalent Fisher Information Matrix}\label{Sec:Evalu_EFIM}

Determining the SPEB requires inverting the FIM $\JTH$ in
\eqref{eq:FIM_JTH_NPrior} and \eqref{eq:FIM_JD_JP}. However, $\JTH$
is a matrix of high dimensions, while only a small submatrix $\left[
\JTH^{-1} \right]_{2 \times 2}$ is of interest. To circumvent
direction matrix inversion and gain insights into the localization
problem, we first introduce the notions of EFI \cite{SheWin:C07, SheWin:C08}.

\begin{defi}[Equivalent Fisher Information Matrix]\label{def:EFIM}
Given a parameter $\B\theta = \left[\, \B\theta_1^\text{T} \;\;
\B\theta_2^\text{T}\right]^\text{T}$ and the FIM $\JTH$ of the form
    \begin{align}\label{eq:JTH_Partition}
        \JTH = \Matrix{cc}
                {   \V{A}       &   \V{B}   \\
                    \V{B}^\text{T}     &   \V{C}   },
    \end{align}
where $\B\theta \in \mathbb{R}^{N}$, $\B\theta_1 \in \mathbb{R}^n$,
$\V{A} \in \mathbb{R}^{n \times n}$, $\V{B} \in \mathbb{R}^{n\times
(N-n)}$, and $\V{C} \in \mathbb{R}^{(N-n) \times (N-n)}$ with $n<N$,
the equivalent Fisher information matrix (EFIM) for $\B\theta_1$ is
given by\footnote{Note that $\JE{\B\theta_1}$ does not
depend on any particular value of $\B\theta_1$ for a random parameter
vector $\B\theta_1$, whereas it is a function of
$\B\theta_1$ for a deterministic parameter vector $\B\theta_1$.}
    \begin{align}\label{eq:Sing_EFIM_FIM}
        \JE{\B\theta_1} \triangleq \V{A} - \V{B} \V{C}^{-1} \V{B}^\text{T}.
    \end{align}
\end{defi}

Note that the EFIM retains all the necessary information to derive
the information inequality for the parameter vector $\B\theta_1$,
since $[\,\JTH ^{-1}]_{n \times n} =
\V{J}_\text{e}^{-1}(\B\theta_1)$,\footnote{The right hand side of
\eqref{eq:Sing_EFIM_FIM} is known as the Schur complement of the
matrix $\V{C}$ \cite{HorJoh:B85}.} and the MSE matrix of the
estimates for $\B\theta_1$ is bounded below by
$\V{J}_\text{e}^{-1}(\B\theta_1)$.\footnote{When $\theta_1 \in
\mathbb{R}^1$, the EFIM has only one element since
$J_\text{e}(\theta_1)$ is a scalar.} For two-dimensional
localization $(n=2)$, we aim to reduce the dimension of the original
FIM to the $2 \times 2$ EFIM.

%
%

\section{Evaluation of EFIM}\label{Sec:Evalu}

In this section, we apply the notion of equivalent Fisher
information (EFI) to derive the SPEB for both the case with and
without a priori knowledge. We also introduce the notion of ranging
information (RI), which turns out to be the basic component of the
SPEB.

\subsection{EFIM without {A Priori} Knowledge}\label{Sec:Evalu_NoPrior}

First consider a case in which a priori knowledge is unavailable. We
apply the notion of EFI to reduce the dimension of the original FIM
in \eqref{eq:FIM_JTH_NPrior}, and the EFIM for the agent's position
is presented in the following proposition.

\begin{pro}\label{thm:EFIM_n-NLOS}
When a priori knowledge is unavailable, an EFIM for the agent's
position is
    \begin{align}\label{eq:Sing_EFIM_NLOS}
        \JE{\V{p}, \left\{\B\kappa_k : k \in \NL \right\}}  =  \frac{1}{c^2}
        \; \HL \LL \HL^\text{T}\,,
    \end{align}
where $\HL$ and $\LL$ are given by \eqref{eq:Apd_HL_HNL} and
\eqref{eq:Apd_Lamda}, respectively.
\end{pro}

\begin{IEEEproof}
Let $\V{A} = \HNL \LNL \HNL^\text{T} + \HL \LL \HL^\text{T}$, $\V{B}
= \HNL \LNL$, and $\V{C} = \LNL$ in \eqref{eq:FIM_JTH_NPrior}.
Applying the notion of EFI in \eqref{eq:Sing_EFIM_FIM} leads to the
result.
\end{IEEEproof}

\begin{remrk}
When a priori knowledge is unavailable, NLOS signals do not
contribute to the EFIM for the agent's position. Hence we can
eliminate these NLOS signals when analyzing localization accuracy.
This observation agrees with the results of \cite{QiSudKob:04}, but
the amplitudes of the MPCs are assumed to be known in their model.
\end{remrk}

Note that the dimension of the EFIM in \eqref{eq:Sing_EFIM_NLOS} is
much larger than $2 \times 2$. We will apply the notion of EFI again
to further reduce the dimension of the EFIM in the following
theorem. Before the theorem, we introduce the notion of the first
contiguous-cluster and ranging information (RI).

\begin{defi}[First Contiguous-Cluster]\label{def:Overlap_cluster}
The first contiguous-cluster is defined to be the set of paths $\{1,
2, \dotsc, j \}$, such that $|\tau_i - \tau_{i+1}| < T_\text{s}$ for
$i = 1, 2, \dotsc, j-1$, and $|\tau_j - \tau_{j+1}| > T_\text{s}$,
where $T_\text{s}$ is the duration of $s(t)$.
\end{defi}

\begin{defi}[Ranging Information]
The ranging information (RI) is a $2 \times 2$ matrix of the form
$\lambda \, \R(\phi)$, where $\lambda$ is a nonnegative number
called the ranging information intensity (RII), and $\R(\phi)$ a $2
\times 2$ matrix called the ranging direction matrix (RDM) with
angle $\phi$, given by
    \begin{align*}
        \renewcommand{\arraystretch}{1.3}
        \R (\phi) \triangleq \Matrix{cc}
                    { \cos^2 \phi           &   \cos \phi \sin \phi \\
                      \cos \phi \sin \phi   &   \sin^2 \phi    }.
    \end{align*}
\end{defi}

\newcounter{MYtempeqncnt}
\begin{figure*}[!b]
  \vspace*{4pt}%
  \hrulefill%
  \normalsize%
  \setcounter{MYtempeqncnt}{\value{equation}}%
  \setcounter{equation}{19}%
      \begin{align}\label{eq:Sing_SPEB}
        \PEB(\V{p}) 
        & = \frac{c^2}{8\pi^2\beta^2} \, \frac{ 2 \, \sum_{k \in \NL} (1-\chi_k) \, \SNR{k}{1}}
            { \sum_{k \in \NL} \sum_{m \in \NL} (1-\chi_k)(1-\chi_m) \, \SNR{k}{1}
            \mathsf{SNR}_m^{(1)} \sin^2(\phi_k - \phi_m) }
    \end{align}%
  \setcounter{equation}{\value{MYtempeqncnt}}
  \vspace*{-6pt}
\end{figure*}

\begin{figure}
    \begin{center}
        \input{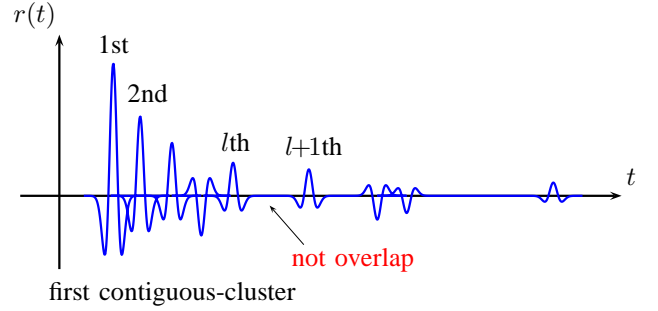}

    \caption{\label{overlap} An illustration of the first contiguous-cluster
    (containing $l$ paths) in a LOS signal.}
    \end{center}
\end{figure}

The first contiguous-cluster is the first group of non-disjoint
paths (see Fig.~\ref{overlap}).\footnote{The first
contiguous-cluster, defined for general wideband received signals,
may contain many MPCs. Two paths that arrive at time $\tau_i$ and
$\tau_j$ are called non-disjointed if $|\tau_i-\tau_j| <
T_\text{s}$.} {The RDM is one-dimensional along the direction $\phi$
with unit intensity, i.e., $\R(\phi)$ has one (and only one)
non-zero eigenvalue equal to $1$ with corresponding eigenvector $\q
= \Matrix{cc}{\cos\phi & \sin\phi}^\text{T}.$}

\begin{thm}\label{Thm:EFIM_NPrior}
When a priori knowledge is unavailable, the EFIM for the agent's
position is a $2 \times 2$ matrix
    \begin{align}\label{eq:Sing_EFIM_overlap}
        \JE{\V{p}} = \sum_{k \in \NL} \lambda_k \, \R \left( \phi_k \right)
        \, ,
    \end{align}
where $\lambda_k$ is the RII from anchor $k$, given by
    \begin{align}\label{eq:Sing_EFIM_strength}
        \lambda_k =  \frac{8\pi^2\beta^2}{c^2}\, (1-\chi_k) \, \SNR{k}{1}
        \, .
    \end{align}
In \eqref{eq:Sing_EFIM_strength}, $0 \leq \chi_k \leq 1$ is given by
\eqref{eq:Apd_R_a'_full},
    \begin{align}\label{eq:Sing_beta}
        \beta & \triangleq \left( \frac{\int_{-\infty}^{+\infty} f^2 \, |S(f)|^2 df}{
        \int_{-\infty}^{+\infty} |S(f)|^2 df}\right)^{1/2}, \\
        \noalign{\noindent and \vspace{\jot}}
        \mathsf{SNR}_{k}^{(l)} & \triangleq \frac{ |\alpha_k^{(l)}|^2 \int_{-\infty}^{+\infty}|S(f)|^2 df}{N_0}
        \, .\label{eq:Sing_SNR}
    \end{align}
Furthermore, only the first contiguous-cluster of LOS signals
contains information for localization.
\end{thm}

\begin{IEEEproof}
See Appendix \ref{apd:Proofs_Evalu_Overlap_cluster}.
\end{IEEEproof}

\begin{remrk}
In Theorem \ref{Thm:EFIM_NPrior}, $\beta$ is known as the
\emph{effective bandwidth} \cite{Sch:C72,Tre:68}, $\chi_k$ is called
\emph{path-overlap coefficient} (POC) that characterizes the effect
of multipath propagation for localization, and
$\mathsf{SNR}_{k}^{(l)}$ is the SNR of the ${l}$th path in $r_k(t)$.
We draw the following observations from Theorem
\ref{Thm:EFIM_NPrior}:
\end{remrk}

\begin{itemize}
\item
The original FIM in \eqref{eq:FIM_JTH_NPrior} can be transformed
into a simple $2 \times 2$ EFIM in a canonical form, given by
\eqref{eq:Sing_EFIM_overlap}, as a weighted sum of the RDM from
individual anchors. Each anchor (e.g. anchor $k$) can provide only
one-dimensional RI along the direction $\phi_k$, from the anchor to
the agent, with intensity $\lambda_k$.\footnote{For notational
convenience, we suppress the dependence of $\phi_k$ and $\lambda_k$
on the agent's position $\V{p}$ throughout the paper.}

\item
The RII $\lambda_k$ depends on the effective bandwidth of $s(t)$,
the SNR of the first path, and the POC. Since $0 \leq \chi_k \leq
1$, path-overlap in the first contiguous-cluster will reduce the
RII, thus leading to a higher SPEB, unless the signal via the first
path does not overlap with others ($\chi_k =0$).

\item
The POC $\chi_k$ in \eqref{eq:Apd_R_a'_full} is determined only by
the waveform $s(t)$ and the NLOS biases of the MPCs in the first
contiguous-cluster. The independence of $\chi_k$ on the path
amplitudes seems counter-intuitive. However, this is due to the fact
that, although large $\A{k}{l}$ causes severe interpath interference
for estimating the TOA $\D{k}{1}$, it increases the estimation
accuracy for $\D{k}{l}$, which in turn helps to mitigate the
interpath interference.
\end{itemize}

We can specialize the above theorem into a case in which the first
path in a LOS signal is completely resolvable, i.e., the first
contiguous-cluster contains only a single component.

\begin{cor}\label{cor:EFIM_n-NLOS_gen}
When a priori knowledge is unavailable and the first
contiguous-cluster of the received waveform from anchor $k$ contains
only the first path, the RII becomes
    \begin{align}\label{eq:Anal_RII_noprior}
        \lambda_k = \frac{8\pi^2\beta^2}{c^2} \, \SNR{k}{1}
        \, .
    \end{align}
\end{cor}

\begin{IEEEproof}
See Appendix \ref{apd:Proofs_Evalu_n-NLOS_gen}.
\end{IEEEproof}

\begin{remrk}
When the first path is resolvable, $\chi_k = 0$ in
\eqref{eq:Sing_EFIM_strength} and hence $\lambda_k$ attains its
maximum value. However, when the signal via other paths overlap with
the first one, these paths will degrade the estimation accuracy of
the first path's arrival time and hence the RII. Corollary
\ref{cor:EFIM_n-NLOS_gen} is intuitive and important: the RII of a
LOS signal depends only on the first path if the first path is
resolvable. In such a case, all other paths can be eliminated, and
the multipath signal is equivalent to a signal with only the first
path for localization.
\end{remrk}

From Theorem \ref{Thm:EFIM_NPrior}, the SPEB can be derived in
\eqref{eq:Sing_SPEB} shown at the bottom of the page.
\addtocounter{equation}{1} When the first paths are resolvable, by
Corollary \ref{cor:EFIM_n-NLOS_gen}, we have all $\chi_k = 0$ in
\eqref{eq:Sing_SPEB} and the corresponding $\PEB(\V{p})$ becomes the
same as those based on single-path signal models in
\cite{QiSudKob:04, GezTiaGiaKobMolPooSah:05}. However, those results
are not  accurate when the first path is not resolvable.

\subsection{EFIM with {A Priori} Knowledge}
\label{Sec:Evalu_WPrior}

We now consider the case where there is a priori knowledge of the
channel parameters, but not of the agent's position. In such cases,
since $\V{p}$ is deterministic but unknown, $\gp(\V{p})$
is eliminated in \eqref{eq:FIM_Prior_Dis}. 
Similar to the analysis in the previous section, we can derive the
$2 \times 2$ EFIM for the corresponding FIM in \eqref{eq:FIM_JD_JP}.

\begin{thm}\label{thm:EFIM_NLOS_gen}
When a priori knowledge of the channel parameters is available and
the sets of channel parameters corresponding to different anchors
are mutually independent, the EFIM for the agent's position is a $2
\times 2$ matrix
    \begin{align}\label{eq:Sing_EFIM_Prior}
        \JE{\V{p}} = \sum_{k \in \NL} \lambda_k \, \R \left( \phi_k \right)
                + \sum_{k \in \NNL} {\lambda}_k \, \R \left( \phi_k \right)
             \, ,
    \end{align}
where $\lambda_k$ is given by \eqref{eq:Apd_Ra_def} for LOS signals
and \eqref{eq:Apd_tRa_def} for NLOS signals.
\end{thm}

\begin{IEEEproof}
See Appendix \ref{apd:Proofs_Evalu_NLOS_gen}.
\end{IEEEproof}

\begin{remrk}
Theorem \ref{thm:EFIM_NLOS_gen} generalizes the result of Theorem
\ref{Thm:EFIM_NPrior} from deterministic to hybrid parameter
estimation.\footnote{This is the case where the agent's position
$\V{p}$ is deterministic and the channel parameters are random.} In
this case, the EFIM can still be expressed in a canonical form as a
weighed sum of the RDMs from individual anchors. Note that due to
the existence of a priori channel knowledge, the RII of NLOS signals
can be positive, and hence these signals contribute to the EFIM as
opposed to the case in Theorem \ref{Thm:EFIM_NPrior}.
\end{remrk}

\begin{cor}\label{cor:prior_knowledge}
A priori channel knowledge increases the RII. In the absence of such
knowledge, the expressions of RII in \eqref{eq:Apd_Ra_def} and
\eqref{eq:Apd_tRa_def} reduce to \eqref{eq:Sing_EFIM_strength} and
zero, respectively.
\end{cor}

\begin{IEEEproof}
See Appendix \ref{apd:Proofs_Consis_RI}.
\end{IEEEproof}

\begin{cor}\label{cor:RII_NLOS_LOS}
LOS signals can be treated as NLOS signals with infinite a priori
Fisher information of $\BB{k}{1}$, i.e., $\BB{k}{1}$ is known.
Mathematically, \eqref{eq:Apd_Ra_def} is equivalent to
\eqref{eq:Apd_tRa_def} with
$\FIM{\B\theta}{\B\theta}{\BB{k}{1}}{\BB{k}{1}} \rightarrow \infty$.
\end{cor}

\begin{IEEEproof}
See Appendix \ref{apd:Proofs_Evalu_NLOS/LOS}.
\end{IEEEproof}

\begin{remrk}
Corollary \ref{cor:prior_knowledge} shows that Theorem
\ref{thm:EFIM_NLOS_gen} degenerates to Theorem \ref{Thm:EFIM_NPrior}
when a priori channel knowledge is unavailable. Moreover, Corollary
\ref{cor:RII_NLOS_LOS} unifies the LOS and NLOS signals under the
Bayesian estimation framework: the NLOS biases $\BB{k}{1}$ $( k \in
\NL)$ can be regarded as random parameters with infinite a priori
Fisher information instead of being eliminated from $\B\theta$ as in
Section \ref{Sec:Model_System}. Hence, all of the signals can be
modeled as NLOS, and infinite a priori Fisher information of
$\BB{k}{1}$ will be assigned for LOS signals.
\end{remrk}

We next consider the case where a priori knowledge of the agent's
position is available in addition to channel parameters. Note that
the topology of the anchors and the agent changes with the position
of the agent. The $2 \times 2$ EFIM is given in
\eqref{eq:Apd_EFIM_WPrior_Pos}, which is more intricate than the
previous two cases. To gain some insights, we consider a special
case where\footnote{This occurs when the agent's {a priori} position
distribution is concentrated in a small area relative to the
distance between the agent and the anchors, so that $g(\V{p})$ is
flat in that area. For example, this condition is satisfied in
far-field scenarios.}
    \begin{align}\label{eq:Appro_Condition}
        \E_\V{p}\left\{ g(\V{p})\right\} = g \left(\bar{\V{p}}\right)\, ,
    \end{align}
in which $\bar{\V{p}} = \E_{\V{p}}\left\{ \V{p} \right\}$ is the
agent's expected position, for some function $g(\cdot)$ involved in
the derivation of the EFIM (see Appendix
\ref{apd:Proofs_Evalu_FarField}).

\begin{pro}\label{thm:EFIM_FarField}
When the a priori position distribution of the agent satisfies
\eqref{eq:Appro_Condition}, and the sets of channel parameters
corresponding to different anchors are mutually independent, the
EFIM for the agent's position is a $2 \times 2$ matrix
    \begin{align}\label{eq:Sing_EFIM_Prior_FF}
        \JE{\V{p}} = \sum_{k \in \NB} \bar\lambda_k \, \R \left(\bar{\phi}_k \right)
                    + \GP
        \, ,
    \end{align}
where $\bar\lambda_k$ is given by \eqref{eq:Apd_RII_Pbar}, and
$\bar{\phi}_k$ is the angle from anchor $k$ to $\bar{\V{p}}$.
\end{pro}

\begin{IEEEproof}
See Appendix \ref{apd:Proofs_Evalu_FarField}.
\end{IEEEproof}

\begin{remrk}
The a priori knowledge of the agent's position is exploited, in
addition to that of the channel parameters, for localization in
Proposition \ref{thm:EFIM_FarField}. The expressions for the EFIM
can be involved in general. Fortunately, if
\eqref{eq:Appro_Condition} is satisfied, the EFIM can be simply
written as the sum of two parts as shown in
\eqref{eq:Sing_EFIM_Prior_FF}: a weighted sum of the RDMs from
individual anchors as in the previous two cases, and the EFIM from
the a priori knowledge of the agent's position. This result unifies
the contribution from anchors and that from the a priori knowledge
of the agent's position into the EFIM. The concept of localization
with a priori knowledge of the agent's position is useful for a wide
range of applications such as successive localization or tracking.
\end{remrk}

%
%

\section{Wideband Localization with Antenna Arrays}
\label{Sec:Array}

In this section, we consider localization systems using wideband
antenna arrays, which can perform both TOA and AOA measurements.
Since the orientation of the array may be unknown, we propose a model to jointly estimate the agent's position and orientation, and
derive the SPEB and the \emph{squared orientation error bound}
(SOEB).

\subsection{System Model and Squared Orientation
Error Bound}\label{Sec:Array_Model}

\begin{figure}
    \begin{center}

    \psfrag{dots}[l][][1.2]{\hspace{5mm}$\cdots$}
    \psfrag{X}[l][][0.8]{\hspace{-2mm}$x$}
    \psfrag{Y}[l][][0.8]{\hspace{-1mm}$y$}
    \psfrag{P}[l][][0.7]{\hspace{-1mm}$\V{p}$}
    \psfrag{P1}[l][][0.7]{\hspace{-1mm}$\V{p}_1^\text{Array}$}
    \psfrag{P2}[l][][0.7]{\hspace{-5mm}$\V{p}_2^\text{Array}$}
    \psfrag{P3}[l][][0.7]{\hspace{-5mm}$\V{p}_3^\text{Array}$}
    \psfrag{P4}[l][][0.7]{\hspace{-1mm}$\V{p}_\Na^\text{Array}$}
    \psfrag{phi}[l][][0.7]{\hspace{-2mm}$\varphi$}

    {\includegraphics[angle=0,width=0.85\linewidth,draft=false]{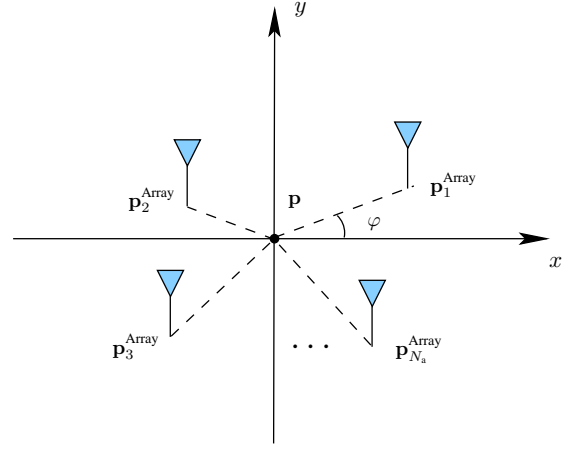}}

    \caption{\label{antenna_array}
    An antenna array is described by the reference point $\V{p}$,
    the orientation $\varphi$, and the relative positions of the antennas.
    }
    \end{center}
\end{figure}

Consider a network where each agent is equipped with an
$\Na$-antenna array,\footnote{Each anchor has only one antenna here.
We will discuss the case of multiple antennas on anchors at the end
of this section.} which can extract both the TOA and AOA information
with respect to neighboring anchors. Let $\NA = \{ 1, 2, \dotsc, \Na
\}$ denote the set of antennas, and let $\V{p}_n^\text{Array}
\triangleq [\,x_n^\text{Array} \; y_n^\text{Array} \,]^\text{T}$
denote the position of the agent's $n$th antenna, which needs to
be estimated. Let $\phi_{n k}$ denote the angle from anchor $k$ to the n k$n$th antenna, i.e.,
    \begin{align*}
        \phi_{n k} = \tan^{-1}\frac{y_n^\text{Array}-y_k}{x_n^\text{Array}-x_k} \, .
    \end{align*}
Since relative positions of the antennas in the array are usually
known, if we denote $\V{p} = [\,x \; y\,]^\text{T}$ as a reference
point and $\varphi$ as the orientation of the array,\footnote{Note
from geometry that the orientation $\varphi$ is independent of the
specific reference point.} then the position of the $n$th antenna
in the array can be represented as (Fig.~\ref{antenna_array})
    \begin{align*}
        \V{p}_n^\text{Array} = \V{p} + \Matrix{c} {\Delta x_n(\V{p},\varphi)  \\
            \Delta y_n(\V{p},\varphi)}, \quad  n \in \NA
        \, ,
    \end{align*}
where $\Delta x_n(\V{p},\varphi)$ and $\Delta y_n(\V{p},\varphi)$
denote the relative distance in $x$ and $y$ direction from the
reference point to the $n$th antenna, respectively.

Since the array orientation may be unknown, we classify the
localization problem into \emph{orientation-aware} and
\emph{orientation-unaware} cases, where $\varphi$ can be thought of
as a random parameter with infinite (orientation-aware) and zero
(orientation-unaware) a priori Fisher information \cite{SheWin:C07}.

The received waveform at the agent's $n$th antenna from anchor
$k$ can be written as
    \begin{align*}
        r_{n k}(t) = \sum_{l=1}^{L_{n k}} \A{n k}{l} \,
        s\left( t - \D{n k}{l} \right)+z_{n k}(t),
        \quad t \in [ \,0, T_{\text{ob}})\, ,
    \end{align*}
where $\A{n k}{l}$ and $\D{n k}{l}$ are the amplitude and delay,
respectively, of the $l$th path, $L_{n k}$ is the number of MPCs,
and $z_{n k}(t)$ represents the observation noise modeled as
additive white Gaussian processes with two-side power spectral
density $N_0/2$. The relationship between the position of the
${n}$th antenna and the delay of the $l$th path is
    \begin{align}\label{eq:Sing_Tau_L_krray}
        \D{n k}{l} = \frac{1}{c} {\Big [} \,  
        \| \V{p}_n^\text{Array} - \V{p}_k \|+ \BB{n k}{l} \,{\Big ]}
        \, .
    \end{align}
The parameters to be considered include the position of the
reference point, the array orientation, and the nuisance multipath
parameter as
    \begin{align}\label{eq:para_vec}
        \B\theta = \Matrix{cccccc}{\V{p}^\text{T} & \varphi &  \check{\B\kappa}_1^\text{T}
        &  \check{\B\kappa}_2^\text{T} & \cdots &  \check{\B\kappa}_{\Na}^\text{T}}^\text{T},
    \end{align}
where $\B{\check\kappa}_n$ consists of the multipath parameters
associated with the received waveforms from all anchors at the
$n$th antenna,
    \begin{align*}
        \check{\B\kappa}_n =
        \Matrix{cccccc}{\B{\kappa}_{n,1}^\text{T} & \B{\kappa}_{n,2}^\text{T}
        & \cdots & \B{\kappa}_{n,\Nb}^\text{T}}^\text{T},
    \end{align*}
and each $\B{\kappa}_{n k}$ consists of the multipath parameters
associated with $r_{n k}(t)$,
    \begin{align*}
        \B{\kappa}_{n k}  = \Matrix{ccccccc}{ b_{n k}^{(1)} & \alpha_{n k}^{(1)}
        &  \cdots & b_{n k}^{(L_{n k})} & \alpha_{n k}^{(L_{n k})} }^\text{T}.
    \end{align*}

Similar to Section \ref{Sec:Model_SPEB}, the overall received
waveforms at the antenna array can be represented, using the KL
expansion, by $\V{r} =  \left[\, \V{r}_1^\text{T} \;\;
\V{r}_2^\text{T} \;\; \cdots \;\; \V{r}_{\Na}^\text{T}
\,\right]^\text{T}$, where
    \begin{align*}
        \V{r}_n = \Matrix{cccc} {\V{r}_{n,1}^\text{T}   &
        \V{r}_{n,2}^\text{T}   & \cdots
        &  \V{r}_{n,\Nb}^\text{T}}^\text{T},
    \end{align*}
in which $\V{r}_{n k}$ is obtained by the KL expansion of
${r}_{n k}(t)$.

\begin{defi}[Squared Orientation Error Bound]
The squared orientation error bound (SOEB) is defined to be
    \begin{align*}
        \PEB(\varphi) \triangleq  \left[ \JTH^{-1} \right]_{3,3}
        \, .
    \end{align*}
\end{defi}

\subsection{EFIM without {A Priori} Knowledge}\label{Sec:Array_WPrior}

We first consider scenarios in which a priori knowledge is
unavailable. Following similar steps in Section
\ref{Sec:Evalu_WPrior}, we have the following theorem.

\begin{thm}\label{thm:EFIM_NLOS_gen_AOA}
When a priori knowledge is unavailable, the EFIMs for the position
and the orientation, using an $\Na$-antenna array, are given
respectively by
    \begin{align}\label{eq:Anal_EFIM_AOA}
        \V{J}_\text{e}^\text{Array}(\V{p}) = 
                \sum_{n \in \NA}  \JE{\V{p}_n^\text{Array}}
                -   \frac{1}{\sum_{n \in \NA} \sum_{k \in \NB} \lambda_{n k} h_{n k}^2}
                \, \V{q} {\V{q}}^\text{T} ,
    \end{align}
and
    \begin{align}\label{eq:Anal_EFIM_AOA_phi}
         J_\text{e}^\text{Array}(\varphi) =
                    \sum_{n \in \NA} \sum_{k \in \NB} \lambda_{n k} h_{n k}^2
                    - {\V{q}}^\text{T}  \left[ \sum_{n \in \NA}
                    \JE{\V{p}_n^\text{Array}} \right]^{-1}\!\! \V{q}
         \,,
    \end{align}
where $\lambda_{n k}$ is given by \eqref{eq:Apd_RII_NoPrior},
$\V{q}_{n k} = \Matrix{cc}{\cos \phi_{n k} &
\sin\phi_{n k}}^\text{T}$, and
    \begin{align}
        \JE{\V{p}_n^\text{Array}} & = \sum_{k \in \NB} \lambda_{n k} \, \R(\phi_{n k})
        \, , \nonumber \\
        \V{q} & =  \sum_{n \in \NA} \sum_{k \in \NB} \lambda_{n k} \,h_{n k}
        \,\q_{n k}\,, \label{eq:q_an}  \\
        \noalign{\noindent and \vspace{\jot}}
        h_{n k} = \frac{d}{d \varphi}\, \Delta & x_n(\V{p},\varphi) \, \cos\phi_{n k}
                  + \frac{d}{d \varphi}\, \Delta y_n(\V{p},\varphi) \, \sin\phi_{n k}
                  \label{eq:h_an}
        \, .
    \end{align}
\end{thm}

\begin{IEEEproof}
See Appendix \ref{apd:Proofs_Array_NLOS_gen_AOA}.
\end{IEEEproof}

\begin{cor}
The EFIM for the position is given by
    \begin{align}\label{eq:Anal_Orien}
        \V{J}_\text{e}^\text{Array}(\V{p})
            = \sum_{n \in \NA} \JE{\V{p}_n^\text{Array}}  \,,
    \end{align}
for orientation-aware localization.
\end{cor}

\begin{IEEEproof}
(Outline) In orientation-aware localization, the angle $\varphi$ is
known and hence excluded from the parameter vector $\B\theta$ in
\eqref{eq:para_vec}. Consequently, the proof of this corollary is
analogous to that of Theorem \ref{thm:EFIM_NLOS_gen_AOA} except that
the components corresponding to $\varphi$ are eliminated from the
FIM in \eqref{eq:Apd_Array_JP} and \eqref{eq:Apd_Array_JD}. One can
obtain \eqref{eq:Anal_Orien} after some algebra.
\end{IEEEproof}

\begin{remrk}
The EFIM $\JE{\V{p}_n^\text{Array}}$ in \eqref{eq:Anal_EFIM_AOA} and
\eqref{eq:Anal_Orien} corresponds to the localization information
from the $n$th antenna. We draw the following observation from
the above theorem.
\begin{itemize}
\item
The EFIM $\V{J}_\text{e}^\text{Array}(\V{p})$ in
\eqref{eq:Anal_EFIM_AOA} consists of two parts: 1) the sum of
localization information obtained by individual antennas, and 2) the
information reduction due to the uncertainty in the orientation
estimate, which is subtracted from the first part.\footnote{For
notational convenience, we suppress the dependence of $h_{n k}$,
$\lambda_{n k}$, and $\V{q}$ on the reference position $\V{p}$.}
Since $\V{q}\, {\V{q}}^\text{T}$ in the second part is a positive
semi-definite $2 \times 2$ matrix and $\sum_{n \in \NA} \sum_{k \in
\NB} \lambda_{n k} h_{n k}^2 $ is always positive, we have the
following inequality
    \begin{align}\label{eq:Anal_EFIM_AOA_Ineq}
        \V{J}_\text{e}^\text{Array}(\V{p}) \preceq  \sum_{n \in \NA} \JE{\V{p}_n^\text{Array}}
        \, .
    \end{align}
The inequality implies that the EFIM for the position, using antenna
arrays, is bounded above by the sum of all EFIMs corresponding to
individual antennas, since the uncertainty in the orientation
estimate degrades the localization accuracy, except for $\V{q} = 0$
or orientation-aware localization (i.e., \eqref{eq:Anal_Orien}).

\item
The EFIM $\V{J}_\text{e}^\text{Array}(\V{p})$ and
$J_\text{e}^\text{Array}(\varphi)$ depend only on the individual RI
between each pair of anchors and antennas (through $\lambda_{n k}$'s
and $\phi_{n k}$'s), and the array geometry (through $h_{n k}$'s).
Hence it is not necessary to \emph{jointly} consider the received
waveforms at the $\Na$ antennas, implying that AOA obtained by
antenna arrays does not increase position accuracy. Though
counterintuitive at first, this finding should not be too surprising
since AOA is obtained indirectly by the antenna array through TOA measurements, whereas the TOA information has already been fully utilized for
localization by individual antennas.

\item
The gain of using antenna arrays for localization mainly comes from
the multiple copies of the waveform received at the $\Na$ antennas
(see \eqref{eq:Anal_EFIM_AOA}),\footnote{In near-field scenarios
where the antenna separation is on the order of the distances
between the array and the anchors, additional gain that arises from
the spatial diversity of the multiple antennas may be possible.} and
its performance is similar to that of a single antenna with $\Na$
measurements. The advantage of using antenna arrays lies in their
ability of simultaneous measurements at the agent.
\end{itemize}
\end{remrk}

The equality in \eqref{eq:Anal_EFIM_AOA_Ineq} is always achieved,
independent of reference point, in orientation-aware localization.
However, only a unique reference point achieves this equality in
orientation-unaware localization. We define this unique point as the
orientation center.

\begin{defi}[Orientation Center]
The orientation center is a reference point $\V{p}^*$ such that
    \begin{align*}
        \V{J}_\text{e}^\text{Array}(\V{p}^*) = \sum_{n \in \NA} \JE{\V{p}_n^\text{Array}} \, .
    \end{align*}
\end{defi}


\begin{pro}\label{thm:orien_center}
Orientation center $\V{p}^*$ exists and is unique in
orientation-unaware localization, and hence for any $\V{p} \neq
\V{p}^*$, 
    \begin{align*}
        \V{J}_\text{e}^\text{Array}(\V{p}) \prec \V{J}_\text{e}^\text{Array}(\V{p}^*) \, .
    \end{align*}
\end{pro}

\begin{IEEEproof}
See Appendix \ref{apd:Proofs_Array_orien_center}.
\end{IEEEproof}

\begin{remrk}
The orientation center $\V{p}^*$ generally depends on the topology
of the anchors and the agent, the properties of the received
waveforms, the array geometry, and the array orientation. Since
$\V{q} = \V{0}$ at the orientation center, the EFIMs for the array
center and the orientation do not depend on each other, and hence
the SPEB and SOEB can be calculated separately. The proposition also
implies that the SPEB of reference points other than $\V{p}^*$ will
be strictly larger than that of $\V{p}^*$. The SPEB for any
reference point is given in the next theorem.
\end{remrk}

\begin{cor}\label{thm:SPEB_Relation}
The SOEB $\PEB(\varphi)$ is independent of the reference point
$\V{p}$, and the SPEB is
    \begin{align}\label{eq:Anal_P_Pstar}
        \PEB(\V{p})  = \PEB(\V{p}^*) + \|\V{p} - \V{p}^*
        \|^2 \cdot \PEB(\varphi) \, .
    \end{align}
\end{cor}

\begin{IEEEproof}
See Appendix \ref{apd:Proofs_Array_SPEB_Relation}.
\end{IEEEproof}

\begin{remrk}
The SOEB does not depend on the specific reference point, which was
not apparent in \eqref{eq:Anal_EFIM_AOA_phi}. However, this is
intuitive since different reference points only introduce different
translations, but not rotations.  On the other hand, different
reference point $\V{p}$ results in different $h_{n k}$'s and hence
different $\V{q}$, which in turn gives different EFIM for position
(see \eqref{eq:Anal_EFIM_AOA}).  We can interpret the relationship
in \eqref{eq:Anal_P_Pstar} as follows: the SPEB of reference point
$\V{p}$ is equal to that of the orientation center $\V{p}^*$ plus
the orientation-induced position error, which is proportional to
both the squared distance from $\V{p}$ to $\V{p}^*$ and the SOEB.
\end{remrk}

\subsection{EFIM with A Priori Knowledge}\label{Sec:Array_MIMO}

We now consider a scenario in which the channel parameter vector
$\B{\kappa}_{n k}$ independent for different $n$'s and $k$'s. The
independence assumption serve as a reasonable approximation of many
realistic scenarios, especially near-field cases. When the different
sets of channel parameters are correlated, our results provide an
upper bound for the EFIM.

\begin{pro}
When {a priori} knowledge of channel parameters is available and the
set of channel parameters corresponding to different anchors and
antennas are mutually independent, the RII $\lambda_{n k}$ becomes
\eqref{eq:Apd_Ra_def_AOA}.
\end{pro}

\begin{IEEEproof}
See Appendix \ref{apd:Proofs_Array_NLOS_gen_AOA}.
\end{IEEEproof}

We then consider the case where a priori knowledge of the agent's
position and orientation is available in addition to channel
knowledge. Note that the topology of the agent's antennas and
anchors changes with the agent's positions and orientations. The
expression of the EFIM can be derived analogous to
\eqref{eq:Apd_EFIM_WPrior_Pos}, which is involved in general. Again
to gain insights about the contribution of a priori position and
orientation knowledge, we consider scenarios under condition
    \begin{align}\label{eq:Appro_Condition_Array}
        \E_{\V{p},\varphi}\left\{ g(\V{p},\varphi)\right\} =
        g(\bar{\V{p}},\bar\varphi) \,,
    \end{align}
where $\bar\varphi=\E_{\varphi}\left\{ \varphi\right\}$, for some
functions $g(\cdot)$ involved in the derivation of the EFIM.

\begin{cor}\label{thm:EFIM_Array_Prior}
When a priori position and orientation distribution of the agent
satisfies \eqref{eq:Appro_Condition_Array}, and the sets of channel
parameters corresponding to different anchors and antennas are
mutually independent, the EFIMs for the position and the
orientation, using an $\Na$-antenna array, are given respectively by
    \begin{align*}
        \V{J}_\text{e}^\text{Array}(\V{p}) & =
                \GP + \sum_{n \in \NA} \sum_{k \in \NB} \bar{\lambda}_{n k}
                \, \R(\bar{\phi}_{n k})  \nonumber\\
                & \hspace{5mm}
                -   \frac{1}{\sum_{n \in \NA} \sum_{k \in \NB} \bar\lambda_{n k} \,
                \bar{h}_{n k}^2 + \Gph} \, \bar{\V{q}}\, {\bar{\V{q}}}^\text{T} \,,
    \end{align*}
and
    \begin{align*}
         J_\text{e}^\text{Array}(\varphi) &=
         \Gph + \sum_{n \in \NA} \sum_{k \in \NB} \bar\lambda_{n k} \bar{h}_{n k}^2
         \nonumber \\
         & \hspace{9mm} - {\bar{\V{q}}}^\text{T}  \left(\sum_{n \in \NA} \sum_{k \in \NB} \bar{\lambda}_{n k}
         \,\R(\bar{\phi}_{n k}) + \GP \right)^{-1}\!\! \bar{\V{q}}
         \,,
    \end{align*}
where $\bar\lambda_{n k}$, $\bar{\phi}_{n k}$, $\bar{h}_{n k}$ and
$\bar{\V{q}}$ are corresponding functions in Theorem
\ref{thm:EFIM_NLOS_gen_AOA} of $\bar{\V{p}}$ and $\bar\varphi$,
respectively, and $\Gph =
\FIM{\B\theta}{\varphi}{\varphi}{\varphi}$.
\end{cor}

\begin{IEEEproof}
(Outline) The proof of this corollary is analogous to that of
Theorem \ref{thm:EFIM_NLOS_gen_AOA}. Note that when condition
\eqref{eq:Appro_Condition_Array} is satisfied, the a priori
knowledge of position and orientation for localization can be
characterized in the EFIM by using the approximation as in the proof
of Proposition \ref{thm:EFIM_FarField}.
\end{IEEEproof}

\subsection{Discussions} \label{Sec:Array_W_Position}

\subsubsection{Far-field scenarios}
The antennas in the array are closely located in far-field
scenarios, such that the received waveforms from each anchor
experience statistically the same propagation channels. Hence we
have $\phi_{n k} = \phi_k$ and $\lambda_{n k} = \lambda_k$ for all
$n$, leading to $\JE{\V{p}_n^\text{Array}} = \JE{\V{p}}$. We define
an important reference point as follows.

\begin{defi}[Array Center]
The array center is defined as the position $\V{p}_0$, satisfying
    \begin{align*}
        \sum_{n \in \NA} \Delta x_n(\V{p}_0,\varphi) = 0
        \quad \text{and} \quad
        \sum_{n \in \NA} \Delta y_n(\V{p}_0,\varphi) = 0
        \,.
    \end{align*}
\end{defi}

\begin{pro}\label{Thm:Array_Center}
The array center becomes the orientation center in far-field
scenarios.
\end{pro}

\begin{IEEEproof}
See Appendix \ref{apd:Proofs_Array_Center}.
\end{IEEEproof}

\begin{remrk}
Since the orientation center has the minimum SPEB, Proposition
\ref{Thm:Array_Center} implies that the array center always achieves
the minimum SPEB in far-field scenarios. Hence the array center is a
well-suited choice for the reference point, since its position can
be determined from the array geometry alone, without requiring the
received waveforms and the knowledge of the anchor's topology.
\end{remrk}

In far-field scenarios, we choose the array center $\V{p}_0$ as the
reference point. The results of Theorem \ref{thm:EFIM_NLOS_gen_AOA}
become
    \begin{align*}
        \V{J}_\text{e}^\text{Array}(\V{p}_0) & = \Na \, \sum_{k \in \NB}
                \lambda_k \, \R(\phi_k)\,,  \nonumber \\
        \noalign{\noindent and \vspace{\jot}}
        J_\text{e}^\text{Array}(\varphi) & = \sum_{n \in \NA} \sum_{k \in \NB}
                \lambda_k  \bar{h}_{n k}^2 \,,
    \end{align*}
where $\bar{h}_{n k}$ is a function of $\V{p}_0$. Similarly, when
the a priori position and orientation knowledge is available and
condition \eqref{eq:Appro_Condition_Array} is satisfied, the results
of Proposition \ref{thm:EFIM_Array_Prior} become
    \begin{align*}
        \V{J}_\text{e}^\text{Array}({\V{p}}_0) & = \Na \, \sum_{k \in \NB}
                \bar{\lambda}_{k} \, \R(\bar{\phi}_{k}) + \GP\,, \nonumber \\
        \noalign{\noindent and \vspace{\jot}}
        J_\text{e}^\text{Array}(\varphi) & = \sum_{n \in \NA} \sum_{k \in \NB}
                {\bar\lambda_k}\,\bar{h}_{n k}^2 + \Gph\,,
    \end{align*}
where $\bar{h}_{n k}$ is a function of $\bar{\V{p}}_0 =
\E_{\V{p}_0}\{\V{p}_0\}$.

Note that the localization performance of an $\Na$-antenna array is
equivalent to that of a single antenna with $\Na$ measurements,
regardless of the array geometry, in far-field scenarios.

\subsubsection{Multiple antennas at anchors}
When anchors are equipped with multiple antennas, each antenna can
be viewed as an individual anchor. In this case, the agent's SPEB
goes down with the number of the antennas at each anchor. Note that
all the antennas of a given anchor provide RI approximately in the
same direction with the same intensity, as they are closely located.

\begin{figure*}[!b]
  \vspace*{4pt}%
  \hrulefill%
  \normalsize%
  \setcounter{MYtempeqncnt}{\value{equation}}%
  \setcounter{equation}{36}%
  \renewcommand{\arraystretch}{1.3}
    \begin{align}\label{eq:Anal_EFIM_TDOA/AOA}
        \V{J}_\text{e}^\text{Array-B} = \Matrix{ccc}{ \sum_{n \in \NA} \sum_{k \in \NB} \lambda_{n k}\, \V{q}_{n k}\, \V{q}_{n k}^\text{T}
                            & \sum_{n \in \NA} \sum_{k \in \NB} h_{n k}\,\lambda_{n k}\, \V{q}_{n k}
                            & \sum_{n \in \NA} \sum_{k \in \NB} \lambda_{n k}\, \V{q}_{n k} \\
                            \sum_{n \in \NA} \sum_{k \in \NB} h_{n k}\, \lambda_{n k} \,\V{q}_{n k}^\text{T}
                            & \sum_{n \in \NA} \sum_{k \in \NB} h_{n k}^2\, \lambda_{n k} + \Xi
                            & \sum_{n \in \NA} \sum_{k \in \NB} h_{n k}\, \lambda_{n k} \\
                            \sum_{n \in \NA} \sum_{k \in \NB} \lambda_{n k} \,\V{q}_{n k}^\text{T}
                            & \sum_{n \in \NA} \sum_{k \in \NB} h_{n k}\, \lambda_{n k}
                            & \sum_{n \in \NA} \sum_{k \in \NB} \lambda_{n k} + \Xi_{\text{B}} }
    \end{align}%
  \setcounter{equation}{\value{MYtempeqncnt}}
  \vspace*{-6pt}
\end{figure*}

\subsubsection{Other related issues}
Other issues related to localization using wideband antenna arrays
include the AOA estimation, the effect of multipath geometry, and
the effect of array geometries. A more comprehensive performance
analysis can be found in \cite{SheWin:J10}.

%
%

\section{Effect of Clock Asynchronism}\label{Sec:ClockBias}

In this section, we consider scenarios in which the clocks of all
anchors are perfectly synchronized but the agent operates
asynchronously with the anchors \cite{LovTow:02}. In such a
scenario, the one-way time-of-flight measurement contains a time
offset between the agent's clock and the anchors' clock.\footnote{We
consider scenarios in which localization time is short relative to
clock drifts, such that the time offset is the same for all
measurements from the anchors.} Here, we investigate the effect of
the time offset on localization accuracy.

\subsection{Localization with a Single Antenna}\label{Sec:ClockBias_Single}

Consider the scenario described in Section \ref{Sec:Model}, where
each agent is equipped with a single antenna. When the agent
operates asynchronously with the anchors, the relationship of
\eqref{eq:Model_Tau_L} becomes
    \begin{align*}
        \D{k}{l} = \frac{1}{c} { \Big [} \, \| \V{p} - \V{p}_k \|
                    + \BB{k}{l} + B \, {\Big ]},
    \end{align*}
where $B$ is a random parameter that characterizes the time offset
in terms of distance, and the corresponding parameter vector
$\B\theta$ becomes
    \begin{align*}
        \B\theta = \Matrix{ccccccc}{\V{p}^\text{T} & B
            & \B{\kappa}_1^\text{T} & \B{\kappa}_2^\text{T}
            & \cdots & \B{\kappa}_{\Nb}^\text{T} }^\text{T} .
    \end{align*}
Similar to Theorem \ref{thm:EFIM_NLOS_gen}, where $\V{p}$ is
deterministic but unknown and the remaining parameters are random,
we have the following result.

\begin{thm}\label{thm:EFIM_NLOS_gen_TDOA}
When a priori knowledge of the channel parameters and the time
offset is available, and the sets of channel parameters
corresponding to different anchors are mutually independent, the
EFIMs for the position and the time offset are given respectively by
    \begin{align}\label{eq:Anal_EFIM_TDOA}
        \V{J}_\text{e}^\text{B}(\V{p})  =
                \sum_{k \in \NB} \lambda_k \, \R(\phi_k)
                - \frac{1}{\sum_{k \in \NB} \lambda_k  + \Xi_{\text{B}}}
                \, \V{q}_\text{B}\, \V{q}_\text{B}^\text{T}
    \end{align}
and
    \begin{align}\label{eq:Anal_EFIM_TDOA_B}
        J_\text{e}(B) = \sum_{k \in \NB} \lambda_k + \Xi_{\text{B}}
            - \V{q}_\text{B}^\text{T}  \left( \sum_{k \in \NB} \lambda_k
            \, \R(\phi_k) \right)^{-1} \!\!\V{q}_\text{B}
        \, ,
    \end{align}
where $\lambda_k$ is given by \eqref{eq:Apd_tRa_def},
$\V{q}_\text{B} = \sum_{k \in \NB} \lambda_k \, \q_k$, and
    \begin{align*}
        \Xi_{\text{B}} \triangleq \FIM{\B\theta}{B}{B}{B} \, .
    \end{align*}
\end{thm}

\begin{IEEEproof}
See Appendix \ref{apd:Proofs_ClockBias_NLOS_gen_TDOA}.
\end{IEEEproof}

\begin{remrk}
Since $ \V{q}_\text{B}\,\V{q}_\text{B}^\text{T} $ is a positive
semi-definite matrix and $\sum_{k \in \NB} \lambda_k$ is positive in
\eqref{eq:Anal_EFIM_TDOA}, compare to Theorem
\ref{thm:EFIM_NLOS_gen}, we always have the inequality
    \begin{align}\label{eq:Anal_TDOA_TOA}
        \V{J}_\text{e}^\text{B}(\V{p})\preceq \V{J}_\text{e}(\V{p})\,,
    \end{align}
where the equality in \eqref{eq:Anal_TDOA_TOA} is achieved for
time-offset-known localization (i.e., $\Xi_{\text{B}} = \infty$), or
time-offset-independent localization (i.e., $\V{q}_\text{B}
=\V{0}$). The former corresponds to the case where accurate
knowledge of the time offset is available, while the latter depends
on the RII from each anchor, and the topology of the anchors and
agent. The inequality of \eqref{eq:Anal_TDOA_TOA} results from the
uncertainty in the additional parameter $B$, which degrades the
localization accuracy. Hence the SPEB in the presence of uncertain
time offset is always larger than or equal to that without a offset
or with a known offset.
\end{remrk}

We next consider the case where a priori knowledge of the agent's
position is available. When the a priori position distribution of
the agent satisfies \eqref{eq:Appro_Condition}, we have the
following corollary.

\begin{cor}
When the a priori position distribution of the agent satisfies
\eqref{eq:Appro_Condition}, and the sets of channel parameters
corresponding to different anchors are mutually independent, the
EFIMs for the position and the time offset are given respectively by
    \begin{align*}
        \V{J}_\text{e}^\text{B}(\V{p}) =
                \sum_{k \in \NB} \bar\lambda_k \, \R(\bar{\phi}_k) +  \GP
                - \frac{1}{\sum_{k \in \NB} \bar\lambda_k  + \Xi_{\text{B}}}
                \, \bar{\V{q}}_\text{B}\, \bar{\V{q}}_\text{B}^\text{T} \, ,
    \end{align*}
and
    \begin{align*}
        J_\text{e}(B) =  \sum_{k \in \NB} \bar\lambda_k + \Xi_{\text{B}}
            - \bar{\V{q}}_\text{B}^\text{T}  \left(\GP +  \sum_{k \in \NB} \bar\lambda_k
            \, \R(\bar{\phi}_k) \right)^{-1}\!\!
            \bar{\V{q}}_\text{B} \,,
    \end{align*}
where $\bar{\phi}_k$ is the angle from anchor $k$ to $\bar{\V{p}}$,
$\bar\lambda_k$ is given by \eqref{eq:Apd_RII_Pbar}, and
$\bar{\V{q}}_\text{B}$ is a function of $\bar{\V{p}}$.
\end{cor}

\begin{IEEEproof}
(Outline) Conditions in \eqref{eq:Appro_Condition} hold in far-field
scenarios, and we can approximate the expectation over random
parameter vector $\V{p}$ using the average position $\bar{\V{p}}$.
By following the steps of Theorem \ref{thm:EFIM_NLOS_gen_TDOA} and
Proposition \ref{thm:EFIM_FarField}, we can derive the theorem after
some algebra.
\end{IEEEproof}

\subsection{Localization with Antenna Arrays} \label{Sec:ClockBias_Array}

Consider the scenario describing in Section \ref{Sec:Array} where
each agent is equipped with an array of $\Na$ antennas.
Incorporating the time offset $B$, \eqref{eq:Sing_Tau_L_krray}
becomes
    \begin{align*}
        \D{n k}{l} = \frac{1}{c} {\Big[} \, \| \V{p}_n^\text{Array} - \V{p}_k \| 
        + \BB{n k}{l} + B \, {\Big ]},
    \end{align*}
and the corresponding parameter vector $\B\theta$ becomes
    \begin{align*}
        \B\theta =  \Matrix{cccccccc}{\V{p}^\text{T} & \varphi & B
                    & \check{\B\kappa}_1^\text{T} &  \check{\B\kappa}_2^\text{T}
                    & \cdots & \check{\B\kappa}_{N}^\text{T} }^\text{T}.
    \end{align*}
Similar to Theorem \ref{thm:EFIM_NLOS_gen_AOA}, where $\V{p}$ and
$\varphi$ are deterministic but unknown and the remaining parameters
are random, we have the following theorem.

\begin{thm}\label{thm:EFIM_NLOS_gen_AOA-B}
When a priori knowledge of the channel parameters is available, and
the sets of channel parameters corresponding to different anchors
and antennas are mutually independent, the EFIM for the position,
the orientation, and the time offset, using an $\Na$-antenna array,
is given by \eqref{eq:Anal_EFIM_TDOA/AOA} shown at the bottom of
the page,\addtocounter{equation}{1}
where $\Xi = \infty$ and $\Xi = 0$ correspond to orientation-aware
and orientation-unaware localization, respectively, and
$\lambda_{n k}$, $\q_{n k}$, and $h_{n k}$ are given by
\eqref{eq:Apd_Ra_def_AOA}, \eqref{eq:q_an}, and \eqref{eq:h_an},
respectively.
\end{thm}

\begin{IEEEproof}
See Appendix \ref{apd:Proofs_ClockBias_NLOS_gen_AOA-B}.
\end{IEEEproof}

\begin{remrk}
Theorem \ref{thm:EFIM_NLOS_gen_AOA-B} gives the overall $4\times4$
EFIM for the position, the orientation, and the time offset, where
individual EFIMs can be derived by applying the notion of EFI again.
\end{remrk}

We finally consider the case where a priori knowledge of the agent's
position and orientation is available. The EFIM in far-field
scenarios is given in the following corollary.

\begin{cor}\label{thm:EFIM_AOA-B_Far}
When a priori knowledge of the agent's position, orientation,
time-offset, and the channel parameters is available, and the sets
of channel parameters corresponding to different anchors and
antennas are mutually independent, in far-field scenarios, the EFIMs
for the position, the orientation, and the time offset, using an
$\Na$-antenna array, are given respectively by
    \begin{align*}
        \renewcommand{\arraystretch}{1.3}
        \V{J}_\text{e}^\text{Array-B}(\V{p}_0)
            & = \Na \, \sum_{k \in \NB} \bar{\lambda}_{k} \,
            \R(\bar{\phi}_{k}) + \GP \nonumber \\
            & \hspace{14mm}
            - \frac{1}{\Na \, \sum_{k \in \NB} \bar\lambda_k + \Xi_{\text{B}} }
            \, \bar{\V{q}}_\text{B}\, \bar{\V{q}}_\text{B}^\text{T}
            \,, \nonumber\\
        J_\text{e}^\text{Array-B}(\varphi)
            & = \sum_{n \in \NA} \sum_{k \in \NB}  \bar\lambda_k \,\bar{h}_{n k}^2
                + \Gph \, ,
    \end{align*}
and
    \begin{align*}
        J_\text{e}^\text{Array-B}(B)
            & = \Na \, \sum_{k \in \NB} \bar\lambda_k + \Xi_{\text{B}}
            \nonumber \\
            & \hspace{7mm} -  \bar{\V{q}}_\text{B}^\text{T} \left( \Na \, \sum_{k \in \NB}
             \bar{\lambda}_{k} \, \R(\bar{\phi}_{k}) + \GP \right)^{-1} \!\! \bar{\V{q}}_\text{B}
            \, ,
    \end{align*}
where $\bar{\V{p}}_0$ is the expected position of the agent's array
center, $\bar\phi_k$ is the angle from anchor $k$ to $\bar{\V{p}}_0$,
and $\bar\lambda_k$, $\bar{\V{q}}_\text{B}$, and $\bar{h}_{n k}$ are
functions of $\bar{\V{p}}_0$.
\end{cor}

\begin{IEEEproof}
See Appendix \ref{apd:Proofs_ClockBias_AOA-B_Far}.
\end{IEEEproof}

%
%

\section{Discussions}\label{Sec:Discus}

In this section, we will provide discussions on some related issues
in the paper. It includes 1) the relations of our results to the
bounds based on signal metrics, 2) the achievability of the SPEB,
and 3) generalization of the results to 3D localization.

\subsection{Relation to Bounds Based on Signal Metrics}

Analysis of localization performance in the literature mainly
employs specific signal metrics, such as TOA, AOA, RSS, and TDOA,
rather than utilizing the entire received waveforms. Our analysis is
based on the received waveforms and exploits all the localization
information inherent in these signal metrics, implicitly or
explicitly. In particular,
TOA and joint TOA/AOA metrics were incorporated in our analysis in
Section \ref{Sec:Evalu} and \ref{Sec:Array}, respectively.
Similarly, TDOA and joint TDOA/AOA metrics were included in the
analysis of Section \ref{Sec:ClockBias}, and the RSS metric has been
implicitly exploited from a priori channel knowledge in Section \ref{Sec:FIM_WPrior}.

\subsection{Achievability of the SPEB}

Maximum a posterior (MAP) and maximum likelihood (ML) estimates
respectively achieve the CRB asymptotically in the high SNR regimes
for both the case with and without a priori knowledge \cite{Tre:68}.
High SNR can be attained using sequences with good correlation
properties \cite{GolSch:65, GolWin:J98, GolGon:B05}, or simply
repeated transmissions. Therefore, the SPEB is achievable.

\subsection{Generalization to 3D Localization}\label{Sec:Evalu_3D}

All results obtained thus far can be easily extended to
three-dimensional case, i.e., $\V{p} = [\, x \;\; y  \;\; z
\,]^\text{T}$ and the RDM becomes
    \begin{align*}
        \R(\varphi_{k},\phi_{k}) = \q_{k} \, \q_{k}^\text{T}
        \, ,
    \end{align*}
where $\varphi_{k}$ and $\phi_{k}$ are the angles in the
coordinates, and
    \begin{align*}
        \q_{k} = \Matrix{ccc} {  \cos\varphi_{k}
        \cos\phi_{k}  & \sin\varphi_{k} \cos\phi_{k}  &  \sin\phi_{k} }^\text{T}.
    \end{align*}
Similarly, we can obtain a corresponding $3 \times 3$ EFIM in the
form of \eqref{eq:Sing_EFIM_Prior}.

%
%

\section{Numerical Results}\label{Sec:Simu}

In this section, we illustrate applications of our analytical
results using numerical examples.  We deliberately restrict our
attention to a simple network to gain insights, although our
analytical results are valid for arbitrary topology with any number
of anchors and any number of MPCs in the received waveforms.

\begin{figure}[t]
    \begin{center}
    \psfrag{A1}[l][][1.3]{\hspace{-2mm}$A_1$}
    \psfrag{A2}[l][][1.3]{\hspace{-2mm}$A_2$}
    \psfrag{A3}[l][][1.3]{\hspace{-2mm}$A_3$}
    \psfrag{A4}[l][][1.3]{\hspace{-1mm}$A_4$}
    {\includegraphics[angle=0,width=0.65\linewidth,draft=false]{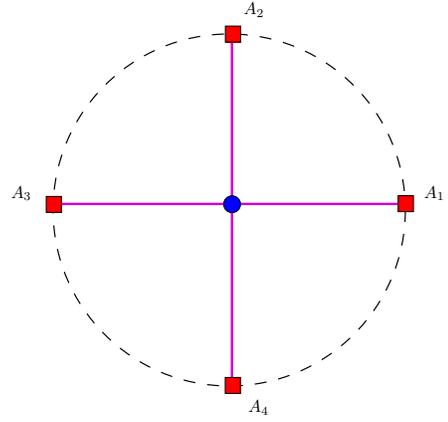}}

    \caption{Network topology: Four anchors are equally spaced
    on a circle with an agent at the center. All signals from
    the anchors to the agent are LOS.
    \label{fig:Sim_network_Noncoop}}
    \end{center}
\end{figure}

\subsection{Effect of Path-Overlap}
\label{Sec:Simu_A}

We first investigate the effect of path-overlap on the SPEB when {a
priori} knowledge is unavailable. In particular, we compare the SPEB
obtained by the \emph{full-parameter model} proposed in this paper
and that obtained by the \emph{partial-parameter model} proposed in
\cite{QiKob:02a}. In the partial-parameter model, the amplitudes of
MPCs are assumed to be known and hence excluded from the parameter
vector.

\begin{figure}
    \begin{center}

    \psfrag{AAAAAAAAAAA1}[l][][1.3]{\hspace{-11.5mm}Full-parameter}
    \psfrag{AAAAAAAAAAA2}[l][][1.3]{\hspace{-12mm}Partial-parameter}
    \psfrag{AAAAAAAAAAA3}[l][][1.3]{\hspace{-12mm}Non-overlap}
    \psfrag{xlabel}[c][][1.3]{Path separation (ns)}
    \psfrag{ylabel}[c][][1.3]{SPEB ($\text{m}^2$)}
    \psfrag{title}[c][][1.3]{}

    {\includegraphics[angle=0,width=1\columnwidth,draft=false]{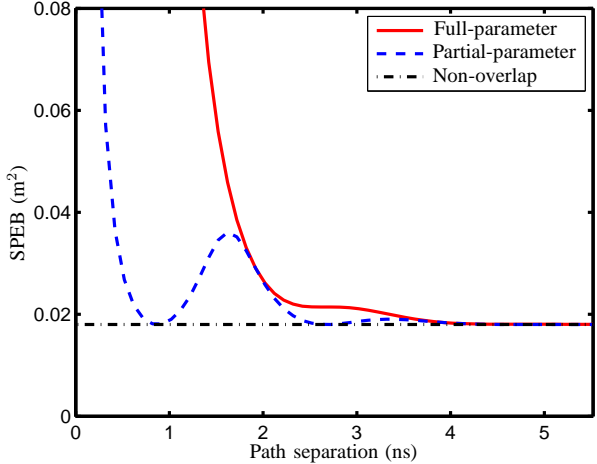}}

    \caption{SPEB as a function of path separation for the full-parameter,
    partial-parameter, and non-overlap models, without a priori
    knowledge.
    \label{fig:peb_with_overlap_anchor}}
    \end{center}
\end{figure}

Consider a simple network with four anchors ($\Nb = 4$) equally
spaced on a circle and an agent at the center receiving all LOS
signals (see Fig.~\ref{fig:Sim_network_Noncoop}). Each waveform
consists of two paths: one LOS path ($\SNR{k}{1} = 0~\text{dB}$) and
one NLOS path ($\SNR{k}{2} = -3~\text{dB}$), and the separations of
the two paths $\D{k}{2} - \D{k}{1}$ are identical for all $k$. 
In addition, the transmitted waveform is a second derivative of
Gaussian pulse with width approximately equal to 4 ns.
Figure~\ref{fig:peb_with_overlap_anchor} shows the SPEB as a
function of path separation $\D{k}{2} - \D{k}{1}$ according to
Theorem \ref{Thm:EFIM_NPrior}.

We can draw the following observations. First of all, path-overlap
increases the SPEB in both models, since it reduces the ability to
estimate the first path and hence decreases the RII. Note that the
shape of the curves depends on the autocorrelation function of the
waveform $s(t)$ \cite{SheWin:C08}. Secondly, when the path
separation exceeds the pulse width (approximately 4 ns), the two
models give the same SPEB, which equals the non-overlapping case. In
such cases, the first contiguous-cluster contains only the first
path, and hence the RII is determined by the first paths. This
agrees with the analysis in Section \ref{Sec:Evalu}. Thirdly,
excluding the amplitudes from the parameter vector incorrectly
provides more RI when the two paths overlap, and hence the
partial-parameter model results in a loose bound. This demonstrates
the importance of using the full-parameter model.

\subsection{Improvement from {A Priori} Channel Knowledge}\label{Sec:Simu_B}

\begin{figure}[t]
    \begin{center}
    \subfigure[SPEB with a priori knowledge of $\A{k}{1}$ and $\A{k}{2}$,
    while $\Xi_{\BB{k}{2}}=0$.]
    { \label{fig:PEB_with_NLOS_A_t}
    \psfrag{AAAAAAAAAAAAA1}[l][][1.2]{\hspace{-16mm}$\Xi_{\A{k}{1}}=\Xi_{\A{k}{2}}=0$}
    \psfrag{AAAAAAAAAAAAA2}[l][][1.2]{\hspace{-16mm}$\Xi_{\A{k}{1}}=\infty, \Xi_{\A{k}{2}}=0$}
    \psfrag{AAAAAAAAAAAAA3}[l][][1.2]{\hspace{-16mm}$\Xi_{\A{k}{1}}=0, \Xi_{\A{k}{2}}=\infty$}
    \psfrag{AAAAAAAAAAAAA4}[l][][1.2]{\hspace{-16mm}$\Xi_{\A{k}{1}}=\Xi_{\A{k}{2}}=\infty$}
    \psfrag{AAAAAAAAAAAAA5}[l][][1.2]{\hspace{-16mm}$\Xi_{\A{k}{1}}=\Xi_{\A{k}{2}}=\infty$}
    \psfrag{AAAAAAAAAAAAA6}[l][][1.2]{\hspace{-16mm}Non-overlap}
    \psfrag{xlabel}[c][][1.3]{Path separation (ns)}
    \psfrag{ylabel}[c][][1.3]{SPEB ($\text{m}^2$)}
    \psfrag{title}[c][][1.3]{}
    \includegraphics[angle=0,width=1\columnwidth,draft=false]{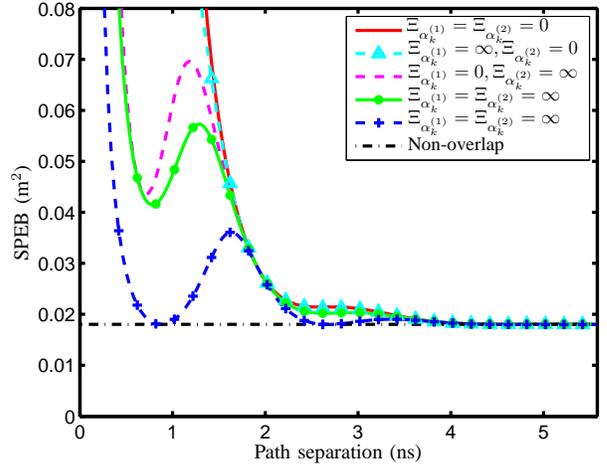}}\\
    \hfill \\
    \psfrag{AA1}[l][][1.2]{\hspace{-3mm}$\Xi_{\BB{k}{2}}=0$}
    \psfrag{AA2}[l][][1.2]{\hspace{-3mm}$\Xi_{\BB{k}{2}}=20$}
    \psfrag{AA3}[l][][1.2]{\hspace{-3mm}$\Xi_{\BB{k}{2}}=\infty$}

    \psfrag{AAAAAAAAAAA1}[l][][1.2]{\hspace{-12mm}Full-parameter}
    \psfrag{AAAAAAAAAAA2}[l][][1.2]{\hspace{-12.5mm}Partial-parameter}
    \psfrag{AAAAAAAAAAA3}[l][][1.2]{\hspace{-12.5mm}Non-overlap}

    \psfrag{xlabel}[c][][1.3]{Path separation (ns)}
    \psfrag{ylabel}[c][][1.3]{SPEB ($\text{m}^2$)}
    \psfrag{title}[c][][1.3]{}

    \subfigure[SPEB with a priori knowledge of $\BB{k}{2}$,
    while $\Xi_{\A{k}{1}} = \Xi_{\A{k}{2}} = 0$.]
    {    \label{fig:PEB_with_NLOS_A_t2}
 \includegraphics[angle=0,width=1\columnwidth,draft=false]{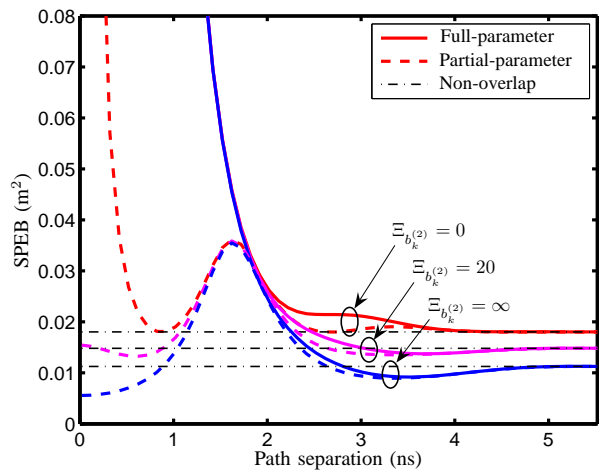}}
    \caption{SPEB with a priori knowledge of the amplitudes and
    the NLOS biases as a function of path separation, respectively.}
    \end{center}
\end{figure}

We then quantify the contribution of the a priori knowledge of
channel parameters to the SPEB. The network topology and channel
parameters are the same as those in Section \ref{Sec:Simu_A}, except
a priori knowledge of $\A{k}{1}$, $\A{k}{2}$ and $\BB{k}{2}$ is now
available. For simplicity, we consider these parameters to be
independent {a priori} and denote the a priori Fisher information of parameter $\theta_1$ by $\Xi_{\theta_1} = \FIM{\B\theta}{\B\theta}{{\theta_1}}{{\theta_1}}$. In Fig.~\ref{fig:PEB_with_NLOS_A_t}, the SPEBs are plotted as functions of the path separation for different {a priori} knowledge of $\A{k}{1}$ and $\A{k}{2}$ (no a priori
knowledge of $\BB{k}{2}$); while in
Fig.~\ref{fig:PEB_with_NLOS_A_t2}, the SPEBs are plotted for
different a priori knowledge of $\BB{k}{2}$ (no a priori knowledge
of $\A{k}{1}$ and $\A{k}{2}$).

We have the following observations. First of all, the SPEB decreases with
the a priori knowledge of the amplitudes and the NLOS biases. This
should be expected since a priori channel knowledge increases the
RII and thus localization accuracy, as indicated in Corollary
\ref{cor:prior_knowledge}. Moreover, the NLOS components are shown
to be beneficial for localization in the presence of a priori biases
knowledge, as proven in Section \ref{Sec:Evalu_WPrior}. Secondly, as
the a priori knowledge of the amplitudes approaches infinity, the
SPEB in Fig.~\ref{fig:PEB_with_NLOS_A_t} obtained using the
full-parameter model converges to that in
Fig.~\ref{fig:peb_with_overlap_anchor} obtained using the
partial-parameter model. This is because the partial-parameter model
excludes the amplitudes from the parameter vector, which is
equivalent to assuming known amplitudes and hence infinite a priori
Fisher information for the amplitudes ($\Xi_{\A{k}{1}} =
\Xi_{\A{k}{2}} = \infty$). Thirdly, it is surprising to observe that,
when the a priori knowledge of the NLOS biases is available,
path-overlap can result in a lower SPEB compared to non-overlapping
scenarios. This occurs at certain regions of path separations,
depending on the autocorrelation function of $s(t)$. Intuitively,
path-overlap can lead to a higher SNR compared to non-overlapping
cases, when a priori knowledge of the NLOS biases is available.

\subsection{Path-Overlap Coefficient}

\begin{figure}
    \begin{center}
    {\psfrag{AAAAAAA1}[l][][1.2]{\hspace{-5.5mm}$L = 2$}
    \psfrag{AAAAAAA2}[l][][1.2]{\hspace{-6mm}$L = 3$}
    \psfrag{AAAAAAA3}[l][][1.2]{\hspace{-6mm}$L = 5$}
    \psfrag{AAAAAAA4}[l][][1.2]{\hspace{-6mm}$L = 50$}
    \psfrag{xlabel}[c][][1.3]{Path inter-arrival time $1/\nu$ (ns)}
    \psfrag{ylabel}[c][][1.3]{Average POC $\chi$}
    \psfrag{title}[c][][1.3]{}

    \includegraphics[angle=0,width=1\columnwidth,draft=false]{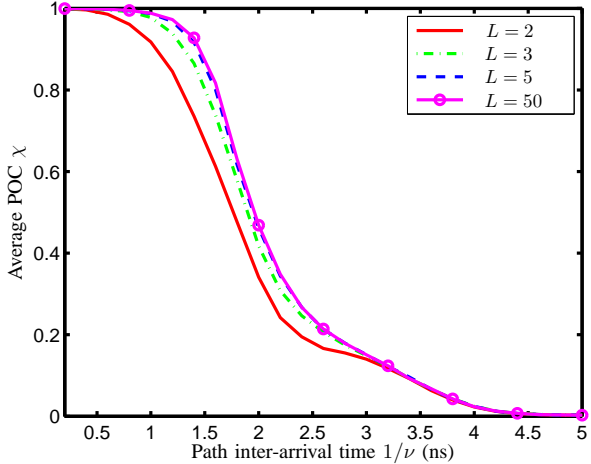}}

    \caption{POC as a function of the path inter-arrival time for different number of MPCs. \label{fig:overlap_coefficient}}
    \end{center}
\end{figure}

We now investigate the dependence of POC $\chi$ on path arrival
rate. We first generate channels with $L$ MPCs according to a simple
Poisson model with a fixed arrival rate $\nu$, and then calculate
$\chi$ according to \eqref{eq:Apd_R_a'_full}.
Figure~\ref{fig:overlap_coefficient} shows the average path-overlap
coefficient as a function of path inter-arrival rate ($1/\nu$) for
different number of MPCs, where the averaging is obtained by
Monte-Carlo simulations.

We have the following observations. First of all, the POC $\chi$ is
monotonically decreasing from 1 to 0 with $1/\nu$. This agrees with
our intuition that denser multipath propagation causes more
interference between the first path and other MPCs, and hence the
received waveform provides less RII. Secondly, for a fixed $\nu$, the
POC increases with $L$. This should be expected as additional MPCs
may interfere with earlier paths, which degrades the estimation
accuracy of the first path and thus reduces the RII. Thirdly, observe
that beyond $L=5$ paths, $\chi$ does not increase significantly.
This indicates that the effect of additional MPCs beyond the fifth
path do not contribute to the RII, regardless of the power
dispersion profile of the received waveforms.

\subsection{Outage in Ranging Ability}

We have observed that the channel quality for ranging is
characterized by the POC.  If the multipath propagation has a larger
POC (close to 1), we may consider the channel in outage for ranging.
We define the ranging ability outage (RAO) as
    \begin{align*}
        p_\text{out}(\chi_\text{th}) \triangleq
        \mathbb{P}\{ \chi > \chi_\text{th} \} ,
    \end{align*}
where $\chi_\text{th}$ is the threshold for the POC. The RAO tells
us that with probability $p_\text{out} (\chi_\text{th})$, the
propagation channel is unsatisfactory for ranging.

\begin{figure}
    \begin{center}
    {
    \psfrag{A1}[l][][1.2]{\hspace{-2mm} $1/\nu$ decreases}
    \psfrag{A3}[l][][1.2]{\hspace{0.5mm} $1/\nu = 3.5$}
    \psfrag{A2}[l][][1.2]{\hspace{3.5mm} $1/\nu = 1.4$}
    \psfrag{xlabel}[c][][1.3]{Threshold $\chi_\text{th}$}
    \psfrag{ylabel}[c][][1.3]{$p_\text{out}(\chi_\text{th})$}
    \psfrag{title}[c][][1.3]{}

    \includegraphics[angle=0,width=1\columnwidth,draft=false]{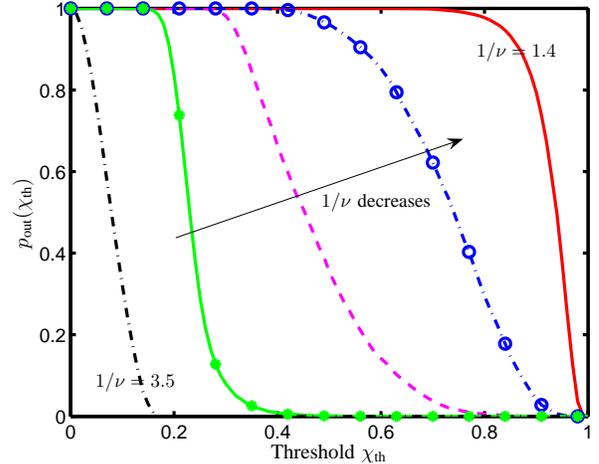} }

    \caption{RAO as a function of the threshold $\chi_\text{th}$ for     different path inter-arrival time $1/\nu$ with $L=50$. The five curves correspond to inter-arrival time $1/\nu = 3.5, 2.5, 2, 1.6, 1.4$ ns, respectively.  \label{fig:outage_prob}}
    \end{center}
\end{figure}

The RAO as a function of $\chi_\text{th}$ for different Poisson
arrival rate is plotted in Fig.~\ref{fig:outage_prob} for a channel
with $L=50$. The RAO decreases from 1 to 0, as the threshold
$\chi_\text{th}$ increases or the path arrival rate $\nu$ decreases.
This should be expected because the probability of path-overlap
decreases with the path arrival rate, and consequently decreases the
RAO. The RAO can be used as a measure to quantify the channel
quality for ranging and to guide the design of the optimal
transmitted waveform for ranging.

\subsection{SPEB and SOEB for Wideband Antenna Array Systems}
\label{Sec:Simu_C}

\begin{figure}[t]
    \begin{center}
    \subfigure[SPEB as a function of the
    reference point-to-array center distance]
    { \label{fig:dif_ref_pos}
    \psfrag{AAAAAA1}[l][][1.2]{\hspace{-6.5mm}$\Gph = 0$}
    \psfrag{AAAAAA2}[l][][1.2]{\hspace{-7mm}$\Gph = 20$}
    \psfrag{AAAAAA3}[l][][1.2]{\hspace{-7mm}$\Gph = 200$}
    \psfrag{AAAAAA4}[l][][1.2]{\hspace{-7mm}$\Gph = \infty$}
    \psfrag{xlabel}[c][][1.3]{Distance from the reference point to the array center (m)}
    \psfrag{ylabel}[c][][1.3]{SPEB $(\text{m}^2)$}
    \psfrag{title}[c][][1.3]{}

    \includegraphics[angle=0,width=1\columnwidth,draft=false]{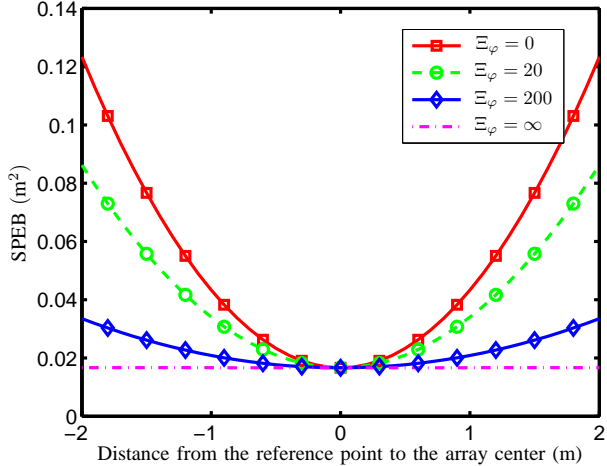}}\\
    \hfill \\
    \subfigure[SOEB as a function of the
    reference point-to-array center distance]
    { \label{Dif_ref_pos_ang}
    \psfrag{AAAAAAAAAAAA1}[l][][1.2]{\hspace{-14.5mm}$\GP \!=\! \Diag{0,0}$}
    \psfrag{AAAAAAAAAAAA2}[l][][1.2]{\hspace{-15mm}$\GP \!=\! \Diag{10,10}$}
    \psfrag{AAAAAAAAAAAA3}[l][][1.2]{\hspace{-15mm}$\GP \!=\! \Diag{10^2,10^2}$}
    \psfrag{AAAAAAAAAAAA4}[l][][1.2]{\hspace{-15mm}$\GP \!=\! \Diag{\infty,\infty}$}
    \psfrag{xlabel}[c][][1.3]{Distance from the reference point to the array center (m)}
    \psfrag{ylabel}[c][][1.3]{SOEB $(\text{rad}^2)$}
    \psfrag{title}[c][][1.3]{}

    \includegraphics[angle=0,width=1\columnwidth,draft=false]{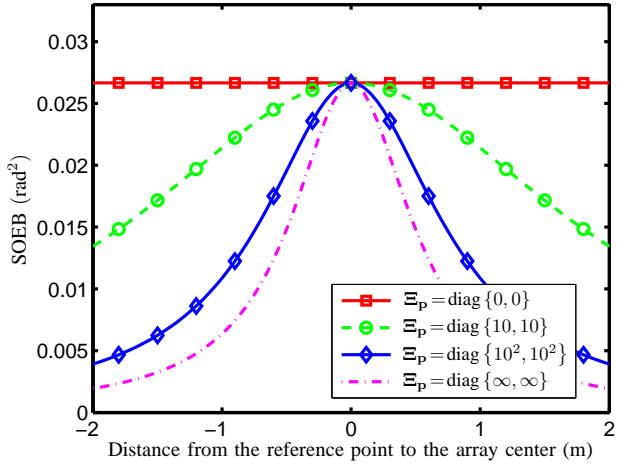}}

    \caption{SPEB and SOEB with different a priori knowledge of agent's position and orientation, respectively}
    \end{center}
\end{figure}

We consider the SPEB and SOEB for different reference points of a uniform linear array (ULA). The numerical results are based on a network with six equally spaced anchor nodes ($\Nb=6$) located on a circle with an agent in the center.  The agent is equipped with a 4-antenna array ($\Na=4$) whose spacing is 0.5 m.  In far-field scenarios, $\lambda_{n k} = \lambda_k = 10$ and $\phi_{n k}=\phi_k$. Fig.~\ref{fig:dif_ref_pos} and Fig.~\ref{Dif_ref_pos_ang} show the SPEB and SOEB, respectively, as a function of different reference point along the ULA for different a priori knowledge of the orientation and reference point.

We have the following observations. First of all, a priori knowledge of the orientation improves the localization accuracy as the SPEB decreases with $\Gph$. The curves for $\Gph = 0$ and $\Gph = \infty$ correspond to the orientation-unaware and orientation-aware cases, respectively. As a counterpart, a priori knowledge of the reference point improves the orientation accuracy as the SOEB decreases with $\GP$. This agrees with both intuition and Theorem \ref{thm:EFIM_NLOS_gen_AOA}. Secondly, the array center has the best localization accuracy, and its SPEB does not depend on $\Gph$, which agrees with Theorem \ref{thm:EFIM_NLOS_gen_AOA}. On the other hand, the array center exhibits the worst orientation accuracy, and its SOEB does not depend on $\GP$. This should be expected since the knowledge for the array center tells nothing about the array orientation. Thirdly, the SPEB increases with both the distance from the reference point to the array center and the SOEB, as predicted by Corollary \ref{thm:SPEB_Relation}. On the contrary, the SOEB decreases as a function of the distance from the reference point to the array center if a priori knowledge of the reference point is available. This observation can be verified by Theorem \ref{thm:EFIM_NLOS_gen_AOA}. Last but not least, the SPEB is independent of specific reference point if $\Gph = \infty$, as referred to orientation-aware localization, and the SOEB is independent of the specific reference point if $\GP=\V{0}$, as shown in Corollary \ref{thm:SPEB_Relation}.

%

%
%
%

\subsection{SPEB with Time Offset and Squared Timing Error Bound}

\begin{figure}[t]
    \begin{center}
    \subfigure[SPEB as a function of Anchor $A_1$'s position]
    { \label{fig:TDOA}
    \psfrag{AAAAA1}[l][][1.2]{\hspace{-5.5mm}$\Xi_{\text{B}} = 0$}
    \psfrag{AAAAA2}[l][][1.2]{\hspace{-6mm}$\Xi_{\text{B}} = 10$}
    \psfrag{AAAAA3}[l][][1.2]{\hspace{-6mm}$\Xi_{\text{B}} = 10^2$}
    \psfrag{AAAAA4}[l][][1.2]{\hspace{-6mm}$\Xi_{\text{B}} = \infty$}
    \psfrag{BB1}[l][][1.2]{\hspace{-2mm}No time offset}
    \psfrag{xlabel}[c][][1.3]{Anchor $A_1$'s position on the
    circle $\phi_1$ (rad)}
    \psfrag{ylabel}[c][][1.3]{SPEB $(\text{m}^2)$}
    \psfrag{title}[c][][1.3]{}

    \includegraphics[angle=0,width=1\columnwidth,draft=false]{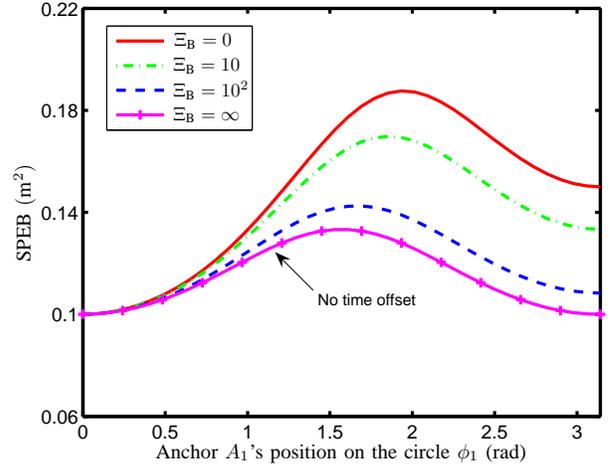}}\\
    \hfill \\
    \subfigure[STEB as a function of Anchor $A_1$'s position]
    {  \label{fig:TDOA2}
    \psfrag{AAAAA1}[l][][1.2]{\hspace{-5.5mm}$\Xi_{\text{B}} = 0$}
    \psfrag{AAAAA2}[l][][1.2]{\hspace{-6mm}$\Xi_{\text{B}} = 10$}
    \psfrag{AAAAA3}[l][][1.2]{\hspace{-6mm}$\Xi_{\text{B}} = 10^2$}
    \psfrag{AAAAA4}[l][][1.2]{\hspace{-6mm}$\Xi_{\text{B}} = \infty$}
    \psfrag{BB1}[l][][1.2]{\hspace{-1mm}No time offset}
    \psfrag{xlabel}[c][][1.3]{Anchor $A_1$'s position on the
    circle $\phi_1$ (rad)}
    \psfrag{ylabel}[c][][1.3]{STEB $(\text{sec}^2)$}
    \psfrag{title}[c][][1.3]{}

    \includegraphics[angle=0,width=1\columnwidth,draft=false]{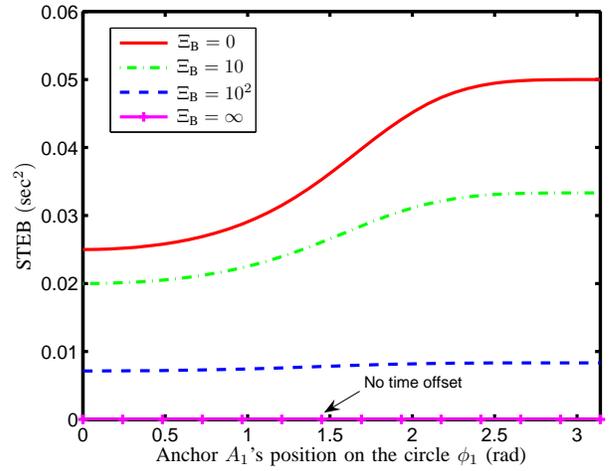}}

    \caption{SPEB and STEB with different a priori knowledge of the time offset,
    and $\Xi_{\text{B}} = 0, 10, 10^2, \infty$ respectively.}
    \end{center}
\end{figure}

We finally investigate the effect of time offset on the SPEB and
squared timing error bound (STEB) for the network illustrated in
Fig.~\ref{fig:Sim_network_Noncoop}.  The RII from each anchor
$\lambda_k=10$, $k \in \{ 1, 2, 3, 4\}$. Initially, four anchors are
placed at $\phi_1 = 0$, $\phi_2 = \pi/2$, $\phi_3 = \pi$, and
$\phi_4 = 3\pi/2$, respectively.  We then vary the position of
Anchor $A_1$ counter-clockwise along the circle.
Figure~\ref{fig:TDOA} and \ref{fig:TDOA2} show the SPEB and STEB,
respectively, as functions of $\phi_1$ for different a priori
knowledge of the time offset.

We have the following observations. First of all, both the SPEB and STEB
decrease with the a priori knowledge of the time offset. The SPEB
for the case $\Xi_{\text{B}} = \infty$ in Fig.~\ref{fig:TDOA}, i.e.,
known time offset, is equal to that of a system without a time
offset. On the other hand, when $\Xi_{\text{B}} = \infty$, the STEB
in Fig.~\ref{fig:TDOA2} is equal to zero regardless of $\phi_1$
since the offset is completely known. Secondly, all the curves in
Fig.~\ref{fig:TDOA} have the same value at $\phi_1 = 0$. The time
offset has no effect on the SPEB at this point, since
$\V{q}_\text{B} = \V{0}$, referred to as
\emph{time-offset-independent} localization. In this case, both the
SPEB and STEB achieve their minimum, implying that location and
timing information of a network are closely related. Third, as
$\phi_1$ increases from 0 to $\pi$, all the curves in
Fig.~\ref{fig:TDOA} first increase and then decrease, whereas all
the curves in Fig.~\ref{fig:TDOA2} increase monotonically. We give
the following interpretations: the estimation error of time offset
in Fig.~\ref{fig:TDOA2} becomes larger when all the anchors tend to
gather on one side of the agent ($\phi_1$ increases from 0 to
$\pi$). In Fig.~\ref{fig:TDOA}, the SPEB first increases since both
the localization information $\sum_{k \in \NB} \lambda_k \,
\R(\phi_k)$ in \eqref{eq:Anal_EFIM_TDOA} and the information for the
time offset becomes smaller. Then the SPEB decreases since the
localization information increases (when $\phi_1>\pi/2$) faster
compared to the decrease of the information for time offset. Note in
Fig.~\ref{fig:TDOA} that although $\phi_1=0$ and $\phi_1=\pi$ result
in the same SPEB in the absence of time offset, $\phi_1=0$ gives a
better performance in the presence of time offset.

%
%

\section{Conclusion}\label{Sec:Conc}

In this paper, we developed a framework to study wideband wireless
location-aware networks and determined their localization accuracy.
In particular, we characterized the localization accuracy in terms of a performance measure called the squared position error bound (SPEB), and derived the SPEB by applying the notion of equivalent Fisher information. This methodology provides insights into the essence of the localization problem by unifying the localization information from the a priori knowledge of the agent's position and information from individual anchors. We showed that the contributions from anchors, incorporating both measurements and a priori channel knowledge, can be expressed in a canonical form as a weighted sum of the ranging direction matrix.
Our results are derived from the received waveforms themselves
rather than the signal metrics extracted from the waveforms.
Therefore, our framework exploits all the information inherent in
the received waveforms, and consequently the results in this paper
serve as fundamental limits of localization accuracy. These results
can be used as guidelines for localization system design, as well as
benchmarks for location-aware networks.


\appendices

\section{Fisher Information Matrix Derivation}
\label{apd:Matrices_for_FIM}

To facilitate the analysis, we consider a mapping from $\B\theta$
into another parameter vector $\B\eta = \Matrix{cccc}
{\B{\eta}_1^\text{T} & \B{\eta}_2^\text{T} & \cdots &
\B{\eta}_{\Nb}^\text{T} }^\text{T}$, where $ \B{\eta}_k =
\Matrix{ccccccc} {\D{k}{1} & \tilde{\alpha}_{k}^{(1)} & \cdots &
\D{k}{L_k} & \tilde{\alpha}_{k}^{(L_k)} }^\text{T}$ with
$\tilde{\alpha}_{k}^{(l)} \triangleq \A{k}{l}/c$. When the agent is
localizable,\footnote{Note that an agent is said to be localizable
if its position can be determined by the signal metrics extracted
from the waveforms received from neighboring anchors, i.e.,
triangulation is possible. This is true when $M \geq 3$, or in some
special cases when $M = 2$.} this mapping is a bijection and
provides an alternative expression for the FIM as
    \begin{align}\label{eq:FIM_TRAN}
        \JTH  =   \V{T} \, \V{J}_{\B\eta} \, \V{T}^\text{T} ,
    \end{align}
where $\V{J}_{\B\eta}$ is the FIM for $\B\eta$, and $\V{T}$ is the
Jacobian matrix for the transformation from $\B\theta$ to $\B\eta$,
given respectively by
    \begin{align}\label{eq:FIM_J_eta}
        \V{J}_{\B\eta} &\triangleq \FIM{\V{r}}{\V{r}|\B\theta}{\B\eta}{\B\eta}
        = \Matrix{cc} {    \LL     &   \V{0}   \\
                           \V{0}   &   \LNL},    \\
        \noalign{\noindent and \vspace{\jot}}
        \V{T}  &\triangleq   \frac{\partial\B\eta}{\partial {\B\theta}}  
               =  \frac{1}{c} \Matrix{cc} { \HL     &   \HNL   \\
                                            \V{0}   &   \V{I} } ,
        \label{eq:FIM_Trans}
    \end{align}
with $\V{0}$ denoting a matrix of all zeros and $\V{I}$ denoting an
identity matrix. The block matrices $\HL$, $\HNL$, $\LL$, and $\LNL$
are given as follows:
    \begin{align}\label{eq:Apd_HL_HNL}
        \HL  & = \Matrix{cccc}{ \V{G}_1 & \V{G}_2 & \cdots & \V{G}_M \\
                              \V{D}_1 & & & \V{0}\\ & \V{D}_2 \\ & & \ddots \\ \V{0}&&&\V{D}_M
                              }, \nonumber \\
        \HNL & = \Matrix{cccc} { \V{G}_{M+1} &   \cdots  &   \V{G}_\Nb \\
                               \V{0}& \cdots &\V{0} } , \\
        \LL & =  \Diag{  \B{\Psi}_1,\B{\Psi}_2,\cdots,\B{\Psi}_M} \,, \nonumber \\
        \noalign{\noindent and \vspace{\jot}}
        \LNL & = \Diag{ \B{\Psi}_{M+1}, \B{\Psi}_{M+2},\cdots,\B{\Psi}_\Nb}\,,
        \label{eq:Apd_Lamda}
    \end{align}
where $\V{D}_k = \Matrix{cc}{\B{0} & \I_{2L_k-1}}$,
    \begin{align}\label{eq:Sing_Ga_def}
        \hspace{7mm} \V{G}_k & =  \q_k \, \V{l}_k^\text{T}
        \;\;\; \text{with}  \;\;\;
        \V{l}_k = {\underbrace{\Matrix{ccccc}{1 & 0 & \cdots & 1 & 0}}_{2L_k
        \;\;\text{components}}}^\text{T} ,\!\!\!
    \end{align}
$\q_k=[\cos\phi_k\;\sin\phi_k]^\text{T}$, and $\B{\Psi}_k \in \mathbb{R}^{2L_k \times 2L_k}$ is given by
    \begin{align}\label{eq:Apd_Psi}
        \B{\Psi}_k \triangleq \FIM{\V{r}}{\V{r}|\B\theta}{\B\eta_k}{\B\eta_k}
        \, .
    \end{align}
Note that elements in $\B{\Psi}_k$ can be expressed as
    \begin{subequations}
            \begin{align*}
                & \hspace{-2mm}\E_{\V{r}} \left\{ -\frac{\partial^2 \ln \FTR }{\partial \D{k}{i}
                \partial \D{k}{j}} \right\} \nonumber \\
                & =  \frac{2\,\A{k}{i}\,\A{k}{j}}{N_0} \int \left| 2 \pi f S(f)
                \right|^2 \exp \left\{ -j2\pi f \, (\D{k}{i}-\D{k}{j}) \right\} d f  \nonumber \\
                & =  \frac{2\,\A{k}{i}\A{k}{j}}{N_0}
                \frac{\partial^2}{\partial\D{k}{i}\partial\D{k}{j}} R_s\left(\D{k}{i}-\D{k}{j}\right)
                \, ,  \\
                & \hspace{-2mm}\E_{\V{r}} \left\{ -\frac{\partial^2 \ln \FTR }{\partial \D{k}{i}
                \partial \tilde{\alpha}_{k}^{(j)}} \right\}
                \nonumber\\
                & =  \frac{2 \, \A{k}{i} \, c}{ N_0} \int j \, 2\pi f \left| S(f)
                \right|^2 \exp\left\{ -j2 \pi f \, (\D{k}{i}-\D{k}{j}) \right\} d f  \nonumber \\
                & =  \frac{2 \, \A{k}{i} \, c }{N_0}
                \frac{\partial}{\partial\D{k}{i}} R_s\left(\D{k}{i}-\D{k}{j}\right)
                \, ,\\
            \noalign{\noindent and \vspace{\jot}}
                & \hspace{-2mm} \E_{\V{r}} \left\{ -\frac{\partial^2 \ln \FTR }{\partial \tilde{\alpha}_{k}^{(i)}
                \partial \tilde{\alpha}_{k}^{(j)}} \right\}  \nonumber\\
                & =  \frac{2 \, c^2}{N_0} \int \left| S(f)
                \right|^2 \exp\left\{ -j2\pi f \, (\D{k}{i}-\D{k}{j}) \right\} d f   \nonumber \\
                & =  \frac{2 \, c^2}{N_0} R_s\left(\D{k}{i}-\D{k}{j}\right)
                \, ,
            \end{align*}
    \end{subequations}
where $R_s(\tau) = \int s(t) s(t-\tau) dt $. In particular,
    \begin{align}\label{eq:Apd_info_1}
        \left[\B\Psi_k \right]_{1,1} =
        \FIM{\V{r}}{\V{r}|\B\theta}{\D{k}{1}}{\D{k}{1}}
        = 8 \pi^2\beta^2 \, \mathsf{SNR}_k^{(1)} \,,
    \end{align}
where $\beta$ and $\mathsf{SNR}_k^{(i)}$ are given by
\eqref{eq:Sing_beta} and \eqref{eq:Sing_SNR}, respectively.
Substituting \eqref{eq:FIM_J_eta} and \eqref{eq:FIM_Trans} into
\eqref{eq:FIM_TRAN}, we have the FIM $\JTH$ in
\eqref{eq:FIM_JTH_NPrior}.

\section{Wideband Channel Model and {A Priori} Channel Knowledge}
\label{Apd:FIM_Channel}

Wideband channel measurements have shown that MPCs follow random
arrival and their amplitudes are subject to path loss, large and
small-scale fading. While our discussion is valid for any wideband
channels described by \eqref{eq:Model_Multipath}, we consider the
model of IEEE 802.15.4a standard for exposition. Specifically, this
standard uses Poisson arrivals, log-normal shadowing, Nakagami
small-scale fading with exponential power dispersion profile (PDP)
\cite{MolCasChoEmaForKanKarKunSchSiwWin:J06}.

\subsection{Path Arrival Time}

The arrival time of MPCs is commonly modeled by a Poisson process
\cite{SalVal:87, MolCasChoEmaForKanKarKunSchSiwWin:J06}. Given the
path arrival rate $\nu$, we have
    \begin{align*}
        g_{\B\tau_k}\left( \D{k}{l} \left|\, \D{k}{l-1}\right.\right) = \nu \, \exp \left\{
        - \nu \left(\D{k}{l} - \D{k}{l-1}\right)\right\} ,
    \end{align*}
for $\D{k}{l} \geq \D{k}{l-1}$ and $l \geq 2$. Using
\eqref{eq:Model_Tau_L}, we obtain
    \begin{align}\label{eq:FIM_Poisson}
        g_{\V{b}_k}\left( \BB{k}{l} \left|\, \BB{k}{l-1} \right.\right) = \frac{\nu}{c} \, \exp
        \left\{ -\frac{\nu}{c} \, \left( \BB{k}{l} - \BB{k}{l-1} \right) \right\} ,
    \end{align}
for $\BB{k}{l} \geq \BB{k}{l-1}$ and $l \geq 1$. Note that we let
$\BB{k}{0} = 0$ for consistency.

\subsection{Path Loss and Large-Scale Fading}

The RSS in dB at the distance $d_k$ can be written as
\cite{MolCasChoEmaForKanKarKunSchSiwWin:J06}
    \begin{align*}
        P_k = P_0 - 10 \varrho \log_{10}\left(\frac{d_k}{d_0}\right) + w
        \, ,
    \end{align*}
where 
$P_0$ is the expected RSS at the reference distance $d_0$, $\varrho$
is the propagation (path gain) exponent, and $w$ is a random
variable (r.v.) that accounts for large-scale fading, or shadowing.
Shadowing is usually modeled with a log-normal distribution, such
that $w$ is a Gaussian r.v. with zero-mean and variance
$\sigma_\text{S}^2$, i.e., $w \sim
N\left(0,\sigma_\text{S}^2\right)$.\footnote{The standard deviation
is typically 1-2 dB (LOS) and 2-6 dB (NLOS) \cite{Mol:05} around the
path gain.} The PDF of the RSS of $r_k(t)$ can then be written as
    \begin{align}\label{eq:FIM_Lognormal}
        g_\text{P}\left(P_k|d_k\right) \propto \exp \left\{
        - \frac{1}{2\sigma_\text{S}^2}\left[ P_k - P_0 + 10 \varrho \log_{10}
        \left(\frac{d_k}{d_0}\right) \right]^2 \right\},
    \end{align}
where $d_k = \| \V{p} - \V{p}_k \|$, and $P_k$ is given by
    \begin{align*}
        P_k = 10 \log_{10} \left[ \sum_{l=1}^{L_k} \E_\text{s} \left\{
        \left|\,\alpha_k^{(l)}\,\right|^2 \right\}
        \right],
    \end{align*}
with $\E_\text{s} \left\{\cdot\right\}$ denoting the average over
small-scale fading.

\subsection{Power Dispersion Profile and Small-Scale Fading}

As in \cite{QueWin:J05, MolCasChoEmaForKanKarKunSchSiwWin:J06},
we consider an exponential PDP given by,\footnote{Note that the
first component of LOS signals can exhibit a stronger strength than
\eqref{eq:FIM_Power_path} in some UWB measurement \cite{CasWinMol:J02}. In such cases, \eqref{eq:FIM_Power_path} and \eqref{eq:FIM_Power_Normal} need to be modified, accordingly.}
    \begin{align}\label{eq:FIM_Power_path}
        \E_\text{s} \left\{ \left|\,\alpha_k^{(l)} \right|^2 \right\}
        = Q_k \exp\left(- \frac{\D{k}{l}}{\gamma_k} \right)
        \triangleq Q_{k}^{(l)}\,,
    \end{align}
where $\gamma_k$ is the decay constant, and $Q_k$ is a normalization
coefficient such that
    \begin{align}\label{eq:FIM_Power_Normal}
        Q_k = \frac{10^{P_k/10}}{\sum_{l=0}^{L_k} \exp\left(-\D{k}{l}/\gamma_k\right)}
        \, .
    \end{align}
In addition, $\alpha_k^{(l)}$ is a Nakagami r.v.~with second moment
given by \eqref{eq:FIM_Power_path}. Specifically, we have
    \begin{align}\label{eq:FIM_Nakagami}
        &\hspace{0mm} g_{\B\alpha_k}\left(\A{k}{l} \left|\, \V{b}_k, d_k, P_k
        \right.\right)\nonumber \\
        & =  g_{\B\alpha_k} \left( \A{k}{l} \left|\, \B{\tau}_k, P_k \right. \right) \nonumber \\
        & =  \frac{2}{\Gamma(m_l)} \left(
        \frac{m_l}{Q_{k}^{(l)} }\right)^{m_l} \left|\,\A{k}{l}\right|^{2m_l-1}
        \exp \left( -\frac{m_l}{Q_{k}^{(l)}} \!\! \left|\,\alpha_k^{(l)}\right|^2  \right),
    \end{align}
where $\Gamma(m_l)$ is the gamma function and $m_l \geq 1/2$ is the
Nakagami $m$-factor, which is a function of $\B\tau_k$
\cite{MolCasChoEmaForKanKarKunSchSiwWin:J06}.

\subsection{{A Priori} PDF for Multipath Parameters}

The joint PDF of the multipath parameters and the RSS, conditioned
on the distance from anchor $k$ to the agent, can be derived as
    \begin{align}\label{eq:FIM_Prior_PDF_t}
        \gk\left(\B\alpha_k, \V{b}_k, P_k\left|d_k\right.\right)
        = g_\text{P}(P_k|d_k)\,& \prod_{l=1}^{L_k}
        g_{\B\alpha_k}\left( \A{k}{l} \left|\, \V{b}_k, d_k, P_k \right.\right) \nonumber \\
        \times & \prod_{l=1}^{L_k} g_{\V{b}_k}\left( \BB{k}{l} | \BB{k}{l-1} \right).
    \end{align}
By integrating over $P_k$, we obtain the PDF of the multipath
parameters of $r_k(t)$ as follows
    \begin{align}\label{eq:FIM_Prior_PDF}
        \gk(\B\kappa_k | d_k)
        & = \gk\left(\B\alpha_k, \V{b}_k  | d_k\right) \nonumber\\
        & = \int_{-\infty}^\infty \gk( \B\alpha_k, \V{b}_k, P_k|d_k) \; d P_k
        \, .
    \end{align}
Equation \eqref{eq:FIM_Prior_PDF} characterizes the a priori
knowledge of channel parameters, and can be obtained, for IEEE
802.15.4a standard, by substituting \eqref{eq:FIM_Poisson},
\eqref{eq:FIM_Lognormal} and \eqref{eq:FIM_Nakagami} into
\eqref{eq:FIM_Prior_PDF_t} and \eqref{eq:FIM_Prior_PDF}. Note that
since $\V{p}_k$ is known, $d_k$ is a function of $\V{p}$ and hence
we have \eqref{eq:FIM_prior_pdf}.

\begin{figure*}[!b]
  \vspace*{4pt}%
  \hrulefill%
  \normalsize%
  \setcounter{MYtempeqncnt}{\value{equation}}%
  \setcounter{equation}{59}%
    \begin{align}\label{eq:Apd_long_partition}
        \Matrix{cc}{\ddot{R}_s(1,1) & \V{t}_k^\text{T} \\ \V{t}_k  &  \B\Upsilon_k}
        \triangleq \Matrix{ccccccc}
        {\ddot{R}_s(1,1) & \dot{R}_s(1,1) &   \ddot{R}_s(1,2)    &  \dot{R}_s(1,2)
         & \cdots &  \ddot{R}_s(1,\tilde{L}_k) &  \dot{R}_s(1,\tilde{L}_k)\\
         \dot{R}_s(1,1)  & R_s(1,1)       &  -\dot{R}_s(1,2)     &  R_s(1,2)
         & \cdots  & - \dot{R}_s(1,\tilde{L}_k)  &  R_s(1,\tilde{L}_k) \\
         \ddot{R}_s(1,2)    &  -\dot{R}_s(1,2) \\
         \dot{R}_s(1,2)     &  R_s(1,2)  \\
         \vdots    & \vdots   \\
         \ddot{R}_s(1,\tilde{L}_k) &  -\dot{R}_s(1,\tilde{L}_k) & \cdots &&&
         \ddot{R}_s(\tilde{L}_k,\tilde{L}_k) &  \dot{R}_s(\tilde{L}_k,\tilde{L}_k)\\
         \dot{R}_s(1,\tilde{L}_k)  &  R_s(1,\tilde{L}_k) & \cdots &&&
         \dot{R}_s(\tilde{L}_k,\tilde{L}_k) &  {R}_s(\tilde{L}_k,\tilde{L}_k)
        }
    \end{align}
  \setcounter{equation}{\value{MYtempeqncnt}}
  \vspace*{-6pt}
\end{figure*}

\section{Proofs of the Results in Section \ref{Sec:Evalu}}
\label{apd:Proofs_Evalu}



\subsection{Proof of Theorem \ref{Thm:EFIM_NPrior}}
\label{apd:Proofs_Evalu_Overlap_cluster}

\begin{IEEEproof}
We first prove that $\JE{\V{p}}$ is given by
\eqref{eq:Sing_EFIM_overlap}. We partition $\V{G}_k$ in
\eqref{eq:Sing_Ga_def} and $\B\Psi_k$ in \eqref{eq:Apd_Psi} as
    \begin{align*}
        \V{G}_k  \triangleq \Matrix{ccc}{\q_k & \breve{\V{G}}_k}
        \;\; \text{and} \;\;
        \B\Psi_k = \Matrix{cc} {   8\pi^2\beta^2 \, \SNR{k}{1}  &   \V{k}_k^\text{T} \\
                                    \V{k}_k                         &   \breve{\B\Psi}_k},
    \end{align*}
where $[\B\Psi_k]_{1,1}$ is obtained by \eqref{eq:Apd_info_1},
$\V{k}_k \in \mathbb{R}^{2L_k-1}$, $\breve{\B\Psi}_k \in
\mathbb{R}^{(2L_k-1)\times(2L_k-1)}$, and
    \begin{align*}
        \breve{\V{G}}_k = \q_k \, {\underbrace{\Matrix{cccccc}{0 & 1
        & 0 & \cdots  &  1  &  0}}_{2L_k-1 \;\;
        \text{components}}}^\text{T}.
    \end{align*}
Using these notations, we can write the EFIM given by
\eqref{eq:Sing_EFIM_NLOS} in Proposition \ref{thm:EFIM_n-NLOS},
after some algebra, in the form of \eqref{eq:JTH_Partition},
    \begin{subequations}
        \begin{align*}
        \V{A} & \triangleq 8\pi^2\beta^2 \sum_{k \in \NL} \SNR{k}{1} \, \q_k\, \q_k^\text{T}
                \nonumber \\
              & \hspace{10.5mm} + \sum_{k \in \NL} \left\{ \breve{\V{G}}_k\V{k}_k \q_k^\text{T}
                    + \q_k \V{k}_k^\text{T} \breve{\V{G}}_k^\text{T}
                    + \breve{\V{G}}_k \breve{\B\Psi}_k \breve{\V{G}}_k^\text{T} \right\}
        , \\
        \V{B}   & \triangleq
                \Matrix{cccc} { \q_1\,\V{k}_1^\text{T} + \breve{\V{G}}_1\,\breve{\B\Psi}_1
                                & \cdots & \q_M\,\V{k}_M^\text{T} + \breve{\V{G}}_M\,\breve{\B\Psi}_M }
        , \\
        \noalign{\noindent and \vspace{\jot}}
        \V{C}   & \triangleq
                \Diag{ \breve{\B\Psi}_1, \breve{\B\Psi}_2, \ldots, \breve{\B\Psi}_M }.
        \end{align*}
    \end{subequations}
Applying the notion of EFI as in \eqref{eq:Sing_EFIM_FIM}, we obtain
the $2\times2$ $\JE{\V{p}}$ as
    \begin{align}\label{eq:Apd_EFIM_NPrior}
        \JE{\V{p}} & =  \frac{8\pi^2\beta^2}{c^2} \sum_{k \in \NL}
        ( 1 - \chi_k) \, \SNR{k}{1} \, \q_k\,\q_k^\text{T}\,,
    \end{align}
where the POC
    \begin{align}\label{eq:Apd_Chi_a_ori}
        \chi_k \triangleq
        \frac{ \V{k}_k^\text{T} \, \breve{\B\Psi}_k^{-1} \, \V{k}_k}{ 8\pi^2\beta^2 \,
        \SNR{k}{1}}\,.
    \end{align}
This completes the proof of \eqref{eq:Sing_EFIM_overlap}.

Next, we show that only the first contiguous-cluster contains
information for localization. Let us focus on $\chi_k$. Define the
following notations for convenience:
    \begin{align*}
        R_s(i,j) & \triangleq R_s(t) |_{t=\D{k}{i}-\D{k}{j}}
        \, , \nonumber \\
        \ddot{R}_s(i,j) &\triangleq - \frac{\partial^2}{\partial t^2 }
        R_s(t)|_{t=\D{k}{i}-\D{k}{j}} \, ,\nonumber \\
        \noalign{\noindent and \vspace{\jot}}
        \dot{R}_s(i,j) & \triangleq \frac{\partial}{\partial t }
        R_s(t)|_{t=\D{k}{i}-\D{k}{j}} = - \dot{R}_s(j,i)
        \, .
    \end{align*}
If the length of the first contiguous-cluster in the received
waveform is $\tilde{L}_k$ where $1\leq\tilde{L}_k\leq L_k$, then $\ddot{R}_s(i,j) = \dot{R}_s(i,j) = R_s(i,j) = 0$ for $i \in \{1, 2, \dotsc, \tilde{L}_k \}$ and $j \in \{
\tilde{L}_k+1, \tilde{L}_k+2, \dotsc, L_k\}$, and\footnote{$\boxtimes$ is a block
matrix that is irrelevant to the rest of the derivation.}
    \begin{align*}
        \V{k}_k \triangleq \Matrix{cc} { \tilde{\V{k}}_k^\text{T}   &
        \B{0}^\text{T}}^\text{T} \quad \text{and} \quad
        \breve{\B\Psi}_k \triangleq
        \Matrix{cc} {   {\tilde{\B\Psi}}_k    &   \B{0}   \\
                        \B{0}                 &   {{\boxtimes}}},
    \end{align*}
where $\tilde{\V{k}}_k \in \mathbb{R}^{2\tilde{L}_k-1}$ and
$\tilde{\B\Psi}_k \in \mathbb{R}^{(2\tilde{L}_k-1) \times (2\tilde{L}_k-1)}$. Hence \eqref{eq:Apd_Chi_a_ori} becomes
    \begin{align}\label{eq:Apd_chi_a}
        \chi_k  =  \frac{\tilde{\V{k}}_k^\text{T} \,
        \tilde{\B\Psi}_k^{-1} \, \tilde{\V{k}}_k}{8\pi^2\beta^2\,
        \SNR{k}{1}} \,,
    \end{align}
which depends only on the first $\tilde{L}_k$ paths, implying that only the first contiguous-cluster of LOS signals contains information for localization.

Finally, we show that $\chi_k$ is independent of $\A{k}{l}$. Note
that $\tilde{\B\Psi}_k$ and $\tilde{\V{k}}_k$ can be written as
    \begin{align}\label{eq:Apd_psi_a'}
        \tilde{\B\Psi}_k =
            \frac{2}{N_0} \, & \Diag{c,\A{k}{2},c,\ldots,\A{k}{L_k},c}
            \, \B\Upsilon_k \nonumber \\
            \times \;\: & \Diag{c,\A{k}{2},c,\ldots,\A{k}{L_k},c},
    \end{align}
and
    \begin{align}\label{eq:Apd_ka}
        \tilde{\V{k}}_k = \frac{2 \, \A{k}{1}}{N_0} \,
        \Diag{c,\A{k}{2},c,\ldots,\A{k}{L_k},c} \, \V{t}_k
        \, ,
    \end{align}
where $\B\Upsilon_k\in \mathbb{R}^{(2\tilde{L}_k-1) \times (2\tilde{L}_k-1)}$ and
$\V{t}_k \in \mathbb{R}^{2\tilde{L}_k-1}$ are given by the matrix partition
in \eqref{eq:Apd_long_partition} shown at the bottom of the page.
\addtocounter{equation}{1}
Substituting \eqref{eq:Apd_psi_a'} and \eqref{eq:Apd_ka} into
\eqref{eq:Apd_chi_a}, we obtain
    \begin{align}\label{eq:Apd_R_a'_full}
        \chi_k =  \frac{1}{4\pi^2\beta^2} \, \V{t}_k^\text{T} \B\Upsilon_k^{-1}
        \V{t}_k  \, ,
    \end{align}
which is independent of all the amplitudes.

Note that $0 \leq \chi_k \leq 1$: $\chi_k$ is nonnegative since it
is a quadratic form and ${\B\Upsilon}_k$ is a positive semi-definite
FIM (hence is ${\B\Upsilon}_k^{-1}$); and $\chi_k \leq 1$ since the
contribution from each anchor to the EFIM in
\eqref{eq:Apd_EFIM_NPrior} is nonnegative.
\end{IEEEproof}

\begin{figure*}[!b]
  \vspace*{4pt}%
  \hrulefill%
  \normalsize%
  \setcounter{MYtempeqncnt}{\value{equation}}%
  \setcounter{equation}{61}%
  \renewcommand{\arraystretch}{1.3}
    \begin{align}\label{eq:Apd_A_longeq}
        \V{A} & \triangleq   \Matrix{cccc}{\sum_{k \in \NB} \V{G}_k \bar{\B{\Psi}}_k \V{G}_k^\text{T} + c^2 \Gpp{k}
                                 & \V{G}_{1}\bar{\B{\Psi}}_{1}\V{D}_1^\text{T} +
                                 c^2 \Gpk{1} &  \cdots
                                 & \V{G}_{M}\bar{\B{\Psi}}_{M}\V{D}_M^\text{T} + c^2 \Gpk{M}  \\
                                 \left( \V{G}_1  \bar{\B{\Psi}}_{1} \V{D}_1^\text{T} + c^2 \Gpk{1} \right) ^\text{T}
                                 & \V{D}_1  \bar{\B{\Psi}}_{1} \V{D}_{1}^\text{T}  + c^2 \Gkk{1} \\
                                 \vdots & & \ddots \\
                                 \left( \V{G}_M  \bar{\B{\Psi}}_{M} \V{D}_M^\text{T} + c^2 \Gpk{M} \right) ^\text{T}
                                 & & & \V{D}_M  \bar{\B{\Psi}}_{M} \V{D}_{M}^\text{T}  + c^2 \Gkk{M}}
    \end{align}
  \setcounter{equation}{\value{MYtempeqncnt}}
  \vspace*{-6pt}
\end{figure*}

\begin{figure*}[!b]
  \vspace*{4pt}%
  \hrulefill%
  \normalsize%
  \setcounter{MYtempeqncnt}{\value{equation}}%
  \setcounter{equation}{62}%
    \begin{align}\label{eq:Apd_JE_WPrior}
        \JE{\V{p}} = \frac{1}{c^2} \, {\Bigg\{} \sum_{k \in \NB} \left( \V{G}_k \bar{\B{\Psi}}_k \V{G}_k^\text{T} + c^2 \Gpp{k} \right)
            & - \sum_{k \in \NL} \left( \V{G}_k\bar{\B{\Psi}}_k\V{D}_k^\text{T} + c^2 \Gpk{k}\right)
            \left( \V{D}_k \bar{\B{\Psi}}_k \V{D}_k^\text{T} + c^2 \Gkk{k}
            \right)^{-1} \!\! \left( \V{G}_k \bar{\B{\Psi}}_k\V{D}_k^\text{T} + c^2 \Gpk{k}\right)^\text{T} \nonumber \\
            & - \sum_{k \in \NNL} \left( \V{G}_k \bar{\B{\Psi}}_k + c^2 \Gpk{k}
            \right)\left( \bar{\B{\Psi}}_k + c^2 \Gkk{k}
            \right)^{-1}\!\!
            \left( \V{G}_k \bar{\B{\Psi}}_k + c^2 \Gpk{k} \right)^\text{T}
            {\Bigg\}}
    \end{align}
  \setcounter{equation}{\value{MYtempeqncnt}}
  \vspace*{-6pt}
\end{figure*}

\begin{figure*}[!b]
  \vspace*{4pt}%
  \hrulefill%
  \normalsize%
  \setcounter{MYtempeqncnt}{\value{equation}}%
  \setcounter{equation}{64}%
  \begin{IEEEeqnarray}[\setlength{\nulldelimiterspace}{1pt}]{rl's}
    &\frac{1}{c^2} {\Big\{} \V{l}_k^\text{T} \bar{\B{\Psi}}_k \V{l}_k + c^2 \tGpp{k}
    - \left( \V{l}_k^\text{T} \bar{\B{\Psi}}_k \V{D}_k^\text{T} + c^2 \tGpk{k}
    \right) \left( \V{D}_k  \bar{\B{\Psi}}_k \V{D}_k^\text{T} + c^2 \Gkk{k}
    \right)^{-1}\!\!
    \left( \V{l}_k^\text{T} \bar{\B{\Psi}}_k \V{D}_k^\text{T} + c^2 \tGpk{k} \right)^\text{T} {\Big\}} ,
    &$k \in \NL$\IEEEyessubnumber \label{eq:Apd_Ra_def}\\*[0\normalbaselineskip]
    \smash{\hspace{-8mm}\lambda_k \triangleq\left\{\IEEEstrut[9\jot][9\jot]\right.}&&
    \nonumber\\*[0\normalbaselineskip]
    &\frac{1}{c^2} {\Big\{} \V{l}_k^\text{T} \bar{\B{\Psi}}_k  \V{l}_k + c^2 \tGpp{k}
    - \left( \V{l}_k^\text{T} \bar{\B{\Psi}}_k + c^2 \tGpk{k}
    \right)\left(  \bar{\B{\Psi}}_k + c^2 \Gkk{k} \right)^{-1}\!\!
    \left( \V{l}_k^\text{T} \bar{\B{\Psi}}_k + c^2 \tGpk{k} \right)^\text{T} {\Big\}},
    & $k \in \NNL$ \IEEEyessubnumber \label{eq:Apd_tRa_def}
    \end{IEEEeqnarray}
  \setcounter{equation}{\value{MYtempeqncnt}}
  \vspace*{-6pt}
\end{figure*}

\subsection{Proof of Corollary \ref{cor:EFIM_n-NLOS_gen}}
\label{apd:Proofs_Evalu_n-NLOS_gen}

\begin{IEEEproof}
This scenario can be thought of as a special case of Theorem
\ref{Thm:EFIM_NPrior} with $\tilde{L}_k = 1$, i.e., the first
contiguous-cluster contains only one path. In this case,
\eqref{eq:Apd_R_a'_full} becomes
    \begin{align*}
        \chi_k =  \frac{1}{4\pi^2\beta^2} \, \frac{ \dot{R}_s^2(1,1) }{R_s(1,1)}
        \, .
    \end{align*}
Since waveform $s(t)$ is continuous and time-limited in realistic
cases, we have
    \begin{align*}
        \left. \dot{R}_s(1,1) = \frac{\partial}{\partial \tau} R_s(\tau) \right|_{\tau=0} = 0
        \, ,
    \end{align*}
implying that $\chi_k = 0$, which leads to
\eqref{eq:Anal_RII_noprior}.
\end{IEEEproof}

\subsection{Proof of Theorem \ref{thm:EFIM_NLOS_gen}}
\label{apd:Proofs_Evalu_NLOS_gen}

\begin{IEEEproof}
When a priori channel knowledge of the channel is available, the FIM
is
    \begin{align*}
        \renewcommand{\arraystretch}{1.3}
        \JTH = \frac{1}{c^2} \Matrix{cc}
            {   \HNL \BLNL \HNL^\text{T} + \HL \BLL  \HL^\text{T}    &   \HNL \BLNL \\
                \BLNL \HNL^\text{T}                           &   \BLNL  }
            + \JP
         \, ,
    \end{align*}
where $\BLNL = \E_{\B\theta} \left\{  \LNL \right\} \triangleq
\Diag{\bar{\B{\Psi}}_1, \bar{\B{\Psi}}_2, \ldots, \bar{\B{\Psi}}_M}$
and $\BLL = \E_{\B\theta} \left\{  \LL \right\} \triangleq
\Diag{\bar{\B{\Psi}}_{M+1}, \bar{\B{\Psi}}_{M+2},  \ldots,
\bar{\B{\Psi}}_\Nb}$. The FIM $\JTH$ can be partitioned as
\eqref{eq:JTH_Partition}, where $\V{A}$ is given by
\eqref{eq:Apd_A_longeq} shown at the bottom of the page,
\addtocounter{equation}{1} and
    \begin{align*}
        \renewcommand{\arraystretch}{1.3}
        \V{B} \! \triangleq  \!\! \Matrix{cccc}{  \V{G}_{M+1}\bar{\B{\Psi}}_{M+1} \!+\! c^2 \Gpk{M+1}  &
                                    \cdots & \V{G}_{\Nb}\bar{\B{\Psi}}_{\Nb} \!+\! c^2 \Gpk{\Nb}  \\
                                    \V{0} & \cdots &  \V{0}},
     \end{align*} and 
	\begin{align*}
        \V{C} & \triangleq  \Diag{\bar{\B{\Psi}}_{M+1} + c^2 \Gkk{M+1},
        \ldots, \bar{\B{\Psi}}_{\Nb} + c^2 \Gkk{\Nb}}
        \, .
    \end{align*}
Apply the notion of EFI, and we have the $2 \times 2$ EFIM, after
some algebra, given by \eqref{eq:Apd_JE_WPrior} at the bottom of
the page. \addtocounter{equation}{1}
From \eqref{eq:FIM_prior_pdf}, we can rewrite $\Gpp{k}$ and
$\Gpk{k}$ in \eqref{eq:FIM_Prior_Matrix} using chain rule as
    \begin{align}\label{eq:Apd_GammaPA}
        \Gpp{k} & = \q_k \, \tGpp{k} \, \q_k^\text{T}
        \qquad \text{and} \qquad
        \Gpk{k} = \q_k \, \tGpk{k}
        \, ,
    \end{align}
where $\tGpp{k} = \FIM{\B\theta}{\B\kappa_k|d_k}{d_k}{d_k}$ and
$\tGpk{k} = \FIM{\B\theta}{\B\kappa_k|d_k}{d_k}{\B\kappa_k}$.
Substituting \eqref{eq:Apd_GammaPA} into \eqref{eq:Apd_JE_WPrior}
leads to \eqref{eq:Sing_EFIM_Prior}, where $\lambda_k$ is given by
\eqref{eq:Apd_Ra_def} and \eqref{eq:Apd_tRa_def} for LOS signals and
NLOS signals, respectively, shown at the bottom of the page.
\addtocounter{equation}{1}
\end{IEEEproof}

\subsection{Proof of Corollary \ref{cor:prior_knowledge}}
\label{apd:Proofs_Consis_RI}

\begin{IEEEproof}
We first show that the a priori channel knowledge increases the RII.
Consider $\lambda_k$ in \eqref{eq:Apd_Ra_def}. Let
    \begin{align*}
    \renewcommand{\arraystretch}{1.3}
        \V{F}_k \triangleq \frac{1}{c^2}
        \Matrix{cc}{\V{l}_k^\text{T} \bar{\B{\Psi}}_k  \V{l}_k + c^2 \tGpp{k}
                    & \V{l}_k^\text{T} \bar{\B{\Psi}}_k \V{D}_k^\text{T} + c^2 \tGpk{k} \\
                    \V{D}_k \bar{\B{\Psi}}_k \V{l}_k  + c^2
                    {\tilde{\B\Xi}_{\text{p},\B\kappa}^{k\,T}}
                    & \V{D}_k  \bar{\B{\Psi}}_k \V{D}_k^\text{T} + c^2 \Gkk{k}},
    \end{align*}
and
    \begin{align*}
    \renewcommand{\arraystretch}{1.3}
        \V{E}_k \triangleq \frac{1}{c^2} \Matrix{cc}{\V{l}_k^\text{T} \bar{\B{\Psi}}_k  \V{l}_k  & \V{l}_k^\text{T} \bar{\B{\Psi}}_k \V{D}_k^\text{T}  \\
                    \V{D}_k \bar{\B{\Psi}}_k \V{l}_k   & \V{D}_k  \bar{\B{\Psi}}_k \V{D}_k^\text{T} }.
    \end{align*}
We have $\V{F}_k \succeq \V{E}_k$, since
    \begin{align*}
    \renewcommand{\arraystretch}{1.3}
        \V{F}_k - \V{E}_k
        = \Matrix{cc}{\tGpp{k} & \tGpk{k} \\
                      \tilde{\B\Xi}_{\V{p},\B\kappa}^{k\,T} & \Gkk{k}}
        = \FIM{\B\theta}{\B\kappa_k|d_k}{\tilde{\B\theta}_k}{\tilde{\B\theta}_k}
        \succeq \V{0}
        \, ,
    \end{align*}
where $\tilde{\B\theta}_k = \Matrix{cc}  {d_k & \B\kappa_k^\text{T}
}^\text{T}$. Hence we have $\lambda_k = 1 / [\,
\V{F}_k^{-1}\,]_{1,1} \geq 1/[\,\V{E}_k^{-1}\,]_{1,1}$, where
$[\,\V{E}_k^{-1}\,]_{1,1}$ equals \eqref{eq:Sing_EFIM_strength}.
This implies that the {a priori} channel knowledge can increase the
RII.

We next show that the RIIs in \eqref{eq:Apd_Ra_def} and
\eqref{eq:Apd_tRa_def} reduce to \eqref{eq:Sing_EFIM_strength} and
zero, respectively, in the absence of a priori channel knowledge.

When a priori channel knowledge is unavailable, $\Gkk{k}$,
$\Gpk{k}$, and $\Gpp{k}$ all equal zero, and the corresponding RII
$\lambda_k$ in \eqref{eq:Apd_Ra_def} and \eqref{eq:Apd_tRa_def}
becomes
    \begin{align*}
        \lambda_k  &  = \frac{1}{c^2} \left\{
            \V{l}_k^\text{T} {\B{\Psi}}_k  \V{l}_k
            - \left( \V{l}_k^\text{T} {\B{\Psi}}_k \V{D}_k^\text{T}  \right)
            \left( \V{D}_k {\B{\Psi}}_k \V{D}_k^\text{T}  \right)^{-1}
            \left( \V{D}_k {\B{\Psi}}_k \V{l}_k  \right)
            \right\} \nonumber \\
            & =  \frac{1}{c^2} \, \V{l}_k^\text{T} \,{\Bigg\{}
                \Matrix{cc}{ 8\pi^2\beta^2 \, \SNR{k}{1} & \V{k}_k^\text{T} \\
                     \V{k}_k & \breve{\B{\Psi}}_k } \nonumber \\
            & \hspace{25mm}
        - \Matrix{c}{\V{k}_k^\text{T} \\ \breve{\B{\Psi}}_k} \breve{\B{\Psi}}_k^{-1}
        \Matrix{cc} {\V{k}_k  & \breve{\B{\Psi}}_k}
        {\Bigg\}}\, \V{l}_k \nonumber \\
        & = \frac{1}{c^2} \left\{ 8\pi^2\beta^2 \, \SNR{k}{1} -  \V{k}_k^\text{T} \breve{\B\Psi}_k^{-1} \V{k}_k
        \right\} \nonumber \\
        & = \frac{8\pi^2\beta^2}{c^2} (1 - \chi_k)\, \SNR{k}{1}\,,
    \end{align*}
for $k \in \NL$, and
    \begin{align*}
        \lambda_k & = \frac{1}{c^2} \left\{
        \V{l}_k^\text{T}  \B\Psi_k  \V{l}_k  -  \V{l}_k^\text{T} {\B{\Psi}}_k
        \, \B\Psi_k^{-1} \, \B\Psi_k \V{l}_k
        \right\} = 0\,,
    \end{align*}
for $k \in \NNL$.
\end{IEEEproof}

\begin{figure*}[!b]
  \vspace*{4pt}%
  \hrulefill%
  \normalsize%
  \setcounter{MYtempeqncnt}{\value{equation}}%
  \setcounter{equation}{66}%
    \begin{align}\label{eq:Apd_EFIM_WPrior_Pos}
        \JE{\V{p}} & = \GP + \sum_{k \in \NB} {\Bigg\{} \E_{\V{p}}\left\{\q_k \, \tGpp{k} \,\q_k^\text{T} \right\}
            +  \frac{1}{c^2} \, \E_{\V{p}} \left\{\q_k \, \V{l}_k^\text{T} \,\bar{\B\Psi}_k\,  \V{l}_k \, \q_k^\text{T}\right\}
            \nonumber \\
            & \hspace{23mm}- \frac{1}{c^2} \, \E_{\V{p}}\left\{ \q_k \left( \V{l}_k^\text{T}
            \bar{\B\Psi}_k + c^2 \, \tGpk{k} \right) \right\}
            \, \E_{\V{p}} \left\{ \bar{\B\Psi}_k + \Gkk{k}
            \right\}^{-1}\!
            \E_{\V{p}} \left\{ \left( \V{l}_k^\text{T} \bar{\B\Psi}_k + c^2 \,
            \tGpk{k} \right)^\text{T} \q_k^\text{T}  \right\} \Bigg\}
    \end{align}
  \setcounter{equation}{\value{MYtempeqncnt}}
  \vspace*{-6pt}
\end{figure*}

\begin{figure*}[!b]
  \vspace*{4pt}%
  \hrulefill%
  \normalsize%
  \setcounter{MYtempeqncnt}{\value{equation}}%
  \setcounter{equation}{67}%
    \begin{align}\label{eq:Apd_RII_Pbar}
        \bar\lambda_k & \triangleq  \frac{1}{c^2}  {\Big\{}
        \V{l}_k^\text{T}\, \bar{\B{\Psi}}_k \V{l}_k + c^2\,\tGpp{k}
        - \left( \V{l}_k^\text{T} \bar{\B{\Psi}}_k + c^2 \tGpk{k}
        \right) \left(  \bar{\B{\Psi}}_k + c^2\, \Gkk{k}
        \right)^{-1} \!\! \left(  \V{l}_k^\text{T}\,\bar{\B{\Psi}}_k + c^2\, \tGpk{k}
        \right)^\text{T}
        {\Big\}}
    \end{align}
  \setcounter{equation}{\value{MYtempeqncnt}}
  \vspace*{-6pt}
\end{figure*}

\subsection{Proof of Corollary \ref{cor:RII_NLOS_LOS}}
\label{apd:Proofs_Evalu_NLOS/LOS}

\begin{IEEEproof}
The block matrices $\Gkk{k}$ and $\tGpk{k}$ in
\eqref{eq:FIM_Prior_Matrix} for NLOS signals can be written as
    \begin{align*}
        \Gkk{k}  & = \Matrix{cc}{  t^2  &  \V{v}_k^\text{T} \\
                                 \V{v}_k &
                                 \breve{\B\Xi}_{\B\kappa,\B\kappa}^k}
        \quad \text{and} \quad
        \tGpk{k} & = \Matrix{cc}{ w & \breve{{\B\Xi}}_{d,\B\kappa}^k },
    \end{align*}
where $\V{v}_k,\, \breve{{\B\Xi}}_{d,\B\kappa}^k \in
\mathbb{R}^{2L_k-1}$, and $\breve{\B\Xi}_{\B\kappa,\B\kappa}^k \in
\mathbb{R}^{(2L_k-1) \times (2L_k-1)}$. Note that $t^2$ corresponds
to the Fisher information of $\BB{k}{1}$.  When the a priori
knowledge of $\BB{k}{1}$ goes to $\infty$, i.e.,
$g_{\V{b}}(\BB{k}{1}) \rightarrow \delta (\BB{k}{1})$, we claim that
    \begin{align}\label{eq:Apd_Equal_NLOS_LOS}
    & \hspace{-5mm} \lim_{t^2 \rightarrow \infty}
        \left[  \bar{\B{\Psi}}_k + c^2 \Gkk{k} \right]^{-1} \nonumber \\
        & = \Matrix{cc} {  0    &    \V{0}^\text{T}    \\    \V{0}
            & \left( \V{D}_k  \bar{\B{\Psi}}_k \V{D}_k^\text{T} + c^2 \breve{\B\Xi}_{\B\kappa,\B\kappa}^k \right)^{-1} }
        .
    \end{align}

To show this, we partition $\bar{\B\Psi}_k$ as
    \begin{align*}
        \bar{\B\Psi}_k = \Matrix{cc}
                        {   u_k^2         &   \V{k}_k^\text{T} \\
                            \V{k}_k       &   \breve{\bar{\B\Psi}}_k},
    \end{align*}
and then the left-hand-side of \eqref{eq:Apd_Equal_NLOS_LOS} becomes
    \begin{align*}
        \text{LHS} & =
        \lim_{t^2 \rightarrow \infty} \Matrix{cc}{ u_k^2 + c^2 t^2  & \V{k}_k^\text{T} + c^2 \V{v}_k^\text{T} \\ \
        \V{k}_k + c^2 \V{v}_k   & \bar{\B{\Psi}}_k' + c^2 \breve{\B\Xi}_{\B\kappa,\B\kappa}^k }^{-1}
        \nonumber \\
        & = \lim_{t^2 \rightarrow \infty} \Matrix{cc} {A & \V{B} \\ \V{B}^\text{T}  &  \V{C}},
    \end{align*}
where
    \begin{align*}
        A & \triangleq {\bigg[}\, u_k^2 + c^2 t^2 \nonumber \\
                & \hspace{7mm}- \left(\V{k}_k + c^2  \V{v}_k\right)^\text{T}\left( \breve{\bar{\B\Psi}}_k
                + c^2 \breve{\B\Xi}_{\B\kappa,\B\kappa}^k \right)^{-1}
                \left(\V{k}_k + c^2  \V{v}_k\right){\bigg]}^{-1} \!, \\
        \V{B} & \triangleq  - \frac{1}{u_k^2 + c^2 t^2} \left(\V{k}_k + c^2 \V{v}_k\right) \, \V{C}^{-1}, \\
        \noalign{\noindent and \vspace{\jot}}
        \V{C} & \triangleq  {\bigg[}\, \breve{\bar{\B\Psi}}_k
                + c^2 \breve{\B\Xi}_{\B\kappa,\B\kappa}^k\nonumber\\
                &  \hspace{12mm}- \frac{1}{u_k^2 + c^2 t^2}
                  \left(\V{k}_k + c^2 \V{v}_k\right) \left(\V{k}_k + c^2  \V{v}_k\right)^\text{T}
                  {\bigg]}^{-1} \! .
    \end{align*}
When $\BB{k}{1}$ is known, i.e., $t^2 \rightarrow \infty$, we have
$\lim_{t^2 \rightarrow \infty} A = 0$, $\lim_{t^2 \rightarrow
\infty} \V{B} = \V{0}$, and $\lim_{t^2 \rightarrow \infty} \V{C} =
\left[ \breve{\bar{\B\Psi}}_k + c^2 \breve{\B\Xi}_{\B\kappa,
\B\kappa}^k \right]^{-1}$. Notice that $\breve{\bar{\B\Psi}}_k =
\V{D}_k  \bar{\B{\Psi}}_k \V{D}_k^\text{T}$. Hence, we proved our
claim in \eqref{eq:Apd_Equal_NLOS_LOS}.

Substituting \eqref{eq:Apd_Equal_NLOS_LOS} into
\eqref{eq:Apd_tRa_def}, we have
    \begin{align*}
        \lim_{t^2 \rightarrow \infty} \lambda_k
        & =  \frac{1}{c^2} {\bigg\{}
        \V{l}_k^\text{T} \bar{\B{\Psi}}_k  \V{l}_k + c^2 \tGpp{k}
        - \left( \V{l}_k^\text{T} \bar{\B{\Psi}}_k \V{D}_k^\text{T} + c^2 \breve{{\B\Xi}}_{d,\B\kappa}^k \right)
        \nonumber \\
        & \hspace{0mm} \times
        \left(  \V{D}_k \bar{\B{\Psi}}_k \V{D}_k^\text{T} + c^2 \breve{\B\Xi}_{\B\kappa,\B\kappa}^k \right)^{-1}
        \!\!
        \left(  \V{l}_k^\text{T} \bar{\B{\Psi}}_k \V{D}_k^\text{T} + c^2 \breve{{\B\Xi}}_{d,\B\kappa}^k \right)^\text{T}
        {\bigg\}} ,
    \end{align*}
for $k \in \NNL$, which agrees with the RII of LOS signals in
\eqref{eq:Apd_Ra_def}.\footnote{Note that the size of $\Gkk{k}$ and
$\tGpk{k}$ for LOS signals and NLOS signals are different for the
same $L_k$. Indeed, $\breve{\B\Xi}_{\B\kappa, \B\kappa}^k$ and
$\breve{{\B\Xi}}_{d,\B\kappa}^k$ are not associated with
$\BB{k}{1}$, and hence they are in the same form as $\Gkk{k}$ and
$\tGpk{k}$ for LOS signals in \eqref{eq:Apd_Ra_def}.} Hence LOS
signals are equivalent to NLOS with infinite a priori knowledge of
$\BB{k}{1}$ for localization.
\end{IEEEproof}

\begin{figure*}[!b]
  \vspace*{4pt}
  \hrulefill%
  \normalsize%
  \setcounter{MYtempeqncnt}{\value{equation}}%
  \setcounter{equation}{68}%
  \renewcommand{\arraystretch}{1.3}
    \begin{align}\label{eq:Apd_Array_JP}
        \JP = \Matrix{ccccc}{\sum_{n \in \NA}\sum_{k \in \NB} \Gpp{n k} & \sum_{n \in \NA}\sum_{k \in \NB} \B\Xi_{\V{p},\varphi}^{n k} & \B\Xi_{\V{p},1} & \cdots & \B\Xi_{\V{p},\Na} \\
                             \sum_{n \in \NA}\sum_{k \in \NB} {\B\Xi_{\V{p},\varphi}^{n k}}^\text{T} & \sum_{n \in \NA}\sum_{k \in \NB} \Xi_{\varphi,\varphi}^{n k} & \B\Xi_{\varphi,1} & \cdots & \B\Xi_{\varphi,\Na} \\
                             \B\Xi_{\V{p},1}^\text{T} & \B\Xi_{\varphi,1}^\text{T} & \B\Xi_{1} &  &  \\
                             \vdots & \vdots &  & \ddots &  \\
                             \B\Xi_{\V{p},\Na}^\text{T} & \B\Xi_{\varphi,\Na}^\text{T} &  &  & \B\Xi_{\Na}  }
    \end{align}
  \setcounter{equation}{\value{MYtempeqncnt}}
  \vspace*{-6pt}
\end{figure*}

\subsection{Proof of Proposition \ref{thm:EFIM_FarField}}
\label{apd:Proofs_Evalu_FarField}

\begin{IEEEproof}
Note that $\q_k$, $\bar{\B\Psi}_k$, $\Gpp{k}$, $\Gkk{k}$, and
$\Gpk{k}$ are functions of $\V{p}$ when a priori knowledge of the
agent's position is available. Hence we need to take expectation of
them over $\V{p}$ in \eqref{eq:FIM_JD_JP}. After some algebra, we
have the EFIM for the agent's position as
\eqref{eq:Apd_EFIM_WPrior_Pos} shown at the bottom of the page.
\addtocounter{equation}{1}

When the condition in \eqref{eq:Appro_Condition} is satisfied for
the functions $g(\V{p})$'s: 1) $\q_k\,\tGpp{k}\,\q_k^\text{T}$, 2)
$\q_k\,\V{l}_k^\text{T}\,\bar{\B\Psi}_k\,\V{l}_k\,\q_k^\text{T}$, 3)
$\q_k \left( \V{l}_k^\text{T} \bar{\B\Psi}_k + c^2 \,\tGpk{k}
\right)$, and 4) $\bar{\B\Psi}_k + \Gkk{k} $, we can approximate the
expectation of each function over $\V{p}$ in
\eqref{eq:Apd_EFIM_WPrior_Pos} by the function value at the expected
position $\bar{\V{p}}$. Hence the EFIM in
\eqref{eq:Apd_EFIM_WPrior_Pos} can be expressed as
    \begin{align*}
        \JE{\V{p}} & = \GP + \frac{1}{c^2}
        \sum_{k \in \NB} {\Bigg\{}
        \q_k\left(\V{l}_k^\text{T}\, \bar{\B\Psi}_k\,
        \V{l}_k + c^2 \,\tGpp{k}\right)\, \q_k^\text{T}
        \nonumber \\
        & \hspace{10mm}- \q_k \left( \V{l}_k^\text{T}\,\bar{\B\Psi}_k + c^2 \, \tGpk{k} \right)
        \,\left( \bar{\B\Psi}_k + \Gkk{k}\right)^{-1} \nonumber \\
        & \hspace{14.5mm} \times
        \left( \V{l}_k^\text{T}\,\bar{\B\Psi}_k + c^2 \, \tGpk{k}\right)^\text{T} \q_k^\text{T}
        {\Bigg\}} \nonumber \\
        & = \GP + \sum_{k \in \NB} \bar\lambda_k \, \R(\bar{\phi}_k)
        \,,
    \end{align*}
where $\bar\phi_k$ is the angle from anchor $k$ to $\bar{\V{p}}$, and
$\bar\lambda_k$ is given by \eqref{eq:Apd_RII_Pbar} shown at the
bottom of the page. \addtocounter{equation}{1}
Note that all functions are evaluated at $\bar{\V{p}}$.
\end{IEEEproof}

\section{Proofs of the Results in Section \ref{Sec:Array}}
\label{apd:Proofs_Array}

\begin{figure*}[!b]
  \vspace*{4pt}%
  \hrulefill%
  \normalsize%
  \setcounter{MYtempeqncnt}{\value{equation}}%
  \setcounter{equation}{69}%
    \begin{align}\label{eq:Apd_Array_JD}
    \renewcommand{\arraystretch}{1.3}
        \JD = \frac{1}{c^2} \Matrix{ccccccc}{
        \sum_{n \in \NA} \V{G}_n \bar{\B\Lambda}_n \V{G}_n^\text{T} & \sum_{n \in \NA} \V{G}_n \bar{\B\Lambda}_n \V{h}_n^\text{T}  & \V{G}_1 \bar{\B\Lambda}_1    &   \cdots  &   \V{G}_\Na \bar{\B\Lambda}_\Na \\
        \sum_{n \in \NA} \V{h}_n \bar{\B\Lambda}_n \V{G}_n^\text{T} & \sum_{n \in \NA} \V{h}_n \bar{\B\Lambda}_n \V{h}_n^\text{T}  & \V{h}_1 \bar{\B\Lambda}_1    &   \cdots  &   \V{h}_\Na \bar{\B\Lambda}_\Na \\
        \bar{\B\Lambda}_1 \V{G}_1^\text{T}     &   \bar{\B\Lambda}_1 \V{h}_1^\text{T}  &  \bar{\B\Lambda}_1\\
        \vdots  &   \vdots  &  & \ddots \\
        \bar{\B\Lambda}_\Na \V{G}_\Na^\text{T}     &   \bar{\B\Lambda}_\Na   \V{h}_\Na^\text{T}  &  &  &
        \bar{\B\Lambda}_\Na}
    \end{align}
  \setcounter{equation}{\value{MYtempeqncnt}}
  \vspace*{-6pt}
\end{figure*}

\begin{figure*}[!b]
  \vspace*{4pt}%
  \hrulefill%
  \normalsize%
  \setcounter{MYtempeqncnt}{\value{equation}}%
  \setcounter{equation}{71}%
    \begin{align}\label{eq:Apd_Ra_def_AOA}
        \lambda_{n k} & \triangleq \frac{1}{c^2} \left\{
        \V{l}_{n k}^\text{T} \, \bar{\B{\Psi}}_{n k}\, \V{l}_{n k} +
        c^2\,\tGpp{n k} -  \left( \V{l}_{n k}^\text{T} \bar{\B{\Psi}}_{n k} + c^2 \tGpk{n k} \right)
        \left(  \bar{\B{\Psi}}_{n k} + c^2 \Gkk{n k}
        \right)^{-1}\!\!
        \left( \V{l}_{n k}^\text{T} \bar{\B{\Psi}}_{n k} + c^2 \tGpk{n k} \right)^\text{T}
        \right\}
    \end{align}
  \setcounter{equation}{\value{MYtempeqncnt}}
  \vspace*{-6pt}
\end{figure*}

\begin{figure*}[!b]
  \vspace*{4pt}%
  \hrulefill%
  \normalsize%
  \setcounter{MYtempeqncnt}{\value{equation}}%
  \setcounter{equation}{72}%
    \begin{align}\label{eq:Apd_RII_NoPrior}
        \lambda_{n k} = \begin{cases}
                          \V{l}_{n k}^\text{T}\, {\Big\{}  \bar{\B{\Psi}}_{n k}
                            - \left( \bar{\B{\Psi}}_{n k} \V{D}_{n k}^\text{T}  \right)
                            \left( \V{D}_{n k} \bar{\B{\Psi}}_{n k} \V{D}_{n k}^\text{T}
                            \right)^{-1}\!\!
                            \left( \V{D}_{n k} \bar{\B{\Psi}}_{n k}  \right)
                            {\Big\}} \, \V{l}_{n k}/c^2 \,, & \text{LOS} \\
                          0\,, & \text{NLOS}
                        \end{cases}
    \end{align}
  \setcounter{equation}{\value{MYtempeqncnt}}
  \vspace*{-6pt}
\end{figure*}

\subsection{Proof of Theorem \ref{thm:EFIM_NLOS_gen_AOA}}
\label{apd:Proofs_Array_NLOS_gen_AOA}

Note that this proof also incorporates the a priori channel
knowledge. In the absence of this knowledge, the corresponding
results can be obtained by removing $\JP$ that characterizes the a
priori channel knowledge.

Since $\V{p}$ and $\varphi$ are deterministic but unknown, the joint
likelihood function of the random vectors $\V{r}$ and ${\B\theta}$
can be written as
    \begin{align*}
        f(\V{r},\B\theta)
            = f(\V{r}|{\B\theta}) \, f (\B\theta)
            = 
            \prod_{n \in \NA}  \prod_{k \in \NB}
            f(\V{r}_{n k}|{\B\theta}) \, f({\B{\kappa}}_{n k}|\V{p}, \varphi)
            \, .
    \end{align*}
Note that $f({\B{\kappa}}_{n k}|\V{p}, \varphi) =
f({\B{\kappa}}_{n k}|d_{n k})$, and the FIM $\JP$ from $f(\B\theta)$
can be expressed as \eqref{eq:Apd_Array_JP} shown at the bottom of the page, \addtocounter{equation}{1}
where $\Gpp{n k} = \q_{n k} \, \tGpp{n k} \, \q_{n k}^\text{T}$,
$\B\Xi_{\V{p},\varphi}^{n k}  = \q_{n k} \, \tGpp{n k} \, h_{n k}$,
and $\Xi_{\varphi,\varphi}^{n k}  = h_{n k}^2 \, \tGpp{n k}$, in
which
    \begin{align*}
        \tGpp{n k} & \triangleq \FIM{\B\theta}{\V{r}_{n k}|\B\theta}{d_{n k}}{d_{n k}}
        \, .
    \end{align*}
Block matrices $\B\Xi_{\V{p},n}$, $\B\Xi_{\varphi,n}$, and
$\B\Xi_{n}$ correspond to the $n$th antenna in the array, and
they can be further decomposed into block matrices corresponding to
each anchor:
    \begin{align*}
        \B\Xi_{\V{p},n} & = \Matrix{cccc} {\Gpk{n,1} & \Gpk{n,2}
        & \cdots  & \Gpk{n,\Nb}  }, \\
        \B\Xi_{\varphi,n} & = \Matrix{cccc} {\Gvk{n,1}  &  \Gvk{n,2}  &  \cdots  & \Gvk{n,\Nb}}, \\
        \noalign{\noindent and \vspace{\jot}}
        \B\Xi_{n} & = \Diag{ \Gkk{n,1},  \Gkk{n,2}, \ldots, \Gkk{n,\Nb}}
        \, ,
    \end{align*}
where $\Gpk{n k} = \V{q}_{n k} \, \tGpk{n k}$ and $\Gvk{n k} =
h_{n k} \, \tGpk{n k}$, and $\Gkk{n k} =
\FIM{\B\theta}{\B\kappa_{n k}|\V{p}, \varphi}{\B\kappa_{n k}}
{\B\kappa_{n k}}$, in which $\tGpk{n k} =
\FIM{\B\theta}{\B\kappa_{n k}|\V{p},\varphi}{d_{n k}}{\B\kappa_{n k}}$.

Similar to the proof of Theorem \ref{thm:EFIM_NLOS_gen} in Appendix
\ref{apd:Proofs_Evalu_NLOS_gen}, the FIM from observation can be
obtained as \eqref{eq:Apd_Array_JD} shown at the bottom of the
page, \addtocounter{equation}{1}
where
    \begin{align*}
        \V{G}_n & =
        \Matrix{cccc}{\V{q}_{n,1} \V{l}_{n,1}^\text{T} & \V{q}_{n,2} \V{l}_{n,2}^\text{T}& \cdots & \V{q}_{n,\Nb} \V{l}_{n,\Nb}^\text{T} }, \\
        \V{h}_n & = \Matrix{cccc}{{h}_{n,1} \V{l}_{n,1}^\text{T} & {h}_{n,2}
        \V{l}_{n,2}^\text{T} & \cdots & {h}_{n,\Nb} \V{l}_{n,\Nb}^\text{T}}, \\
        \noalign{\noindent and \vspace{\jot}}
        \bar{\B\Lambda}_n & = \Diag{\bar{\B\Psi}_{n,1}, \bar{\B\Psi}_{n,2}, \ldots, \bar{\B\Psi}_{n,\Nb}}
    \end{align*}
correspond to the $n$th antenna as defined in \eqref{eq:Apd_Psi}.

The overall FIM $\JTH$ is the sum of \eqref{eq:Apd_Array_JP} and \eqref{eq:Apd_Array_JD}. By applying the notion of EFI, we have the
$3 \times 3$ EFIM for the position and the orientation as follows
    \begin{align}\label{eq:Apd_proof_AOA_JE}
    \renewcommand{\arraystretch}{1.3}
        \JE{\V{p},\varphi}
            = \sum_{n \in \NA} \sum_{k \in \NB}
                \Matrix{cc}
                {  \lambda_{n k} \, \q_{n k}\, \q_{n k}^\text{T}
                &  \lambda_{n k}\, h_{n k}\,  \q_{n k}  \\
                \lambda_{n k}\, h_{n k} \, \q_{n k}^\text{T}
                & \lambda_{n k}\, h_{n k}^2
                },
    \end{align}
where $\lambda_{n k}$ is given by \eqref{eq:Apd_Ra_def_AOA} shown at
the bottom of the page. \addtocounter{equation}{1}

Note that in the absence of a priori channel knowledge, the above
result is still valid, with the RII of \eqref{eq:Apd_Ra_def_AOA}
degenerating to \eqref{eq:Apd_RII_NoPrior} shown at the bottom of
the page, \addtocounter{equation}{1}
where $\V{D}_{n k} = \Matrix{ccc}{\V{0} & \V{I}_{2L_{n k}-1}}$.

\subsection{Proof of Proposition \ref{thm:orien_center}}
\label{apd:Proofs_Array_orien_center}

Since $\V{q}\, \V{q}^\text{T} $ is always positive semi-definite, we
need to simply prove that there exists a unique $\V{p}^*$ such that
$\V{q}^* = \V{0}$.

\begin{IEEEproof}
Let $\V{p}$ be an arbitrary reference point, and
    \begin{align*}
        \V{p}^* = \V{p} + \V{g}(\varphi) \,,
    \end{align*}
where $\V{g}(\varphi) = [\,g_x(\varphi) \;\; g_y(\varphi)
\,]^\text{T}$, and $g_x(\varphi)$ and $g_y(\varphi)$ denote the
relative distance in $x$ and $y$ directions, respectively.  Then,
$h_{n k}$ corresponding to $\V{p}$ can be written as a sum of two
parts
    \begin{align*}
        h_{n k} = {h}_{n k}^* + \tilde{h}_{n k}\,,
    \end{align*}
where ${h}_{n k}^*$ corresponds to $\V{p}^*$
    \begin{align*}
        {h}_{n k}^* =
            \frac{d}{d \varphi} \Delta x_n(\V{p}^* , \varphi) \, \cos\phi_{n k}
            + \frac{d}{d \varphi} \Delta y_n(\V{p}^*, \varphi) \, \sin\phi_{n k}
        \,,
    \end{align*}
and
    \begin{align*}
        \tilde{h}_{n k}
        & =   \frac{d}{d \varphi} g_x(\varphi) \, \cos\phi_{n k}
        + \frac{d}{d \varphi} g_y(\varphi) \, \sin\phi_{n k}
        \nonumber \\
        & \triangleq  \dot{g}_x \, \cos\phi_{n k} + \dot{g}_y \, \sin\phi_{n k}
        = \dot{\V{g}}^\text{T} \, \V{q}_{n k}
        \,.
    \end{align*}
Hence, $\V{q}$ corresponding to the reference position $\V{p}$ is
given by
    \begin{align}\label{eq:Apd_q}
        \V{q} = \underbrace{\sum_{n \in \NA} \sum_{k \in \NB} \lambda_{n k}
        \,{h}_{n k}^*\,\q_{n k}}_{\triangleq \, \V{q}^*}
        + \underbrace{\sum_{n \in \NA} \sum_{k \in \NB} \lambda_{n k}
        \,\tilde{h}_{n k}\,\q_{n k}}_{\triangleq
        \, \tilde{\V{q}}} \,,
    \end{align}
and $\tilde{\V{q}}$ can be written as
    \begin{align}\label{eq:Apd_tilde_q}
        \tilde{\V{q}}
        & = \sum_{n \in \NA} \sum_{k \in \NB} \q_{n k}^\text{T} \,
        \dot{\V{g}} \, \lambda_{n k} \, \q_{n k}
        \nonumber \\
        & = \sum_{n \in \NA} \sum_{k \in \NB} \lambda_{n k} \,
        \q_{n k} \,\q_{n k}^\text{T} \, \dot{\V{g}}
        = \sum_{n \in \NA} \V{J}_{\text{e},n} \, \dot{\V{g}} \,.
    \end{align}
Since $\sum_{n \in \NA} \V{J}_{\text{e},n} \succ \V{0}$, we have
$\V{q}^* = \V{0}$ if and only if
    \begin{align*}
        \dot{\V{g}} = \left( \sum_{n \in \NA} \V{J}_{\text{e},n} \right)^{-1}
        \!\!\V{q} \, ,
    \end{align*}
implying that there exists only one $\dot{\V{g}}$, and hence only
one $\V{g}(\varphi)$, such that $\V{q}^* = \V{0}$.  Therefore, the
orientation center $\V{p}^*$ is unique.
\end{IEEEproof}

\begin{figure*}[!b]
  \vspace*{4pt}%
  \hrulefill%
  \normalsize%
  \setcounter{MYtempeqncnt}{\value{equation}}%
  \setcounter{equation}{77}%
    \begin{align}\label{eq:Apd_TDOA_JTH}
    \renewcommand{\arraystretch}{1.3}
        \JTH  = \frac{1}{c^2} \Matrix{ccccc}{
        \sum_{k \in \NB} \V{G}_k \bar{\B\Psi}_k \V{G}_k^\text{T}         &   \sum_{k \in \NB} \V{G}_k \bar{\B\Psi}_k \V{l}_k  &   \V{G}_1 \bar{\B\Psi}_1     & \cdots &  \V{G}_{\Nb} \bar{\B\Psi}_{\Nb} \\
        \sum_{k \in \NB} \V{l}_k^\text{T} \bar{\B\Psi}_k \V{G}_k         &   \sum_{k \in \NB} \V{l}_k^\text{T} \bar{\B\Psi}_k \V{l}_k  & \V{l}_1^\text{T} \bar{\B\Psi}_1   & \cdots &  \V{l}_{\Nb}^\text{T} \bar{\B\Psi}_{\Nb}\\
        \bar{\B\Psi}_1 \V{G}_1^\text{T}     & \bar{\B\Psi}_1 \V{l}_1   &   \bar{\B\Psi}_1\\
        \vdots  & \vdots    & & \ddots \\
        \bar{\B\Psi}_\Nb \V{G}_\Nb^\text{T} & \bar{\B\Psi}_\Nb \V{l}_\Nb
        & & & \bar{\B\Psi}_\Nb
        } + \JP
    \end{align}
  \setcounter{equation}{\value{MYtempeqncnt}}
  \vspace*{-6pt}
\end{figure*}

\begin{figure*}[!b]
  \vspace*{4pt}%
  \hrulefill%
  \normalsize%
  \setcounter{MYtempeqncnt}{\value{equation}}%
  \setcounter{equation}{78}%
    \begin{align}\label{eq:Apd_TDOA_JTH_AOA-B}
    \renewcommand{\arraystretch}{1.3}
        \JTH = \frac{1}{c^2} \Matrix{ccccccc}{
        \sum_{n \in \NA} \V{G}_n \bar{\B\Lambda}_n \V{G}_n^\text{T} & \sum_{n \in \NA} \V{G}_n \bar{\B\Lambda}_n \V{h}_n^\text{T} & \sum_{n \in \NA} \V{G}_n \bar{\B\Lambda}_n \V{l}_n^\text{T} & \V{G}_1 \bar{\B\Lambda}_1     &   \cdots  &   \V{G}_\Na \bar{\B\Lambda}_\Na \\
        \sum_{n \in \NA} \V{h}_n \bar{\B\Lambda}_n \V{G}_n^\text{T} & \sum_{n \in \NA} \V{h}_n \bar{\B\Lambda}_n \V{h}_n^\text{T} & \sum_{n \in \NA} \V{h}_n \bar{\B\Lambda}_n \V{l}_n^\text{T} & \V{h}_1 \bar{\B\Lambda}_1     &   \cdots  &   \V{h}_\Na \bar{\B\Lambda}_\Na \\
        \sum_{n \in \NA} \V{l}_n \bar{\B\Lambda}_n \V{G}_n^\text{T} & \sum_{n \in \NA} \V{l}_n \bar{\B\Lambda}_n \V{h}_n^\text{T} & \sum_{n \in \NA} \V{l}_n \bar{\B\Lambda}_n \V{l}_n^\text{T} & \V{l}_1 \bar{\B\Lambda}_1     &   \cdots  &   \V{l}_\Na \bar{\B\Lambda}_\Na\\
        \bar{\B\Lambda}_1 \V{G}_1^\text{T}     &   \bar{\B\Lambda}_1 \V{h}_1^\text{T}  & \bar{\B\Lambda}_1 \V{l}_1^\text{T} & \bar{\B\Lambda}_1\\
        \vdots  &   \vdots  &  \vdots & & \ddots \\
        \bar{\B\Lambda}_\Na \V{G}_\Na^\text{T}     &   \bar{\B\Lambda}_\Na   \V{h}_\Na^\text{T}  & \bar{\B\Lambda}_\Na \V{h}_\Na^\text{T} &  &  &
        \bar{\B\Lambda}_\Na}  + \JP
    \end{align}
  \setcounter{equation}{\value{MYtempeqncnt}}
  \vspace*{-6pt}
\end{figure*}

\subsection{Proof of Corollary \ref{thm:SPEB_Relation}}
\label{apd:Proofs_Array_SPEB_Relation}

\begin{IEEEproof}
We first prove that the SOEB is independent of the reference point
$\V{p}$. It is equivalent to show that the EFI for the orientation
given by \eqref{eq:Anal_EFIM_AOA_phi} equals the EFI for the
orientation based on $\V{p}^*$, given by
    \begin{align*}
        J_\text{e}^*(\varphi)
        = \sum_{n \in \NA} \sum_{k \in \NB} \lambda_{n k}\, {h_{n k}^{*2}}
        \, .
    \end{align*}

Let $\V{J} = \sum_{n \in \NA} \V{J}_{\text{e},n}$. From
\eqref{eq:Apd_q} and \eqref{eq:Apd_tilde_q}, we have $\V{q} =
\tilde{\V{q}} = \V{J} \, \dot{\V{g}}$, and hence
    \begin{align*}
        \V{q}^\text{T} \V{J}^{-1} \V{q}
        & = \tilde{\V{q}}^\text{T} \V{J}^{-1} \tilde{\V{q}}
        = \tilde{\V{q}}^\text{T}  \, \dot{\V{g}}
        = \sum_{n \in \NA} \sum_{k \in \NB} \lambda_{n k}\, \tilde{h}_{n k}^2
        \,.
    \end{align*}
On the other hand, we also have
    \begin{align*}
        \sum_{n \in \NA} \sum_{k \in \NB} \lambda_{n k} {h}_{n k}^* \, \tilde{h}_{n k}
        = \V{q}^* \, \dot{\V{g}}
        = 0
        \,.
    \end{align*}
Therefore, we can verify that the EFI for the orientation in
\eqref{eq:Anal_EFIM_AOA_phi}
    \begin{align}\label{eq:Apd_proof_SOEB_2}
         J_\text{e}(\varphi)
         & = \sum_{n \in \NA} \sum_{k \in \NB}  \lambda_{n k} ( {h}_{n k}^* + \tilde{h}_{n k} )^2
                    - \tilde{\V{q}}^\text{T} \V{J}^{-1} \tilde{\V{q}} \nonumber \\
         & = \sum_{n \in \NA} \sum_{k \in \NB}  \lambda_{n k} {h}_{n k}^{*2}
         + 2 \sum_{n \in \NA} \sum_{k \in \NB}  \lambda_{n k} {h}_{n k}^{*}
         \tilde{h}_{n k}\nonumber\\
         & = J_\text{e}^*(\varphi)
         \, .
    \end{align}
This shows that the EFI for the orientation is independent of the
reference point, and thus is the SOEB.

We next derive the SPEB for any reference point given in
\eqref{eq:Anal_P_Pstar}. The $3\times3$ EFIM in
\eqref{eq:Apd_proof_AOA_JE} can be written, using \eqref{eq:Apd_q}
and \eqref{eq:Apd_proof_SOEB_2}, as
    \begin{align*}
        \JE{\V{p},\varphi} = \Matrix{cc}
            {   \V{J}               &    \tilde{\V{q}}  \\
                \tilde{\V{q}}^\text{T}     &    J_\text{e}(\varphi)
                + \tilde{\V{q}}^\text{T} \V{J}^{-1} \tilde{\V{q}}  }.
    \end{align*}
Using the equation of Shur's complement \cite{HorJoh:B85}, we have
    \begin{align}\label{eq:Apd_AOA_IJe}
        \V{J}^{-1}_\text{e}(\V{p}) & = \V{J}^{-1} + \frac{1}{J_\text{e}(\varphi)}
        \, \left( \V{J}^{-1} \tilde{\V{q}} \right)  \left( \V{J}^{-1} \tilde{\V{q}}
        \right)^\text{T} \nonumber \\
        & = \V{J}^{-1} + \frac{1}{J_\text{e}(\varphi)} \,
        \dot{\V{g}}\,\dot{\V{g}}^\text{T} \, .
    \end{align}
Since the translation $\V{g}(\varphi)$ can be represented as
    \begin{align*}
        \V{g}(\varphi) = \|\V{p} - \V{p}^*\| \,
        \Matrix{c}{\cos(\varphi+\varphi_0) \\ \sin(\varphi+\varphi_0) },
    \end{align*}
where $\varphi_0$ is a constant angle, we have $\| \dot{\V{g}} \| =
\|\V{p} - \V{p}^*\|$ . Then, by taking the trace of both sides of
\eqref{eq:Apd_AOA_IJe}, we obtain
    \begin{align*}
        \PEB(\V{p})
        & = \PEB(\V{p}^*) + \frac{\dot{\V{g}}^\text{T}\, \dot{\V{g}}}{J_\text{e}(\varphi)}
        \nonumber \\
        & = \PEB(\V{p}^*) + {\|\V{p} - \V{p}^* \|^2}\cdot {\PEB(\varphi)}
        \, .
    \end{align*}
\end{IEEEproof}

\subsection{Proof of Proposition \ref{Thm:Array_Center}}
\label{apd:Proofs_Array_Center}

\begin{IEEEproof}
Take the array center $\V{p}_0$ as the reference point, and we have
    \begin{align*}
        \sum_{n \in \NA} h_{n k}
        & = \sum_{n \in \NA} \frac{d}{d \varphi} \Delta x_n(\V{p}_0,\varphi) \, \cos\phi_{n k}
        \nonumber \\
        & \hspace{15mm} + \sum_{n \in \NA} \frac{d}{d \varphi} \Delta y_n(\V{p}_0,\varphi) \, \sin\phi_{n k} \nonumber \\
        & =  \frac{d}{d \varphi} \left(\sum_{n \in \NA} \Delta x_n(\V{p}_0,\varphi) \right) \, \cos\phi_{n k}
        \nonumber \\
        & \hspace{15mm} + \frac{d}{d \varphi} \left( \sum_{n \in \NA} \Delta y_n(\V{p}_0,\varphi) \right) \, \sin\phi_{n k}
        \nonumber \\
        & =  0
        \, .
    \end{align*}
Consequently,
    \begin{align*}
        \V{q} = \sum_{n \in \NA} \sum_{k \in \NB} \lambda_{k} h_{n k} \, \q_{k}
                = \sum_{k \in \NB} \left( \sum_{n \in \NA} h_{n k} \right)  \lambda_{k} \,
                \q_{k} = 0
        \, ,
    \end{align*}
implying $\V{p}_0=\V{p}^*$, i.e., the array center is the
orientation center.
\end{IEEEproof}

\begin{figure*}[!b]
  \vspace*{4pt}%
  \hrulefill%
  \normalsize%
  \setcounter{MYtempeqncnt}{\value{equation}}%
  \setcounter{equation}{79}%
    \begin{align}\label{eq:Apd_Array_JP_AOA-B}
    \renewcommand{\arraystretch}{1.3}
        \JP = \Matrix{cccccc}{
            \sum_{n \in \NA}\sum_{k \in \NB} \Gpp{n k} & \sum_{n \in \NA}\sum_{k \in \NB} \B\Xi_{\V{p},\varphi}^{n k}                               &  \V{0}    & \B\Xi_{\V{p},1} & \cdots & \B\Xi_{\V{p},\Na} \\
            \sum_{n \in \NA}\sum_{k \in \NB} {\B\Xi_{\V{p},\varphi}^{n k}}^\text{T} &   \sum_{n \in \NA}\sum_{k \in \NB} \Xi_{\varphi,\varphi}^{n k}    &  {0}      & \B\Xi_{\varphi,1} & \cdots & \B\Xi_{\varphi,\Na} \\
            \V{0}^\text{T}    & 0         &       \Xi_{\text{B}}    &   \V{0}^\text{T}    &   \cdots      & \V{0}^\text{T}\\
            \B\Xi_{\V{p},1}^\text{T} & \B\Xi_{\varphi,1}^\text{T}   &   \V{0}           &   \B\Xi_{1}   &  &  \\
            \vdots & \vdots  &   \vdots & & \ddots &  \\
            \B\Xi_{\V{p},\Na}^\text{T} & \B\Xi_{\varphi,\Na}^\text{T} &  \V{0}  & &  & \B\Xi_{\Na}
            }
    \end{align}
  \setcounter{equation}{\value{MYtempeqncnt}}
  \setcounter{MYtempeqncnt}{\value{equation}}%
  \vspace*{4pt}%
  \hrulefill%
  \setcounter{equation}{80}%
    \begin{align}\label{eq:Apd_TDOA_Array1}
    \renewcommand{\arraystretch}{1.3}
        \V{J}_\text{e}^\text{Array-B}
            =  \Matrix{ccc}{ \sum_{n \in \NA} \sum_{k \in \NB} \bar\lambda_{n k} \V{q}_{n k} \V{q}_{n k}^\text{T} + \GP
                            & \sum_{n \in \NA} \sum_{k \in \NB} \bar\lambda_{n k} h_{n k}  \V{q}_{n k}
                            & \sum_{n \in \NA} \sum_{k \in \NB} \bar\lambda_{n k} \V{q}_{n k} \\
                            \sum_{n \in \NA} \sum_{k \in \NB} \bar\lambda_{n k} h_{n k}  \V{q}_{n k}^\text{T}
                            & \sum_{n \in \NA} \sum_{k \in \NB} \bar\lambda_{n k} h_{n k}^2  +
                            \Gph
                            & \sum_{n \in \NA} \sum_{k \in \NB} \bar\lambda_{n k} h_{n k}  \\
                            \sum_{n \in \NA} \sum_{k \in \NB} \bar\lambda_{n k} \V{q}_{n k}^\text{T}
                            & \sum_{n \in \NA} \sum_{k \in \NB} \bar\lambda_{n k} h_{n k}
                            & \sum_{n \in \NA} \sum_{k \in \NB} \bar\lambda_{n k} + \Xi_{\text{B}} }
    \end{align}
  \setcounter{equation}{\value{MYtempeqncnt}}
\end{figure*}

\section{Proofs of the Results in Section \ref{Sec:ClockBias}}
\label{apd:Proofs_ClockBias}

\subsection{Proof of Theorem \ref{thm:EFIM_NLOS_gen_TDOA}}
\label{apd:Proofs_ClockBias_NLOS_gen_TDOA}

In the presence of a time offset, the FIM can be written as
\eqref{eq:Apd_TDOA_JTH} shown at the bottom of the page,
\addtocounter{equation}{1} where
    \begin{align*}
        \renewcommand{\arraystretch}{1.3}
        \JP = \Matrix{ccccc}{\sum_{k \in \NB} \Gpp{k} & \V{0} & \Gpk{1} & \cdots & \Gpk{\Nb} \\
                             \V{0}^\text{T} & \Xi_{\text{B}} & \V{0}^\text{T} & \cdots & \V{0}^\text{T} \\
                             \Gpk{1\;\,\text{T}} & \V{0} & \Gkk{1} &  &  \\
                             \vdots & \vdots &  & \ddots &  \\
                             \Gpk{\Nb\;\,\text{T}} & \V{0} &  &  & \Gkk{\Nb}}.
    \end{align*}
Applying the notion of EFI, we obtain the $3 \times 3$ EFIM
    \begin{align*}
        \renewcommand{\arraystretch}{1.3}
        \JE{\V{p},B} = \Matrix{cc}
                {   \sum_{k \in \NB} \lambda_k \, \q_k \,\q_k^\text{T}  &  \sum_{k \in \NB} \lambda_k \, \q_k \\
                    \sum_{k \in \NB} \lambda_k \, \q_k^\text{T}       &  \sum_{k \in \NB} \lambda_k + \Xi_{\text{B}}
                },
    \end{align*}
where $\lambda_k$ is given by \eqref{eq:Apd_tRa_def}, and another
step of EFI leads to \eqref{eq:Anal_EFIM_TDOA} and
\eqref{eq:Anal_EFIM_TDOA_B}.

\subsection{Proof of Theorem \ref{thm:EFIM_NLOS_gen_AOA-B}}
\label{apd:Proofs_ClockBias_NLOS_gen_AOA-B}

We consider orientation-unaware case, whereas orientation-aware case
is a special case with a reduced parameter set. The FIM using an
antenna array can be written as \eqref{eq:Apd_TDOA_JTH_AOA-B} shown
at the bottom of the page, \addtocounter{equation}{1}
where $\V{l}_n  = \Matrix{cccc}{\V{l}_{n,1}^\text{T} &
\V{l}_{n,2}^\text{T} & \cdots & \V{l}_{n,\Nb}^\text{T} }$, and $\JP$
is given by \eqref{eq:Apd_Array_JP_AOA-B} shown at the bottom of
the page. \addtocounter{equation}{1}
Applying the notion of EFI to $\JTH$, we obtain the $4 \times 4$
EFIM in \eqref{eq:Anal_EFIM_TDOA/AOA}.

\subsection{Proof of Corollary \ref{thm:EFIM_AOA-B_Far}}
\label{apd:Proofs_ClockBias_AOA-B_Far}

We incorporate the a priori knowledge of the array center and
orientation into \eqref{eq:Anal_EFIM_TDOA/AOA}, and obtain the EFIM
in far-field scenarios as \eqref{eq:Apd_TDOA_Array1} shown at the
bottom of the page. \addtocounter{equation}{1}
Recall that in far-field scenarios, $\V{p}_0 = \V{p}^*$, implying
that $\sum_{n \in \NA} \sum_{k \in \NB} \lambda_{n k} h_{n k}
\V{q}_{n k} =\V{0}$ and $\sum_{n \in \NA} \sum_{k \in \NB}
\lambda_{n k} h_{n k} = 0$. Also, we have $\bar\lambda_{n k} =
\bar\lambda_k$ and $\bar\phi_{n k} = \bar\phi_k$ for all $n$, and
hence the EFIM can be written as \eqref{eq:Apd_TDOA_Array2} shown at
the top of the next page, \addtocounter{equation}{1}
where $\bar{h}_{n k}$ and $\bar{\V{q}}_k$ is a function of
$\bar{\V{p}}_0$.

\begin{figure*}[!t]
  \vspace*{4pt}%
  \normalsize%
  \setcounter{MYtempeqncnt}{\value{equation}}%
  \setcounter{equation}{81}%
  \renewcommand{\arraystretch}{1.3}
  \begin{align}\label{eq:Apd_TDOA_Array2}
  \renewcommand{\arraystretch}{1.3}
      \V{J}_\text{e}^\text{Array-B}
          =  \Matrix{ccc}{ \Na \, \sum_{k \in \NB} \bar\lambda_{n k} \R(\bar\phi_k)  + \GP   & \V{0}
                          & \Na \sum_{k \in \NB} \bar\lambda_k \bar{\V{q}}_k \\
                          \V{0}^\text{T}
                          & \sum_{n \in \NA} \sum_{k \in \NB}  \bar\lambda_k \bar{h}_{n k}^2 +
                          \Gph
                          & 0 \\
                          \Na \sum_{k \in \NB} \bar\lambda_k \bar{\V{q}}_k^\text{T}
                          & 0
                          & \Na \sum_{k \in \NB} \bar\lambda_k + \Xi_{\text{B}} }
  \end{align}
  \hrulefill%
  \setcounter{equation}{\value{MYtempeqncnt}}
  \vspace*{-6pt}
\end{figure*}

%
%

\section*{Acknowledgments}\label{Sec:Achn}
The authors would like to thank R. G. Gallager, A. Conti, H.
Wymeersch, W. M. Gifford, and W. Suwansantisuk for their valuable
suggestion and careful reading of the manuscript. We would also like
to thank the anonymous reviewers for their constructive comments.

\bibliographystyle{IEEEtran}

\begin{IEEEbiographynophoto}
{Yuan Shen} (S'05) received his B.S.\ degree (with highest honor) from Tsinghua University, China, in 2005, and S.M.\ degree from the Massachusetts Institute of Technology (MIT) in 2008, both in electrical engineering. 

Since 2005, he has been with Wireless Communications and Network Science Laboratory at MIT, where he is now a Ph.D.\ candidate. He was with the Hewlett-Packard Labs, CA, in winter 2009, the Corporate R\&D of Qualcomm Inc., CA, in summer 2008, and the Intelligent Transportation Information System Laboratory, Tsinghua University, China, from 2003 to 2005. His research interests include communication theory, information theory, and statistical signal processing. His current research focuses on wideband localization, cooperative networks, and ultra-wide bandwidth communications.

Mr.~Shen served as a member of the Technical Program Committee (TPC) for the IEEE Global Communications Conference (GLOBECOM) in 2010, 
the IEEE International Conference on Communications (ICC) in 2010, and the IEEE Wireless Communications \& Networking Conference (WCNC) in 2009 and 2010. He received the Ernst A.~Guillemin Thesis Award (first place) for the best S.M.~thesis from the Department of Electrical Engineering and Computer Science at MIT in 2008, the Roberto Padovani Scholarship from Qualcomm Inc.~in 2008, the Best Paper Award from the IEEE WCNC in 2007, and the Walter A.~Rosenblith Presidential Fellowship from MIT in 2005.
\end{IEEEbiographynophoto} 


\begin{IEEEbiographynophoto}
{Moe Z. Win} 
(S'85-M'87-SM'97-F'04)
received both the Ph.D.\ in Electrical Engineering and M.S.\ in Applied Mathematics
as a Presidential Fellow at the University of Southern California (USC) in 1998.
He received an M.S.\ in Electrical Engineering from USC in 1989, and a B.S.\ ({\em magna cum laude})
in Electrical Engineering from Texas A\&M University in 1987.

Dr.\ Win is an Associate Professor at the Massachusetts Institute of Technology (MIT).
Prior to joining MIT, he was at AT\&T Research
Laboratories for five years and at the Jet Propulsion Laboratory for seven years.
His research encompasses developing fundamental theories, designing algorithms, and
conducting experimentation for a broad range of real-world problems.
His current research topics include location-aware networks,
time-varying channels, multiple antenna systems, ultra-wide bandwidth
systems, optical transmission systems, and space communications systems.

Professor Win is an IEEE Distinguished Lecturer and
        elected Fellow of the IEEE, cited for ``contributions to wideband wireless transmission.''
He was honored with
        the IEEE Eric E. Sumner Award (2006), an IEEE Technical Field Award for
        ``pioneering contributions to ultra-wide band communications science and technology.''
Together with students and colleagues, his papers have received several awards including
        the IEEE Communications Society's Guglielmo Marconi Best Paper Award (2008)
    and the IEEE Antennas and Propagation Society's Sergei A. Schelkunoff Transactions Prize Paper Award (2003).
His other recognitions include
        the Laurea Honoris Causa from the University of Ferrara, Italy (2008),
        the Technical Recognition Award of the IEEE ComSoc Radio Communications Committee (2008),
        Wireless Educator of the Year Award (2007),
        the Fulbright Foundation Senior Scholar Lecturing and Research Fellowship (2004),
        the U.S. Presidential Early Career Award for Scientists and Engineers (2004),
        the AIAA Young Aerospace Engineer of the Year (2004),
    and the Office of Naval Research Young Investigator Award (2003).

Professor Win has been actively involved in organizing and chairing
a number of international conferences. He served as
    the Technical Program Chair for
        the IEEE Wireless Communications and Networking Conference in 2009,
        the IEEE Conference on Ultra Wideband in 2006,
        the IEEE Communication Theory Symposia of ICC-2004 and Globecom-2000,
        and
        the IEEE Conference on Ultra Wideband Systems and Technologies in 2002;
    Technical Program Vice-Chair for
        the IEEE International Conference on Communications in 2002; and
    the Tutorial Chair for
        ICC-2009 and
        the IEEE Semiannual International Vehicular Technology Conference in Fall 2001.
He was
    the chair (2004-2006) and secretary (2002-2004) for
        the Radio Communications Committee of the IEEE Communications Society.
Dr.\ Win is currently
    an Editor for {\scshape IEEE Transactions on Wireless Communications.}
He served as
    Area Editor for {\em Modulation and Signal Design} (2003-2006),
    Editor for {\em Wideband Wireless and Diversity} (2003-2006), and
    Editor for {\em Equalization and Diversity} (1998-2003),
        all for the {\scshape IEEE Transactions on Communications}.
He was Guest-Editor
        for the
        {\scshape Proceedings of the IEEE}
        (Special Issue on UWB Technology \& Emerging Applications) in 2009 and
        {\scshape IEEE Journal on Selected Areas in Communications}
        (Special Issue on Ultra\thinspace-Wideband Radio in Multiaccess
        Wireless Communications) in 2002.
\end{IEEEbiographynophoto}

\end{document}